\documentclass[11pt,letterpaper]{article}
\pdfoutput=1
\usepackage{jheppub}
\usepackage{amsmath}
\usepackage{bm}
\usepackage{comment}
\usepackage{rotfloat}
\usepackage{amssymb}
\usepackage{extarrows}
\usepackage{tikz-cd}
\usepackage{bbm}

\hypersetup{
    colorlinks=true,       
    linkcolor=red,          
    citecolor=blue,        
    filecolor=magenta,      
    urlcolor=blue           
}

\def\B{{\cal B}}
\def\F{{\cal F}}
\def\N{{\cal N}}

\def\tr{\,{\rm tr}\,}

\def\nf{n_{\rm f}}

\def\half{\tfrac{1}{2}}

\def\R{\mathbb{R}}

\def\Z{\mathbb{Z}}

\def\1{\mathbbm{1}}
\renewcommand{\bar}[1]{\overline{#1}}

\title{Graded Hilbert spaces,  quantum distillation  and  connecting  SQCD to  QCD}

\author[a]{Mithat \"Unsal}
\emailAdd{unsal.mithat@gmail.com}
\affiliation[a]{Department of Physics, North Carolina State University, Raleigh, NC, USA}

\abstract{The dimension of the Hilbert space of QFT scales exponentially with the  volume of the space in which the theory lives, yet in supersymmetric theories, one can define a graded dimension (such as the supersymmetric index) that counts just the number of bosonic minus fermionic ground states. Can we make this observation useful in non-supersymmetric QFTs in four dimensions? 
In this work, 
we construct {\it symmetry graded state sums} for a variety of non-supersymmetric theories. 
 Among the theories we consider is one that is remarkably close to QCD: Yang--Mills theory with $N_f = N_c$ fundamental  Dirac fermions and one  adjoint  Weyl fermion, QCD(F/adj). This theory can be obtained from SQCD by decoupling scalars and carry exactly the same anomalies. 
 Despite the  existence of fundamental fermions, 
the theory possess an exact 0-form  color-flavor center (CFC)  symmetry for a particular grading/twist  under which Polyakov loop is a genuine order parameters.  
 By a two-loop analysis, we prove that CFC-symmetry remains unbroken at small $\beta $ due to  grading. 
 Chiral symmetry  is spontaneously broken  within the domain of validity of semi-classics on $\R^3 \times S^1$  in a pattern identical to $N_f=N_c$ SQCD on $\R^4$ and the two regimes are  adiabatically connected. 
 The vacuum structures of the theory on $\R^4$ and   $\R^3 \times S^1$ are controlled by the same  mixed 't Hooft anomaly condition, implying a remarkable persistent order.  
}

\begin{document}

\maketitle

\noindent

\section{The general  idea of quantum distillation and summary}
This work  is a collection of ideas to determine  the non-perturbative   dynamics and phase structure  of a class of  non-supersymmetric QFTs,   including QCD.  The story exploits the  mapping  between 
the Hilbert space  and Hamiltonian  formalisms by constructing fairly exotic looking {\it  symmetry graded state sums}, and their images in  the  path integral  formalism. 
Despite the fact that the theories of interest here are non-supersymmetric and in the thermo-dynamic limit,  
our line of thinking can be traced back to  two sources which possess qualities opposed to what we desire:  
Supersymmetric indices \cite{Witten:1982df},   and  standard singularity theorems in the theory of phase transitions \cite{ Yang:1952be, Lee:1952ig, Fisher}. 
Our intention is to construct  graded state sums that avoid Lee--Yang-Fisher type singularities  in generalized partition functions in non-supersymmetric theories. This class of ideas goes under the umbrella term  ``adiabatic continuity" \cite{Unsal:2008ch, Dunne:2016nmc}.  Adiabatic continuity   is  a stronger version of what is called persistent order in condensed matter physics, or persistent mixed 't Hooft anomaly in QFT.

\vspace{3mm} 
Thermal phase transitions in quantum field theory are probed with the singularity structure of the  partition function
\begin{align}
{\cal Z}(\beta)= \tr [e^{- \beta H}]
\label{partitionf}
\end{align}
where $\beta$ is the inverse temperature, and $H$ is the Hamiltonian.  Let ${\cal H}$ denote the Hilbert space. 
 In the $\beta \rightarrow \infty$ limit, ${\cal Z}(\beta)$  receives dominant contributions from  ground states and low lying states.  It is an uncontaminated quantity, but strongly coupled QFTs are not  usually amenable to analytic treatment in this limit. As $\beta \rightarrow 0 $,  asymptotically free theories become weakly coupled, 
  but the state sum  is extremely contaminated. It  essentially  receives contributions on the same footing from  entire Hilbert space $\cal H$ .  
 In this regime, it is impossible to isolate and understand the role of ground states and low lying states.   Furthermore, 
 there is in general a phase transition at some $\beta_c$. One may think that this phase transition is due to the growth in the density of states in ${\cal H}$.

\vspace{3mm} 
A   well-known way to avoid phase transition in supersymmetric gauge theories is to construct a $\Z_2$-graded state sum \cite{Witten:1982df}:
\begin{align}
I(\beta )= \tr [(-1)^F e^{-\beta  H}] 
\label{reg-W}
\end{align} 
where $F$ is fermion number modulo two, which counts the number of supersymmetric vacua in supersymmetric QFTs .  
One way to interpret the supersymmetric index  \eqref{reg-W} is as follows: If we were to evaluate $\tr [1]$ over the Hilbert space $\cal H= B \oplus F$, 
we would in fact count the dimension of the Hilbert space which grows exponentially with the volume of the space  
in which the theory  lives.  
Yet $\lim_{\beta  \rightarrow 0} I(\beta )$  may be interpreted as a {\it graded dimension} of the Hilbert space. 
If  the spectrum of the supersymmetric QFT is rendered discrete, supersymmetry guarantees that all states with $E> 0$ are Bose-Fermi paired and their  contribution to the graded sum is zero. Hence, \eqref{reg-W} is just a pure number independent of $\beta $,  and merely counts the bosonic minus fermionic ground states up to a sign.   Importantly, \eqref{reg-W} no longer scales with ${\rm dim} \; [{\cal H}]$.

\vspace{3mm}
In general,  we cannot carry  over the precise level-by-level  Bose-Fermi  cancellations  pertinent to supersymmetric theory to non-supersymmetric theory.\footnote{There are remarkable exceptions. See first example in the next subsection.
  }
There are no such luxuries in real life.  We will, however, state what $(-1)^F$ achieves differently and aim to carry it over  to non-supersymmetric theory: 
\begin{itemize}
\item {$(-1)^F$ is a grader over the Hilbert space  ${\cal H}$ and the graded state sum  distills a sub-set of states, such that $\tr [(-1)^F e^{-\beta  H}]$  is an analytic function of $\beta $.   }
\end{itemize}

\vspace{3mm}
The idea of {\it quantum distillation} of Hilbert space aims to generalize the partition function without changing the Hamiltonian $H$  and Hilbert space ${\cal H}$ and obtain  new state sums which   {\it i)} reduce the state sum in magnitude due to cancellations between the states, and 
 {\it ii)}  effectively represent a subset of states in the Hilbert space \cite{Kanazawa:2019tnf, Dunne:2018hog}. This  is not a ``projecting out"  procedure, since ${\cal H}$ is still the same. Rather,  quantum distillation  is a  useful sign problem over the state sum in the Hamiltonian formulation. 
We define the subset of states in the Hilbert space  which do survive after  graded summation    as Distill$[{\cal H}$].   
This is ultimately 
tied with the manipulation of singularities of the partition function.    
 Consider a generalized partition function:  
\begin{align} 
{ \cal Z}(\beta , \epsilon_1, \epsilon_2,  \ldots ) = \tr \Big[ e^{-\beta  H}  \prod_{a} e^{i \epsilon_a Q_a}  \Big] 
\label{trial}
\end{align} 
where $Q_a, a=1, 2, \ldots   $ are some charges associated with the QFT, $[H, Q_a]=0$, and $H$ is Hamiltonian of the theory on $\R^3$.   The main question is the following. 
\begin{itemize}
\item[] {\bf Main question:} 
  Assume that  thermal partition function  ${\cal Z}(\beta ) \equiv  {\cal Z}(\beta ,0 \ldots, 0) $ possesses singularities in  $\beta  \in [0, \infty)$ (which is generically the case). Does there exist  a grading $ \prod_a e^{i \epsilon_a Q_a}$  over the Hilbert space $\cal H$  such that  ${\cal Z}(\beta ,  \epsilon_1, \ldots, \epsilon_N)  $ is  an analytic function of $\beta $? 
\end{itemize} 
Our goal is to answer this question in the positive for a broad class of non-supersymmetric  quantum field theories intimately related to QCD with fundamental fermions. 
Three complementary perspectives about  the process of quantum distillation are:
\begin{itemize}
\item  {\bf   Grading of Hilbert space, and quantum distillation:}  This  is the process of  drastically reducing  the magnitude of the state-sums without changing the Hamiltonian or Hilbert space by using symmetry grading over the Hilbert space. 
\item   {\bf  Path integrals with generalized boundary conditions} (or equivalently, path integrals in the background of global symmetry holonomies).  This allows for the evaluation of gauge-holonomy potentials in the  presence of global-symmmetry holonomy  backgrounds, and sometime admits reliable semi-classical analysis of dynamics and study of phase transitions. 
\item {\bf Graded-thermodynamics}: 
 Thermodynamics is the  thermal worth of the Hilbert space ${\cal H}$ through the thermal partition function  $Z(\beta ) = \tr [ e^{-\beta  H}  ]$. 
Graded-thermodynamics may be viewed as the thermal worth of the distilled Hilbert space Distill$[{\cal H}]$. 
\end{itemize}

\subsection{Trivial and   non-trivial examples in 1d,  2d and 4d}
\begin{figure}[t]
\begin{center}
\includegraphics[width = .70\textwidth]{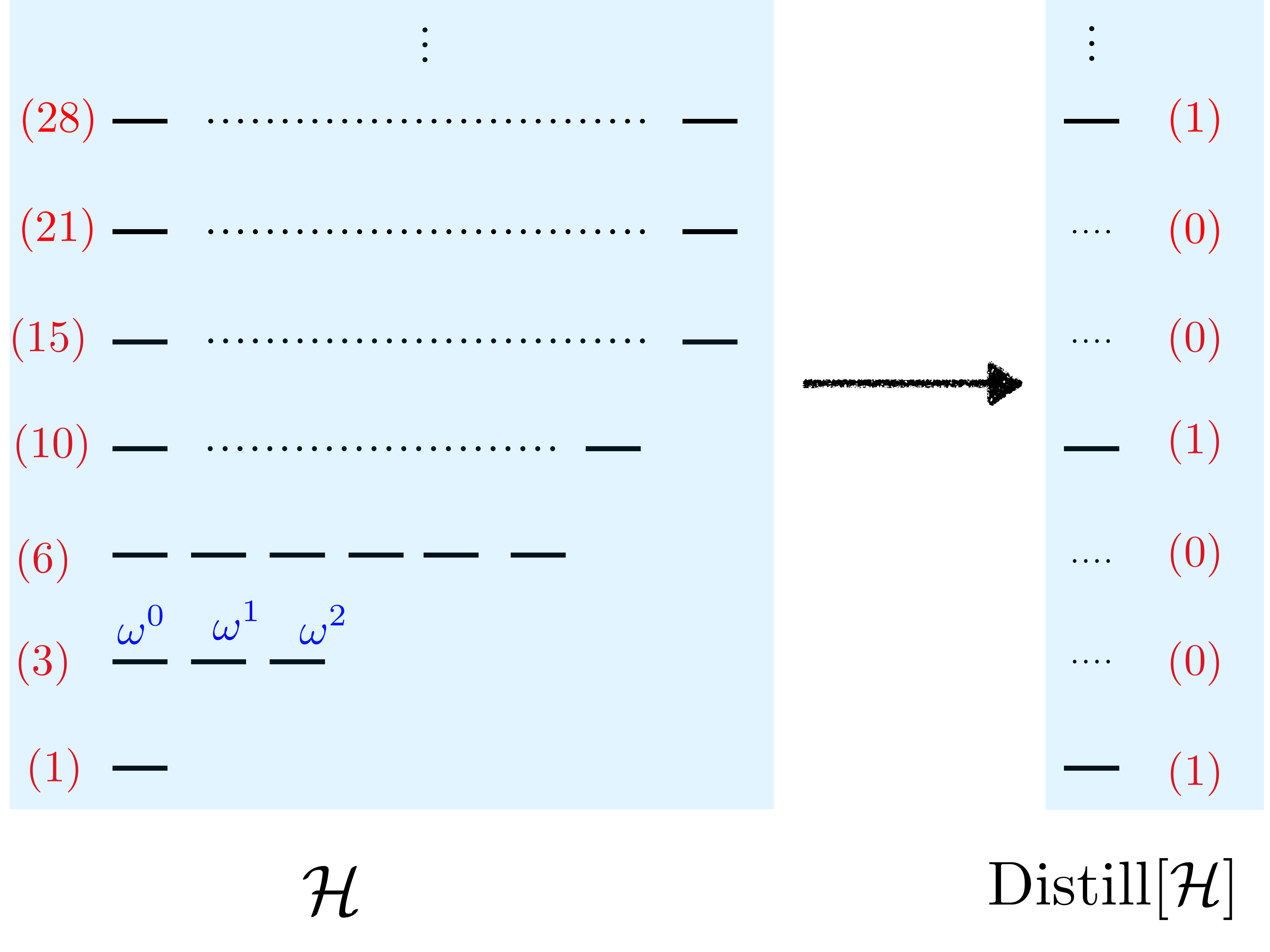}
\vspace{-0.5cm}
\caption{Hilbert space of 3d simple harmonic oscillator and its  $SU(3)$ symmetry graded distillation. The terms inside parenthesis indicate degeneracies.  Despite  its simplicity,  this  picture morally captures what takes place in 2d   sigma models  and 
4d QCD(F/adj).  In the $N$-dimensional oscillator  as $N\rightarrow \infty$, only ground state contribute to symmetry graded partition function. This is also the case in the large-$N$ limits of $\mathbb {CP}^{N-1}$ sigma model and principle chiral model,  as we review, where symmetry graded state sum exhibits spectral cancellations (similar to  Witten index in supersymmetric theories), but now story can take place   in  purely bosonic theories. 
  }
\label{fig:distill}
\end{center}
\end{figure}

Perhaps, with the principle that a picture says thousand words, we can start with a simple system in quantum mechanics and discuss the concept of quantum distillation there, see Fig.\ref{fig:distill}. Despite the fact that this is a   trivial  non-interacting system, extremely similar  phenomena does  occur in the non-trivial asymptotically free QFT in  any  dimensions. After this example, 
we  construct  two types of perfect quantum distillation in 2d QFT, which will be  useful when we build a similar structure  in 4d QCD with fundamental and adjoint fermions.

\vspace{3mm}
\noindent
{\bf 1d QM:}  Consider  $N$-dimensional simple harmonic oscillator.  This system has $U(N)$ global symmetry, and harmonic level $k$   is in the   totally symmetric $k$-index representations of $SU(N)$ and have degeneracies given in \eqref{deg-deg}.   We can construct a  symmetry graded state sum   
\begin{align}
 {\cal Z}_{\Omega_F^0} (\beta) = \tr (e^{-\beta H} 
\prod_{j=1}^{N} e^{ { i {2\pi\over N}j    \widehat Q_j} } ), \qquad Q_j = \widehat{a}^{\dagger}_j \widehat{a}_j 
\end{align}
    where $Q_j$  the number operator for the $j^{\rm th}$ oscillator.   In the graded  state sum,  
  many degenerate states cancel among themselves due to phases attached to them.    After cancellations, even arbitrarily  large degeneracy factors maps  to  graded degeneracy factors, which is  $0$ or $1$:
\begin{align}
{\rm deg}(k) = { N+k-1 \choose k}   \longmapsto  (\underbrace{1, 0, \ldots, 0,}_{N}  \underbrace{1, 0, \ldots, 0,}_{N }1, \ldots)
\label{deg-deg}
\end{align}
See Fig.\ref{fig:distill} for $N=3$ case.  Clearly, in the $N \rightarrow \infty$ limit, the contribution of the whole Hilbert space reduces to just a single state  (the ground state)  for any finite value of $\beta$!

\vspace{3mm}
\noindent
{\bf 2d QCD(adj):} 
 In non-supersymmetric  $SU(N)$  QCD(adj) in 2d, there is a mixed anomaly between  $(\Z_2)_F$ and $ (\Z_2)_\chi $ for $N$ even,  and the partition function on the torus with periodic boundary conditions on both cycles vanishes identically, 
 \begin{align}
  {\cal Z}_{++} (\beta, L)= \tr [(-1)^F e^{-\beta H_L}]=0.
  \end{align}
at any $\beta$ and $L$.  This indeed implies that   
 \begin{align}
 { \rm Distill} [{\cal H}]= \emptyset 
  \end{align}
  due to exact pairwise cancellations.    In the Hilbert space formulation, this is due to exact level-by-level   cancellation between  bosonic and fermionic states.  
  The  exact bose-fermi degeneracy in this case   is protected by  mixed  anomaly rather than supersymmetry \cite{Cherman:2019hbq}.  
Generalization and other examples  of this phenomena, that exact Bose-Fermi spectral degeneracy can be protected by mixed anomalies involving $(-1)^F$,    can be found in   recent interesting work \cite{Delmastro:2021xox}.\footnote{
The cancellation we describe in this paper  is less perfect than exact Bose-Fermi degeneracy  \cite{Cherman:2019hbq, Delmastro:2021xox} which is protected  by   mixed anomaly  (rather than supersymmetry). 
In the present  case, our construction  has Bose-Bose, Bose-Fermi and Fermi-Fermi  type cancellations. 
\cite{Delmastro:2021xox}  also provides examples of exact Bose-Fermi degeneracy in 3d examples, but not the 4d ones. 
In our 4d QCD(F/adj),  we demonstrate  that the  effective density of states (after all the cancellation in the Hilbert space takes place) takes the form of a 2d QFT in large-$N$ limit, same as supersymmetric theories \cite{DiPietro:2014bca}.
Our construction   in 4d is {\it not} a perfect spectral cancellation, but a  sufficiently  good one, which allows to adiabatically continue between small and large $\R^3 \times S^1$ without phase transitions.  }
  Perhaps, more interestingly,  for odd $N$,   there is no mixed anomaly and  there is no exact  Bose-Fermi degeneracy for finite odd values.  But smoothness of the large $N$ limit requires spectral  degeneracies to emerge in  the limit.  In this sense,  $N$ odd theories is one nice  realization of the quantum distillation idea in a strongly coupled QFT where spectral cancellation is not guaranteed by a mixed anomaly but is there nevertheless.

\vspace{3mm}
\noindent
{\bf 4d QCD(adj):} 
 Another  example is 4d QCD(adj)  in the large-$N_c$ limit \cite{Basar:2013sza,Cherman:2018mya}. This theory, defined through the graded partition function, 
$ \tr [(-1)^F e^{-\beta H}]  = Z_{\cal B}(\beta) - Z_{\cal F}(\beta)$, avoids all phase transitions as a function of $\beta$ and satisfies volume independence.
For $N_f=1$, the theory is supersymmetric $\N=1$ SYM and this is just the supersymmetric index \cite{Witten:1982df}.   For $N_f >1$,  where the microscopic theory is non-supersymmetric, Ref.\cite{Cherman:2018mya} showed that  powerful Bose-Fermi cancellations over the Hilbert space of the theory take place.  In particular, the relative density of states $\rho_B(E) -\rho_F(E)$  for the theory defined on a curved 3-manifold has the scaling of a 2d QFT, identical to supersymmetric theories in a similar set-up \cite{DiPietro:2014bca}.

\vspace{3mm}
\noindent
{\bf 2d sigma models:} 
One may be tempted to think that quantum distillation  should not be possible in  purely bosonic   field theory. 
However,  two  powerful  and generalizable counter-examples are the following. 
In the $\mathbb {CP}^{N-1}$ model,  a judiciously graded partition function 
\begin{align}
{\cal Z}_{\Omega_F^0} (\beta) = \tr (e^{-\beta H} 
\prod_{j=1}^{N} e^{ { i {2\pi\over N}j    \widehat Q_j} } ) 
\end{align}
 can yield powerful cancellations between all higher states.   
In the  $N\rightarrow \infty$ limit of the bosonic $\mathbb {CP}^{N-1}$ model, one can obtain 
\begin{align}
{\rm Distilled}[\cal H]= \{\rm ground \; state(s)\} 
\label{DHS0}
\end{align}
at arbitrary $\theta$.  (See Appendix~\ref{app3} for full details.)
Another useful  example is an asymptotically free  matrix model,  the principal chiral model in $d=1+1$.  In this case,  powerful spectral conspiracy can take place with symmetry graded state sums  as  described in Appendix.\ref{app4}, and all but the ground state  cancel.  The implication of these exact cancellations is large-$N$ volume independence \cite{Cherman:2014ofa, Dunne:2016nmc, Dunne:2012ae, Sulejmanpasic:2016llc}. 

\vspace{3mm} 
\noindent
{\bf Perfect quantum distillation:} We refer to quantum distillations that can prevent all phase transitions in infinite volume or large-$N$ 
  thermodynamic  limits on $\R^{d-1} \times S^1$  as the radius $\beta$ is dialed as perfect quantum distillations.   For $d \leq 2$,  to achieve the thermodynamic limit,  it is necessary to take  $N\rightarrow \infty$.

\vspace{3mm}
In conceptually simpler  cases, quantum distillations may result from supersymmetry or anomalies.  Supersymmetric vector spaces are the most obvious graded Hilbert space construction. In non-supersymmetric theories, anomalies are sometimes equally powerful, implying exact Bose-Fermi degeneracy as mentioned above. In general, gauging global symmetries always acts in the direction of diluting the Hilbert space, as one removes non-gauge invariant states from the Hilbert space in the  gauged theory.  However, the fuller story of quantum distillation is not restricted to these special cases. 
As mentioned above, in a purely bosonic theory, it is capable of generating graded state sums that lead to equally powerful cancellations as in the case of the supersymmetric index, forcing the state sum to just the ground states.  On the other hand, what to choose for the symmetry graded sums is not quite  obvious and requires some guesswork. Yet, we can actually test whether a given guess works or not via explicit computation.  Our construction in 
4d QCD(adj/F) is in this category.

\vspace{3mm}
\noindent
\subsection{4d SQCD, QCD-like theories and  QCD} 

If we wish to implement  the idea of  (perfect) quantum distillation  in 4d non-supersymmetric gauge theories, a class of theories arises naturally.  
This is 
 $SU(N_c)$ gauge theory  with $N_f $ fundamental representation Dirac fermions $\psi^a$  and one adjoint representation Weyl fermion, QCD(F/adj) with content  $(A_{\mu}, \psi^a, \lambda)$. A  special emphasis  is given to 
$N_f=N_c$ 
theory for arbitrary $N_c$.    At    large $N_c$, this becomes  a slight generalization of  the  Veneziano  limit with  
$x= \frac{N_f}{N_c}$ fixed \cite{Veneziano:1976wm}. The matter content of this class of theories is the one of $N_f$-flavor  supersymmetric SQCD   with decoupled scalars $m_{q^a}  \rightarrow \infty$ \cite{ Affleck:1983mk, Aharony:1995zh}.

\vspace{3mm}
The QCD(F/adj) theory with mixed representation matter fields allows us to interpolate between different theories as we dial 
a  flavor-symmetric  mass $m_\psi \geq 0$ for fundamental fermions $\psi_a$  
and  a  mass $m_{\lambda} \geq 0$ for the adjoint fermion.  With the decoupling of the adjoint fermion, the theory reduces to the flavor symmetric limit of real QCD for $N_f=N_c=3$. 
With the decoupling of fundamental fermions, the theory reduces to $\N=1$ SYM. Decoupling all fermions, the theory reduces to pure YM: 
\begin{equation}
\begin{tikzcd}
 { \rm SQCD }  \arrow[r, "m_{q^a} \rightarrow \infty " red]    & { \rm QCD( F/adj) }  \arrow[r, "m_{\psi^a} \rightarrow \infty " red] \arrow[d,   "m_\lambda \rightarrow \infty"   red]
    & { \N=1\; \rm SYM} \arrow[d, "m_\lambda \rightarrow \infty" red] \\
&  { \rm QCD(F)}   \arrow[r, black, "m_{\psi^a} \rightarrow \infty"   red] & {\rm YM}
 \end{tikzcd} 
 \label{square1}
\end{equation}
%
We will take advantage of the limits of  square  \eqref{square1}. 
We construct a graded partition function   for QCD(F/adj) by using global symmetries of the theory: 
\begin{align} 
{\cal Z}(\beta, \epsilon_a )  = \tr \Big[ e^{-\beta H} (-1)^F 
 \prod_{a=1}^{N_f} e^{i \epsilon_a Q_a}  \Big] 
\label{GPF}
\end{align} 
where $Q_a= \int d^3x  \; \bar \psi_a \gamma^0 \psi_a$ are charges associated with the maximal abelian subgroup of $U(N_f)_V$. 
This object is a mixture of quantum distillations that worked perfectly  in 2d QCD(adj) in which we implemented a $(-1)^F$ grading   \cite{Cherman:2019hbq}  and 2d   
 $\mathbb {CP}^{N-1}$ model where we implemented a flavor-symmetry grading \cite{Dunne:2012ae, Sulejmanpasic:2016llc}. 
In path integral formalism,  this corresponds to periodic boundary conditions for  $\lambda$ and flavor-twisted boundary conditions 
on $\psi_a$. We have 
\begin{align} 
{\cal Z} (\beta, \epsilon_a )= \int_{
\tiny{
\begin{array}{l} 
A(\beta)=+A(0) \\
 \lambda(\beta) = + \lambda(0) \\
\psi(\beta)= + \psi(0) \overline \Omega_F e^{i \pi} 
\end{array}
}
}   DA_{\mu} D\psi D\lambda  \;\; e^{- S[  A_{\mu}, \psi,  \lambda] }
\label{GPF2}
\end{align} 
There are in principle many choices for boundary conditions $\Omega_F$ for fundamental fermions. Restricting to flavor rotations in a vector-like subgroup, 
$\Omega_F e^{i \pi} $ is a general $U(N_f)_V$ matrix.  

\vspace{3mm}
There are   three  independent procedures which  pick a {\it unique} boundary condition  (up to an over-all $U(1)_V$ factor), or equivalently grading operator, in Hamiltonian  formalism: 
\begin{itemize} 
\item  maximizing the graded free energy (or minimizing the graded pressure),
\item  demanding the presence  of color-flavor center symmetry upon  compactification  on $\R^3 \times S^1$,
\item or demanding  that certain mixed anomalies present on $\R^4$ persist on $\R^3 \times S^1$
\end{itemize} 
picks   a unique  configuration for flavor-twist  or grading operator: 
 \begin{align}
 \Omega_F^0 = \textrm{diag}
(1, \omega, \cdots, \omega^{N_f-1}),  \qquad  \omega = e^{2\pi i /N_f}
\label{flavor-hol-0}
\end{align} 
We will explain the reasoning behind all three conditions in detail.

\vspace{5mm}
\noindent
\subsection{Color-flavor center symmetry}
First, we note an unconventional and  fairly recently understood aspect of flavor-twisted boundary conditions. It is well-known that 
theories with only adjoint representation fields, such as 
pure YM and $\N=1$  SYM, have $\Z_{N_c}^{[1]}$ one-form center-symmetry.  When the theory is compactified on a circle $\R^3 \times S^1$, part of the one-form symmetry becomes a zero-form symmetry, 
for which the order parameter is the Polyakov loop, i.e., a Wilson line wrapping the $S^1$ direction. 
 
 \vspace{3mm}
Once dynamical fundamental fermions are introduced, the one-form  center-symmetry is explicitly broken, 
and in particular, the Polyakov loop  is no longer a good order parameter; see  \cite{Gross:1980br} or standard texts on thermal field theory \cite{Kapusta:2006pm,Laine:2016hma}.    However,  it turns out that one can actually salvage the situation  \cite{Cherman:2017tey}:
 both QCD(F) and QCD(F/adj)  can have  an {\bf exact} $\Z_{{\rm gcd}(N_f, N_c)}^{[0]}$ (gcd stands for greatest common divisor) zero-form center symmetry which acts non-trivially on Polyakov loops, despite the absence of one-form center symmetry in the theory.  
The  $\Z_{{\rm gcd}(N_f, N_c)}^{[0]}$  symmetry lives in 
the  center of $SU(N_c)$  and in a cyclic  permutation subgroup $\Gamma_S \subset SU(N_f)_V$.  This was appreciated as a genuine symmetry fairly recently in  \cite{Cherman:2017tey} where it was shown to be an invariance of the partition function,  ${\cal Z}_{\Omega_F^0} \mapsto {\cal Z}_{\Omega_F^0} $. This symmetry is called color-flavor center (CFC) symmetry and Wilson lines around the $S^1$ circle are charged under it  \cite{Cherman:2017tey}. An earlier lattice study indeed shows a sharp phase transition associated with  CFC \cite{Iritani:2015ara}. 
For QCD-like  theories in  \eqref{square1} on $\R^3 \times S^1$,  there is a CFC  or  zero-form part of center-symmetry acting on Polyakov loops as follows: 
\begin{equation} 
\begin{tikzcd}
 {  \Z_{{\rm gcd}(N_f, N_c)} \;  \rm CFC}  \arrow[r, "m_\psi \rightarrow \infty " red] \arrow[d,   "m_\lambda \rightarrow \infty"   red]
    & {  \Z_{N_c}} \arrow[d, "m_\lambda \rightarrow \infty" red] \\
  { \Z_{{\rm gcd}(N_f, N_c)} \;  \rm CFC}   \arrow[r, black, "m_\psi \rightarrow \infty"   red]
& {\Z_{N_c} }
 \end{tikzcd} 
\label{eq:CFC} 
\end{equation}
 A Wilson line  wrapping  the $S^1$ circle is a good order parameter at any point in the $(m_\lambda, m_\psi)$ plane provided ${\rm gcd}(N_f, N_c) >1 $. 
Therefore, we can examine the phase structure of these theories according to the  CFC or zero-form center-symmetries,  and pose questions about analyticity of the graded partition function as a function of $\beta$. The CFC plays a major role in our construction. 

\vspace{3mm}
The appearance of CFC is  correlated with two other effects.   Turning on a background for $SU(N_f)_V$, or equivalently, 
a flavor twisted boundary condition $ \Omega_F^0 $, explicitly breaks the global symmetry of the theory ${\bf G}$ given in 
\eqref{symmetry}   to its maximal abelian subgroup  
${\bf G}_{\rm max-ab}$  given in \eqref{MAG}.   At the same time, this is the origin of the flavor part of the 
 quantum distillation operator in \eqref{GPF}.

   \subsection{Color-holonomy potentials in the flavor-holonomy  background and surprises}

The image of quantum distillation of ${\cal H}$ in the path integral formalism  is  reflected in the flavor-holonomy dependence of the gauge-holonomy potential. Clearly,  flavor-holonomy $\Omega_F$  is a {\it choice}; it is intrinsically non-dynamical. It corresponds to boundary conditions for fundamental fermions. On the other hand, gauge-holonomy  $\Omega$ is  dynamical and its vacuum expectation value  is determined by the minimum of the gauge-holonomy potential, which in turn 
 determines some  properties  of the ground states or thermal equilibrium states. It turns out that the minimum of the Wilson line potential is inherently tied to the choice of  flavor-holonomy background  in pleasantly surprising ways. 

\vspace{3mm}
We calculate the gauge-holonomy potential  for the Polyakov loop:
\begin{align} 
\Omega = e^{i \oint dx_4  a_4 } \equiv 
\text{diag}(e^{ i v_1}, \ldots, e^{ i v_{N_c}} )
\label{vev}
\end{align} 
at two-loop order  with the boundary conditions \eqref{GPF2}.\footnote{Two-loop order suffices to determine the realization of CFC-symmetry for any flavor-holonomy background.}   
 To do so, we  generalized  the tour de force
 thermal studies  of Refs.~\cite{KorthalsAltes:1993ca,KorthalsAltes:1999cp} and  \cite{Guo:2018scp} to incorporate flavor-holonomy backgrounds.  
The center-breaking gauge boson contribution is completely undone by one adjoint fermion with periodic boundary condition  to {\it all orders} in perturbation theory, similar to $\N=1$ SYM. 
 \begin{align}
V^{\rm  gauge} + V^{\rm \lambda}&=0 \qquad  \text {all orders in perturbation  theory}   
	\label{1-loop-allorders} 
\end{align} 
The story therefore  boils down to what the fundamental fermions with twisted boundary conditions do.   The 
two-loop result which carries many new insights is:
\begin{align}
V_{ \text{1-loop}, \Omega_F} + 	V_{ \text{2-loop}, \Omega_F}  
 &=  {\textstyle \frac{2}{\pi^2 \beta^4} \left( 1 -  
	\frac{3 g^2 }{ 16  \pi^2}  \left(\frac{N^2_c-1}{N_c}\right) \right)}  \sum_{n=1}^{\infty}\frac{(-1)^n}{n^4}[\text{Tr}(\bar{\Omega}_F^n)\text{Tr}(\Omega^n)+\text{c.c.}]   \cr 	
	&+ {\textstyle \frac{ g^2 N_f}{8  \pi^4 \beta^4} } \sum_{n=1}^{\infty}\frac{1}{n^4}       \big| \text{Tr} (\Omega^n) \big|^2  \,.
	\label{2-loop-full} 
\end{align} 
where $(g^2)^0$ and $(g^2)^1$ 
are   one- and  two-loop terms, respectively. Here is the sharp contrast   between two choices of flavor holonomy $\Omega_F$:
\begin{itemize} 
\item For $\Omega_F={\bf 1}_{N_f}$,  there are terms like $\tr (\Omega^n)$ in the potential that are manifestly non-invariant under center-transformation, hence  the center-symmetry is explicitly broken. 
The minimum of gauge-holonomy potential is at 
\begin{align}
\Omega \big|_{\rm min}={\bf 1}_{N_c},
\label{centerbrok}%
\end{align}
the theory is in the chirally symmetric high-temperature phase, separated from low temperature chirally broken phase. 

\item For $\Omega_F=\Omega_F^0$ given in \eqref{flavor-hol-0}, the terms in the potential  that transform non-trivially under $  \Z_{{\rm gcd}(N_f, N_c)} $  drop out as they must.  
Not only is $\Z_{N_c}$  CFC symmetry an exact symmetry of the $N_f=N_c$ theory, it remains   unbroken for any value of $\beta\Lambda \lesssim 1$ and   for any $N_c$.  
The minimum of the gauge-holonomy potential is  now at 
\begin{align}
\Omega \big|_{\rm min}&=  \text{diag}(1,\omega,\cdots,\omega^{N_c-1}), \qquad  \qquad N_c~\text{odd}, \cr
	\Omega \big|_{\rm min} & = \omega^{1/2}\text{diag}(1,\omega,\cdots,\omega^{N_c-1}), \qquad N_c~\text{even}.
	\label{centersym0}%
\end{align}
as reported in \cite{Kanazawa:2019tnf}.    The center-stability holds to all loop orders in perturbation theory as well as non-perturbatively as such effects are suppressed compared to our two-loop result and cannot alter the minimum of the potential. 
\end{itemize}
The center-stability also holds away from chiral limit, for $m_\lambda=0, m_\psi \geq 0$  and 
  $m_\lambda \leq m_\lambda^*,  \;  m_\psi = 0$, and even in cases where both fermions may have non-zero masses.  These limits will  be used to access  flavor limit of QCD.

\vspace{3mm}
\noindent
{\bf Quantum distillation in  the Hilbert space of QCD(F/adj): } What is happening with the potential is quite striking. Essentially, we changed the small-$\beta$ limit of QCD(F/adj) from a center-non-invariant  configuration 
given in \eqref{centerbrok} to a center-symmetric configuration  given in \eqref{centersym0} by inserting $(-1)^F   \prod_{a=1}^{N_f} e^{i \epsilon_a Q_a} $ into the trace without 
changing the Hilbert space.  One of our goals in this paper is to benefit from the combined thinking of these two-process distillings of Hilbert space and their image in  the holonomy potentials as well as corresponding graded thermodynamics.

\vspace{3mm}
In the graded state sum, there are cancellations of  different nature; bosons  cancel against bosons (e.g. mesons),  
fermions cancel against fermions (e.g. baryons and  fermionic mesons), and 
 in certain limits,  bosons cancel against fermions (eg. 
 glueballs  against gluino-balls).   We refer to  these cancellations as 
$\cal B B,  \cal F F,   \cal B  F $ cancellations, respectively.  

 \subsection{Volume independence in Veneziano-type  limits} 
 In the Veneziano type large-$N_c$ limit of QCD where $N_f$ scales with $N_c$\cite{Veneziano:1976wm},  it is impossible to satisfy volume (or temperature) independence \cite{Kovtun:2007py} even in the confined phase. This is because even the contribution of the mesons  to the free energy density  is of order ${ \cal F } \approx -N_c^2 T^4$, explicitly violating  temperature  independence at leading order in $N_c$. 
 The graded partition function provides the first realization of large-$N_c$ volume independence in this limit. 
 The gauge-holonomy potential  \eqref{2-loop-full}  in the large-$N_c$  Veneziano  limit of QCD(F/adj) reduces to: 
 \begin{align} 
	V_{\text{1-loop}, \Omega_F} + V_{\text{2-loop}, \Omega_F}=   +  x  \frac{(g^2N_c) }{ 8 \pi^4 \beta^4}
	\sum_{n=1}^{\infty}\frac{\big| \text{Tr} (\Omega^n) \big|^2}{n^4}  \,, 
	\label{V12largeN-1} \qquad x= N_f/N_c
\end{align} 
 Clearly, \eqref{V12largeN-1} has a center-symmetric minimum.  Quite strikingly, the fundamental fermions' contribution which is normally of the form  $x N_c \sum_{n=1}^{\infty} \frac{1}{n^4}  \tr \Omega^n + {\rm c.c.}$ turns into a quantity of order $1 \over N_c^3$ vanishing in the large-$N_c$ limit. 
 One may be tempted to think that 
 all information about the existence of microscopic fundamental fermions  
  is ``forgotten", but the truth is subtler.  The center-stabilizing double trace operator comes exactly from the fundamental fermions with  $\Omega_F^0$ twisted boundary conditions!  In other words, fundamental fermions, for the purpose of center-symmetry, act exactly as 
$x (g^2 N_c)$ many  adjoint fermions with periodic boundary conditions, hence stabilizing center-symmetry.

\vspace{3mm}
  At the $N_c=\infty$ limit, volume independence implies that the graded partition function avoids Hagedorn singularities  in all hadronic channels,  mesons, glue as well as baryons \cite{PandoZayas:2003yb} due to powerful spectral cancellation.   This statement can be proven by studying  QCD(F/adj)  on  $S^3 \times S^1$    \cite{Basar:2014jua, Cherman:2018mya}.

   \subsection{Unified mechanism of chiral symmetry breaking in QCD(F) and $\N=1$ SYM} 
 Probably, the most important insight that the idea of quantum distillation brings is  a unified understanding of chiral symmetry breaking  in   QCD(F)  and $\N=1$ SYM.   Consider  $N_f=N_c$ theory  with the insertion of the $\Omega_F^0$-twisted boundary condition in path integral. Then, as 
 \eqref{centersym0} implies, the gauge holonomy in the small circle regime  is $\Z_{N_c}$ symmetric. Due to dynamical abelianization, there are $N_c$ types of monopoles  each with action 
  $S_0=  \frac{8\pi^2}{g^2N}$. In QCD(F/adj), each monopole possess  four fermion zero modes, two  adjoint and two fundamental fermion zero mode  due to  index theorem for Dirac operator in monopole-background \cite{Nye:2000eg, Poppitz:2008hr}: 
   \begin{align}
 {\cal M}_i    \sim e^{-S_i} 
    e^{ - \frac{4\pi}{g^2}  \alpha_i \cdot \phi +   i \alpha_i \cdot \sigma } 
      (\psi_{Ri} \psi_{L}^i)  (\alpha_i \cdot  \lambda)^2     \qquad {\rm QCD(F/adj)} 
      \label{mon-op-0}
 \end{align} 
 Turning on a mass term for adjoint (fundamental)  quark lifts the adjoint (fundamental)  zero modes, leaving only two  fundamental (adjoint) zero modes per monopole: 
    \begin{align}
 {\cal M}_i   & \sim e^{-S_i} 
    e^{ - \frac{4\pi}{g^2}  \alpha_i \cdot \phi +   i \alpha_i \cdot \sigma } 
      (\psi_{Ri} \psi_{L}^i)      \qquad {\rm QCD(F) \;  +  \;massive  \; adjoint} \cr 
       {\cal M}_i   & \sim e^{-S_i} 
    e^{ - \frac{4\pi}{g^2}  \alpha_i \cdot \phi +   i \alpha_i \cdot \sigma } 
       (\alpha_i \cdot  \lambda)^2    \qquad  \N=1 \; {\rm  SYM  \; + \; massive \; fund.}
            \label{mon-op-1}          
 \end{align} 
   In the first case, the chiral symmetry which commutes with the quantum distillation operator is a maximal abelian subgroup (or maximal torus)   $ {\bf G}_{\rm max-ab} $  of the global chiral symmetry $ SU(N_f)_L \times SU(N_f)_R $ of QCD.  In the second case, only  $\Z_{2N_c}$  discrete chiral symmetry is present,  exactly as in $\N=1$.  The monopole operator is invariant  under  $ {\bf G}_{\rm max-ab} $  and  $\Z_{2N_c}$, respectively. 
 Remarkably, 
the gauge field part of the monopole-operator is capable of soaking up chiral charge (either discrete or continuous)  \cite{Cherman:2016hcd}.  This is 
  due to  intertwining of chiral and topological-shift symmetry which is a  generalization of the result of  Ref.\cite{Affleck:1982as} on 
   $\R^3$ to locally four-dimensions on    $\R^3 \times S^1$.   As a result, an interesting phenomenon  occurs. The gauge fluctuations (and flux  part of the monopole-operator) acquire chiral charge under continuous and  discrete   chiral symmetry, respectively.\footnote{Truth is stranger than fiction.}

The spontaneous breaking of the chiral symmetry occurs due to flux part of the monopole operator acquiring a vacuum expectation value.
 On the vacuum,  exactly the same vev 
\begin{align}
\langle { \rm VAC} | e^{   \alpha_i \cdot  z  }  | {\rm  VAC} \rangle  \neq 0 
 \end{align}
 generates continuous chiral symmetry breaking in QCD(F) and discrete chiral symmetry breaking in $\N=1$ SYM.  In the first case, we can actually prove that dual photons remain gapless non-perturbatively,  
  and they  are the semi-classical realization of Nambu--Goldstone bosons.   In the latter case, this mechanism produce $N_c$ isolated vacua  associated with  discrete chiral symmetry breaking, and 
  there are  domain walls  between vacua \cite{Davies:2000nw, Davies:1999uw}.  \footnote{ 
  In the standard interpretation in the literature,  $\chi$S is asserted to be  broken even  on small $\R^3 \times S^1$ due to  fermion bilinear $\tr \lambda \lambda$ or 
  $(\psi_{Ra} \psi_{L}^b)$ acquiring a vev.  This interpretation is {\it not} quite correct.  It is  the vacuum expectation value of magnetic flux part of   monopole operator that breaks the symmetry.  Once this operator acquires a vev, a chiral symmetry breaking mass term is induced for fermions.   See Sec. \ref{sec:what}  for details. }

 \subsection{Mixed anomaly (persistent order) vs. adiabatic continuity}

\begin{figure}[t]
\vspace{-1.5cm}
\begin{center}
\includegraphics[width = 1.0\textwidth]{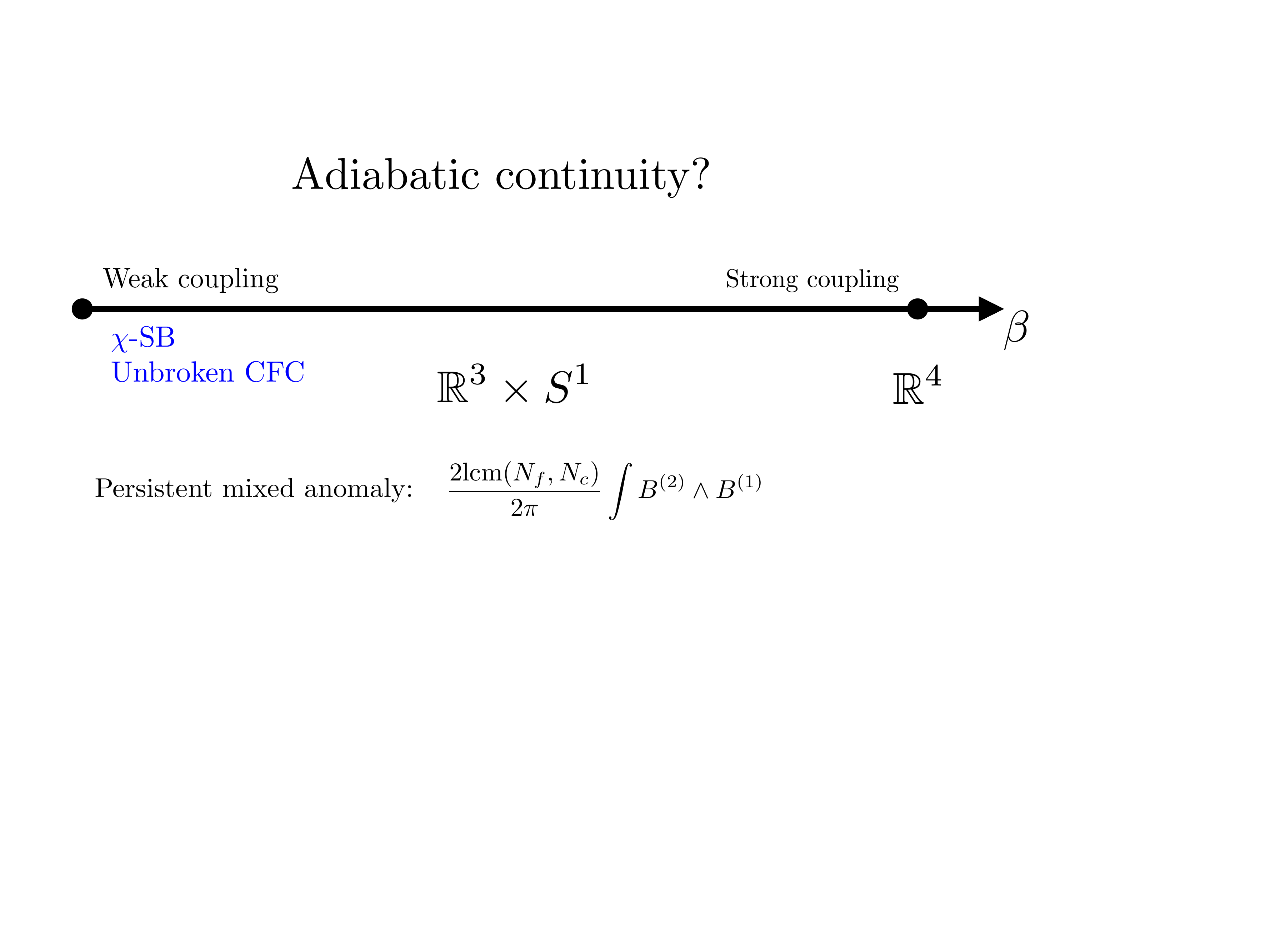}
\vspace{-5.5cm}
\caption{By employing  graded partition function \eqref{GPF}, we prove chiral symmetry breaking at weak coupling for various  $ SU(N_c)$ gauge theories with $N_f$ massless fundamental Dirac fermions and one massive adjoint fermion.  This is done 
 by using semi-classical methods and index theorem for Dirac operator on $\R^3 \times S^1$ in non-trivial gauge $\Omega$  and flavor  $\Omega_F^0$ holonomy background. 
There exists a mixed anomaly polynomial that persists at any compactification scale $\beta$ (both at weak and strong coupling)   if and only if one uses  $\Omega_F^0$ background. 
The combination of the two methods does not suffice  to analytically prove $\chi$-SB at strong coupling  and $\R^4$. 
   }
\label{fig:continuity}
\end{center}
\end{figure}

What is the implication of spontaneously broken chiral symmetry on small $S^1 \times \R^3$ for the theory on $\R^4$? 
Is it possible that the theory remains in the same phase at arbitrary radius?  This is  logically possible, and in fact, for the $N_f=N_c$ theory
  it is very  likely the case. This idea is called adiabatic continuity, and has been achieved via double-trace deformations in Yang--Mills theory \cite{ Unsal:2008ch}. See  \cite{Bonati:2018rfg,  Athenodorou:2020clr} for lattice tests of this idea. 
But  unfortunately, we cannot prove  adiabatic continuity analytically.  What we can prove is a weaker (but still non-trivial) statement that the mixed anomaly condition on  $\R^3 \times S^1$  and $\R^4$ are the same.

\vspace{3mm}
We can say more about the possible vacuum structures of the theory both on  $\R^4$ and  $\R^3 \times S^1$ 
by using recent progress on mixed anomalies involving discrete symmetries \cite{Gaiotto:2017yup, Gaiotto:2017tne}. 
There is a  mixed anomaly on $\R^4$ between  $SU(N_f)_V/ \Z_{  {\rm gcd}(N_f, N_c) }$  symmetry and $  \Z_{  2 N_f }$ subgroup of the chiral symmetry \eqref{eq:non_ab2}. 
The partition function in the presence of background fields $(A, B)$ corresponding to $SU(N_f)_V/ \Z_{  {\rm gcd}(N_f, N_c) }$   fails to be  invariant under $h  \in    \Z_{  2 N_f }$ chiral transformations. As  dictated by the anomaly polynomial,
  \begin{align}
{\cal Z}(h(A, B) ) &  = 
 \exp       \left[  - i  { 2 {\rm lcm}(N_f, N_c)   \over 4\pi } \int 
  B  \wedge B    \right]  {\cal Z} ((A, B) )    \cr
  & =  \exp       \left[  - i  2 \pi  { 2 {\rm lcm}(N_f, N_c)   \over   \left( {\rm gcd} (N_f, N_c) \right)^2  } \right] {\cal Z}((A, B) )   
  \label{poly}
 \end{align}  
 which means that the ground state cannot be a unique, gapped (i.e. trivial) state provided  
 $ \frac{  2 {\rm lcm}(N_f, N_c)}  { \left( {\rm gcd} (N_f, N_c) \right)^2 }\in   {\mathbb Q \backslash \mathbb Z}$.  
 Possibilities include spontaneous (chiral) symmetry breaking or a CFT on 
 $\R^4$. 
 
 \vspace{3mm}
 Remarkably, the same anomaly condition persists  on  $\R^3 \times S^1$ {\it if and only if} one uses the twisted 
 boundary conditions \eqref{flavor-hol-0} for fundamental fermions.  For similar discussions in QCD(adj) and QCD(F),  see \cite{Komargodski:2017smk, Tanizaki:2017qhf, Shimizu:2017asf, Cherman:2017dwt}.
 This gives a   
  triple  mixed anomaly  between shift symmetry $\Gamma_S \subset SU(N_f)_V$, abelianized flavor symmetry $U(1)_V^{N_f-1}/\Z_{{\rm gcd}(N_f, N_c)}$,  and the discrete chiral symmetry   $\Z_{2N_f}$  
  as in provided  $\R^4$.  In this case, spontaneous symmetry breaking scenario includes   chiral symmetry and color-flavor center symmetry.  We cannot have a phase in which none of the symmetries is unbroken. This notion is referred to as persistent order. 

\vspace{3mm}
Therefore, the anomaly conditions that constrain the vacuum structure on $\R^4$ and  $\R^3 \times S^1$ are one and the same. The chirally broken phase is a realization of an anomaly permitted  phase. This is certainly not as strong as the
adiabatic continuity conjecture which states that only one mode of the mixed anomaly permitted phase is operative at any 
radius. However, it is also encouraging that the ground states are to be chosen by dynamics among just a few anomaly-permitted possibilities.

 \section{QCD(F/adj) and SQCD: general considerations} 
\label{sec:general}
Consider  QCD  with    $SU(N_c)$ gauge group,  $N_f $ flavors of fundamental massless Dirac fermions $\psi^{a}$,   
 $a = 1,  \cdots, N_f$ and one   adjoint Weyl  
fermion $\lambda$ with Euclidean Lagrangian\footnote{We use Wess and Bagger convention with spinors which is commonly used in supersymmetry literature \cite{Wess:1992cp}.  So, our Dirac spinor is $\psi  = { \psi_L \choose \bar   \psi_R }$, and mass term  is $ \bar \psi  \psi =  \psi_R \psi_L +  
\bar   \psi_L \bar   \psi_R $. }:
\begin{align} 
{\cal L}= \frac{1}{2g^2} \tr F_{\mu \nu}^2 + \sum_{a=1}^{N_f}  \bar \psi_a \gamma_\mu D_{\mu} \psi^a +   2 \tr  \bar \lambda \bar \sigma_\mu D_{\mu}  \lambda
\label{Lag}
\end{align} 
where  $ D_{\mu} \psi^a   = \partial_\mu  \psi^a    + i a_{\mu}  \psi^a $   and  $ D_{\mu} \lambda   = \partial_\mu  \lambda     + i [a_{\mu}, \lambda] $. These theories are one-parameter family deformations of the general SQCD  \cite{Seiberg:1994bz} with a  supersymmetry breaking  mass term for the 
scalar  quark  field  (squark,  the super-partner of $\psi_a$ fundamental
fermion) $m_{\Phi_a}$ turned on, and taken to decoupling limit \cite{Aharony:1995zh}.  However, we investigate \eqref{Lag} without any reference to supersymmetry, with semi-classical and mixed anomaly tools that usefully apply to non-supersymmetric theories.

We consider both massless theory  \eqref{Lag}  as well as it 
  its  mass deformations,  with a common  mass   $m_{\psi_1}  =  \ldots =m_{\psi_{N_f}} \equiv m_\psi \geq 0$ and with $m_{\lambda} \geq 0$. 
 If both mass terms are turned on,  then the  $\theta$ angle also becomes a physical parameter in the Lagrangian. Hence, 
\begin{align} 
\delta{\cal L}= m_{\psi} \bar \psi_a \psi^a + m_{\lambda} (\lambda \lambda + \bar \lambda \bar \lambda) + i \frac{\theta }{16\pi^2}   \tr F_{\mu \nu}  \tilde  F^{\mu \nu}
\end{align}

\vspace{3mm}
We refer to this theory as QCD(F/adj).  When the adjoint fermion decouples and $m_\psi=0$,   it corresponds to  the chiral limit of  QCD(F).   
We will see that this theory carries  many new insights into non-perturbative dynamics of QCD(F). 
 It also possesses calculable examples of chiral and color-flavor center symmetry changing phase transitions in thermodynamic limit. 
We will study the dynamics of this theory on $\R^3 \times S^1$ as a function of  $(m_\lambda, m_{\psi}, \beta)$ parameters.  First, let us discuss the  global symmetries of this theory.

\subsection{Global symmetries}
 In the chiral limit,  $(m_\lambda, m_{\psi})= (0,0)$,  
the classical theory possesses the global   symmetry 
\begin{eqnarray}
{\bf G}_{\rm classical}  
&=& {SU(N_f)_L \times SU(N_f)_R \times U(1)_V \times U(1)_{A_\psi}  \times U(1)_{A_\lambda}\over \mathbb{Z}_{N_c}\times (\mathbb{Z}_{N_f})_L\times (\mathbb{Z}_{N_f})_R\times (\mathbb{Z}_2)_\psi}. 
  \label{eq:non_ab}
\end{eqnarray}
where  $\mathbb{Z}_{N_c} $ is the center of gauge group $SU(N_c)$,  which  is not a global symmetry but just local gauge structure  which has to be removed. The other discrete groups   in the denominator are there to  prevent double counting 
of the symmetries. 
However, due to ABJ anomaly, the    classical   abelian axial symmetry  is reduced: 
\begin{align}
U(1)_{A_\psi}  \times U(1)_{A_\lambda} \longrightarrow U(1)_{A_D} \times \mathbb{Z}_{2\mathrm{gcd}(N_c,N_f)}
\end{align}
The symmetry of the quantum theory is
\begin{align}
{\bf G}={SU(N_f)_L \times SU(N_f)_R \times U(1)_V  \times U(1)_{A_D} \times \mathbb{Z}_{2\mathrm{gcd}(N_c,N_f)}\over \mathbb{Z}_{N_c}\times (\mathbb{Z}_{N_f})_L\times (\mathbb{Z}_{N_f})_R\times (\mathbb{Z}_2)_\psi}.
\label{symmetry}
\end{align}
To see this, note that  the classical  axial $U(1)_{A_\psi}$ and $U(1)_{A_\lambda}$  currents are  non-conserved as:
\begin{align}
&\partial_{\mu} J^{\mu 5}_{\psi} = 2N_f \frac{1 }{16\pi^2}   \tr F_{\mu \nu}  \tilde  F^{\mu \nu}, \qquad 
&\partial_{\mu} J^{\mu 5}_{\lambda} = 2N_c \frac{1 }{16\pi^2}   \tr F_{\mu \nu}  \tilde  F^{\mu \nu}. 
\end{align}
and a  diagonal  subgroup of  $U(1)_{A_\psi}  \times U(1)_{A_\lambda} $  (call it $ U(1)_{A_D}$) survives in the quantum theory. 
 \begin{align}
&\partial_{\mu}  \left( N_c J^{\mu 5}_{\psi } - N_f J^{\mu 5}_{\lambda} \right) =0
\end{align}
There is also a discrete remnant. To determine it, consider another  linear combination of  classical axial currents $k_1 J^{\mu 5}_{\psi} +  k_2 J^{\mu 5}_{\lambda} $.
 The charge associated with this current is  conserved modulo $ k_1  (2 N_f) + k_2  (2N_c) $.  According to   Bezout identity in elementary number theory,     there exists $k_1, k_2 \in \Z$ such that 
\begin{align} 
N_f  k_1    +    N_c  k_2 = {\rm gcd}(N_f, N_c)
\end{align} 
Therefore, the charge associated with this current is conserved  modulo  $2 {\rm gcd}(N_f, N_c)$.  This is the 
discrete chiral symmetry which cannot be undone by continuous chiral rotations  $ U(1)_{A_D}$.   The action of 
$U(1)_{A_D} \times \Z_{2{\rm gcd}(N_f, N_c) } $ on the fermions is given by 
\begin{align}
\label{diagonal}
U(1)_{A_D} \times  \Z_{2{\rm gcd}(N_f, N_c) }  :& \;\;\;\;     \psi_{L/R} \rightarrow e^{ - i  \frac{N_c}{ N_f}   \gamma}   \;   e^{ \frac{2 \pi i}{ 2 {\rm gcd}(N_f, N_c)} k_1} \;  \psi_{L/R}   \cr  \cr
:& \;\;\;\;    \lambda    \longrightarrow   \;\;\;\;   e^{+i  \gamma}    \;\;\;  \;  e^{ \frac{2 \pi i}{2 {\rm gcd}(N_f, N_c)} k_2}       \;    \lambda
\end{align}
The ABJ anomaly also  manifests itself in the  instanton amplitude \cite{tHooft:1976fv}, which  for the QCD(F/adj) takes the form: 
\begin{align}
\label{eq:instanton_fermions}
\mathcal{I}_{4d}  & \sim   e^{- \frac{8 \pi^2}{g^2}}  (\tr \lambda \lambda)^{N_c}
  \,\,  \det_{a,b=1}^{N_f}  \left[ \psi_{Ra} \psi_{L}^{b}  \right]   
\end{align}

\vspace{3mm}
To summarize, the transformation properties of the microscopic fermions under gauge structure and continuous global 
symmetry is given by:\footnote{In \eqref{diagonal},  \eqref{table1}, we chose $U(1)_{A_D}$ charges of $\lambda,  \psi_{Ra},  \psi_{L}^{a}$ to match SQCD literature.  In the context of SQCD,  $U(1)_{A_D}$ is called $U(1)_{R}$ symmetry.}
  \begin{align}
  \label{table1}
\begin{tabular}{ |c|c||c|c|c|c| } 
 \hline
&  $SU(N_c)$ &  $SU(N_f)$  & $SU(N_f)$ & $U(1)_V$ & $U(1)_{A_D}$  \\ 
\hline \hline
 $\psi_L^a$ &  $  \Box$  &     $ \bar  \Box$   &    {\bf 1} &  +1  & $-\frac{N_c}{N_f}$  \\ 
  $\psi_{Ra}$ &  $ \bar  \Box$  &      {\bf 1} &    $   \Box$    &  -1  & $-\frac{N_c}{N_f}$  \\ 
$\lambda$  &  {\bf adj} &      {\bf 1} &    {\bf 1}   &  0   & $ +1$   \\
\hline
\end{tabular}
\end{align}

The global  symmetry \eqref{symmetry}  coincides with the bosonic symmetry of the $SU(N_c)$  SQCD with $N_f$ quarks \cite{Affleck:1983mk}, see also \cite{Peskin:1997qi} for a  pedagogical introduction.

\vspace{3mm}
\noindent 
{\bf Turning on masses for fermions and global symmetry:}
We will  consider turning on two types of mass terms in  QCD(F/adj):   
Correspondingly, global 0-form symmetries reduce to: 
\begin{align}
& (m_\psi =0, m_\lambda>0): \qquad  && {\bf G}=  \frac{SU(N_f)_L \times SU(N_f)_R \times U(1)_V \times (\Z_{2N_f})_{\psi}   }{ \Z_{N_c} \times (\Z_{N_f})_R \times (\Z_{N_f})_L  \times  \Z_{2} }
 \cr \cr
& (m_\psi >0, m_\lambda=0): \qquad && {\bf G}=  \frac{SU(N_f)_V \times U(1)_V }{  \Z_{N_c} \times \Z_{N_f}} \times  
 (\Z_{2N_c})_{\lambda} \cr  \cr 
 & (m_\psi >0, m_\lambda>0): \qquad && {\bf G}=  \frac{SU(N_f)_V \times U(1)_V }{  \Z_{N_c} \times \Z_{N_f}}  \cr  \cr 
& (m_\psi =\infty, m_\lambda=0): \qquad && {\bf G}=  (\Z_{2N_c})_{\lambda} 
  \label{eq:non_ab2}
\end{align}
The first one of these is the   correct global symmetry of massless QCD(F), and the last is the one of  $\N=1$ SYM. In each case,   {\bf G} has the faithful representation on the physical Hilbert space ${\cal H}$  of corresponding theory \cite{Tanizaki:2018wtg}.

\subsection{Expectations on  $\R^4$  in $N_f=N_c$ theory  and relation to SQCD} 
\label{sec:expectations}
In this section, we focus our attention to $N_f=N_c$  QCD(F/adj). 
There are
two  physically well motivated possibilities for the  behaviour of $N_f=N_c$  QCD(F/adj) theory on $\R^4$ concerning  the realization of continuous global symmetry \eqref{symmetry}.

\vspace{3mm}
\noindent 
{\bf 1)} The chiral global symmetry  \eqref{symmetry}  can be broken down to vector-like subgroup:
\begin{align}
{\bf G} \rightarrow  \frac{SU(N_f)_V \times
U(1)_V }{  \Z_{N_c} \times \Z_{N_f}} 
\label{pattern1}
\end{align}
This is QCD-type behaviour where fermion bilinears
condense 
\begin{align}
\langle \psi_{Ra} \psi_{L}^{b} \rangle =c_1 \Lambda^3 \delta^{a}_{b}, \qquad \langle  \tr \lambda \lambda \rangle   = c_2 \Lambda^3 
\label{condensate1}
\end{align}

\vspace{3mm}
\noindent 
{\bf 2)} The chiral global symmetry can be broken down to a subgroup which possess a chiral $U(1)_{A_D}$ part \begin{align}
{\bf G} \rightarrow  \frac{SU(N_f)_V \times
U(1)_V \times U(1)_{A_D} }{  \Z_{N_c} \times \Z_{N_f} \times \Z_2}
\label{pattern2}
\end{align} 
Unbroken $U(1)_{A_D}$ along with broken non-abelian chiral symmetry 
$SU(N_f)_L \times SU(N_f)_R   \rightarrow
SU(N_f)_V $ is possible if fermion bilinears do not acquire a vev
(because they are charged under
both chiral symmetries), but only if a $ U(1)_A$ singlet four-fermion
operator acquires a vev:
\begin{align}
\langle \psi_{R b} \psi_{L}^{a}  \rangle = 0, \qquad  \langle  \tr \lambda
\lambda \rangle = 0, \qquad  \langle   \psi_{R b}   \psi_{L}^{a}    \tr \lambda \lambda
\rangle = c_3 \delta_{b}^{a}  \Lambda^6
\label{condensate2}
\end{align}

In the first  case, there will be $N_f^2$ gapless NG-bosons,  instead of $N_f^2-1$ as in  QCD(F) since there is one extra axial charge generator.   Therefore, the IR theory will  also 
possess an exactly  gapless  $\eta'$ boson, and will be described by  the chiral Lagrangian of $N_f^2$ NG-bosons. In particular, since the symmetry breaks to vector-like subgroup, there are no massless fermions in the spectrum just like chiral limit of QCD(F).

The second case  is  strongly motivated by SQCD 
 as  described in \cite{Aharony:1995zh}.   Since a part of the chiral symmetry remains unbroken, the IR  theory must possess  both exactly massless  bosons and composite fermions,  and it  must satisfy  standard (zero form) 't Hooft  anomaly matching conditions between UV and IR degrees of freedom \cite{tHooft:1979rat}.   The reason that the second possibility is not ruled out immediately is because it satisfies non-trivial 't Hooft anomaly matching conditions in the non-supersymmetric QCD(F/adj) just like supersymmetric SQCD.

 \vspace{3mm}
Although   chiral symmetry breaking without quark bilinear condensate is ruled out in QCD(F) by Tanizaki \cite{Tanizaki:2018wtg}, this is still a perfectly viable (mixed anomaly allowed)  option in QCD(F/adj). 

 \vspace{3mm}
 \noindent
 {\bf Anomalies in $N_f=N_c$ SQCD  and  QCD(F/adj) are the same.}
 Let us first   describe why the second  option is as viable as first option starting with SQCD. 
 The space of ground states of $N_f= N_c$  SQCD theory   is a  quantum moduli space,  
described in terms of  composite superfields, mesons $M^{a}_{b}$  and baryon  $B$, obeying  \cite{Seiberg:1994bz}.
\begin{align}
 \det M - B\bar B= \Lambda^{2N_c}.
 \end{align}
At the point $B=\bar B =0$,  we have    $\det M = \Lambda^{2N_c}$,   corresponding to 
 \begin{align}
 M^{a}_{b}=Q_{Ra}Q_{L}^{b} = \Lambda^2 \delta_{a}^{b} 
 \label{pattern-5}
 \end{align}
and the chiral symmetry is broken  as in \eqref{pattern2}. At this point, 
 gaugino condensate  for adjoint quark vanishes,   $ \langle \tr \lambda \lambda \rangle  =0 $. 
 Ref.\cite{Aharony:1995zh} showed that  with a supersymmetry breaking soft mass for the 
 scalar   field  (squark,  the super-partner of $\psi_a$ fundamental
fermion) $m_{q_a} $ turned on, this pattern persists. With a small $m_{q_a} $, 
$N_f=N_c$  SQCD breaks its non-abelian  chiral symmetry, but not
the $U(1)_{A_D}$ part.    Increasing  $m_{q_a} $, there
are two possibilities. Unbroken $U(1)_{A_D}$  may persist to decoupling
limit or
there may be a  $U(1)_{A_D}$ breaking phase transition at some critical
value of  $m_{q_a} ^{\rm cr}$. If unbroken  $U(1)_{A_D}$  persists, the pattern of the chiral condensate in the decoupling limit must be 
given by \eqref{condensate2}. 
In this case, infrared physics can be described in terms of   $N_f^2-1$ NG bosons,  and  composite massless fermions which saturate the anomalies  associated with unbroken $U(1)_{A_D}$.

\vspace{3mm}
Since QCD(F/adj) has a fermionic matter content identical to SQCD, the UV anomalies coincide precisely.  
In the IR,  
we can construct fermionic mesons and baryons in QCD(F/adj) by using   Fradkin-Shenker complementarity \cite{Fradkin:1978dv} to SQCD. 
To see this, 
 denote supersymmetric chiral matter multiplets as: 
\begin{align}
 Q_L^a &= q_L^a + \theta \psi_L^a + \ldots  \cr 
 Q_{Rb}&= q_{Rb}  + \theta \psi_{Rb} + \ldots 
 \end{align}
 The fermionic component of the composite meson multiplet can be expressed as:  
 \begin{align}
 M^{a}_{b}=Q_{Ra}Q_{L}^{b} = \ldots  + \theta ( q_{Ra}  \psi_{L}^{b} + \psi_{Ra}  q_{L}^{b} ) +    \ldots
 \end{align}
 The gauge and global quantum numbers of the $q_{Ra}$ scalar and  $(\psi_{Ra} \lambda)$  composite agree with each other. Therefore, we can view the mesino in QCD(F/adj) as continuation of mesino in SQCD and use the replacement: 
 \begin{align}
 q_{Ra} \longleftrightarrow   (\psi_{Ra} \lambda), \qquad q_{L}^{b}   \longleftrightarrow  \lambda \psi_L^b 
 \label{FS-comp}
 \end{align}
resulting in the fermionic meson   (or mesino) $\psi_{Ma}^{b}$ given by 
\begin{align} 
\psi_{Ma}^{\;\;\;\;\; b}= 
 \psi_{Ra} \lambda \psi_L^b  
  \qquad a, b=1, \ldots, N_f 
  \label{mesino}
\end{align} 
Note that  the natural continuation of scalar-quark bilinear \eqref{pattern-5}  to the QCD-like regime  where scalar decouples is given by
four-fermi operators: 
 \begin{align}
q_{Ra}q_{L}^{b} \longleftrightarrow  \langle   \psi_{R b}   \psi_{L}^{a}    \tr \lambda \lambda \rangle
 \label{pattern-6}
 \end{align}
the expression given in \eqref{condensate2}. 

\vspace{3mm}
Similarly,  the  baryon multiplet  in $N_f= N_c$  SQCD and its fermionic component are   given by 
\begin{align}
 B_L=  \epsilon_{a_1 \ldots a_{N_f} } Q_L^{a_1} \ldots Q_L^{a_{N_f}} 
 =  \ldots  + \theta  \epsilon_{a_1 \ldots a_{N_f} } q_L^{a_1} \ldots q_L^{a_{N_f-1}}  \psi_L^{a_{N_f}} + \ldots 
 \end{align}
 Therefore, the fermionic baryon $\psi_{B_L}$  (and $\psi_{B_R}$)   can be written as\footnote{Note that in QCD(F), baryons are fermionic for $N_c=$ odd
 and bosonic for $N_c=$ even. In QCD(F/adj), we can have fermionic baryons for either choice of $N_c$.} 
 \begin{align} 
\psi_{B_L}&= \epsilon_{a_1  \ldots a_{N_f} }   (\lambda \psi_L^{a_1}) \ldots   (\lambda \psi_L^{a_{N_f-1}})   \psi_L^{a_{N_f}}\cr
\psi_{B_R}&= \epsilon^{a_1 \ldots a_{N_f} }   (\lambda \psi_{Ra_1}) \ldots   (\lambda \psi_{Ra_{N_f-1}})   \psi_{Ra_{N_f}}
  \label{baryon}
\end{align} 
The quantum numbers of these composite fermions under   unbroken symmetry 
\eqref{pattern2} are given by 
  \begin{align}
\begin{tabular}{ |c|c|c|c| } 
 \hline
&  $SU(N_f)_V$ & $U(1)_V$ & $U(1)_{A_D}$  \\ 
\hline \hline
 $\psi_{Ma}^{\;\;\;\;\; b}$ &  {\bf adj}    &  0   & -1 \\ 
  $\psi_{B_L}$ &  {\bf 1} & $N_c$ &-1 \\
  $\psi_{B_R}$  &   {\bf 1} & $-N_c$ &-1 \\
\hline
\end{tabular}
\label{composite}
\end{align}
The non-trivial anomalies in the UV   and IR are (set $N_c=N_f$ in all formulas: )
\begin{align} 
U(1)_{A_D}: \qquad & -2 N_f N_c  + (N_c^2 -1)  = - (N_f^2-1)  -2 \cr  \cr
U(1)_{A_D}^3: \qquad&  -2 N_f N_c  + (N_c^2 -1)  = - (N_f^2-1)  -2  \cr \cr
U(1)_V^2 \times  U(1)_{A_D}: \qquad&   -2 N_f N_c     =-2 N_c^2  \cr \cr
SU(N_f)_V^2 \times  U(1)_{A_D} \qquad &  - N_f d^{(2)}(\Box)  - N_f d^{(2)}(\bar \Box) = -d^{(2)}({\bf adj})
\label{0-anomaly}
\end{align}
 where $d^{(2)}(\Box)=\half,  d^{(2)}({\bf adj})= N_f$ are the  corresponding quadratic $SU(N_f)$  Casimir operators. 
 Needless to say, these are the anomaly matching conditions for $N_f=N_c$ SQCD as well \cite{Seiberg:1994bz} as they must be, because we can obtain QCD(F/adj) by decoupling the scalar in supersymmetric  theory.  This point is also emphasized in the analogous  discussion of $\N=2$ SYM theory and  its non-supersymmetric deformation to QCD(adj)  with $n_f=2$ flavors \cite{Cordova:2018acb}.

 \vspace{3mm}
The  0-form  't Hooft  anomalies of the UV theory are matched  by the massless composite IR fermions, $N_f^2 -1$ fermionic meson and massless baryon. This suggests that apart from the NG-bosons and the composite fermions \eqref{composite}, there should not be any other gapless degrees of freedom.

\vspace{3mm}
Note one crucial distinction from   QCD(F).  Since both adjoint as well as fundamental fermion bilinears are charged 
under  $U(1)_{A_D}$,  a constituent quark mass cannot be created 
with the chiral symmetry breaking pattern  \eqref{pattern2} in QCD(F/adj).

\subsection{Exact color-flavor center  symmetry on $\R^3 \times S^1$} 
\label{CFC-sec}
 It is well-known that introducing fundamental fermions  in $SU(N_c)$ gauge theory breaks 1-form  $\Z_{N_c}^{[1]}$ center-symmetry explicitly and completely, as all Wilson lines become endable   on quarks. 
  Obviously,   in the theory compactified on a circle  $\R^3 \times S^1$, it also breaks the 0-form center-symmetry that acts on the Wilson line (Polyakov loop)  wrapping $S^1$ circle.
   It is recently understood that in QCD(F) with $N_f$ fermions,   it is actually possible to preserve  a $\Z_{{\rm gcd}(N_f, N_c)}$  sub-group of 0-form center-symmetry acting non-trivially on Polyakov loop  by paying an appropriate price.  
   This procedure does  not restore  a 1-form    center  symmetry on $\R^3$. Below, we describe the appearance of 
 0-form center-symmetry in compactified theory.  
 
  \vspace{3mm}
The center-symmetry in pure Yang--Mills theory is a 1-form symmetry acting on Wilson line operators on $\R^4$.  On the 
theory compactified on $\R^3 \times S^1$, it decompose into a  0-form symmetry acting non-trivially on   Polyakov loops 
(which becomes a point operator from the $\R^3$ point of view),  and 1-form symmetry acting on line operators on $\R^3$. 
Traditional way to think about   0-form center-symmetry is sufficient for our purpose. 
0-form center-symmetry may be  associated with  gauge transformations $g(x_4)$ aperiodic up to an element  of center group, $g(x_4+ \beta) = \omega^{-1} g(x_4), \; \omega^{N_c}=1$.  Polyakov loop transforms under it as 
$ \tr   e^{i \int_0^{\beta} a_4 \; dx_4}   \rightarrow \tr  ( g(0)  e^{i \int_0^{\beta} a_4 \; dx_4} g^{\dagger}(\beta)) =  \omega  \tr   e^{i \int_0^{\beta} a_4 \; dx_4} $. Therefore, 
\begin{align}
\Z_{N_c}:  \;    \tr \Omega ({\bf x})\equiv  \tr  e^{i \int_0^{\beta} a_4 \; dx_4}  \mapsto   \omega   \tr \Omega ({\bf x}) 
\end{align}
For example,  the holonomy  potential for  pure Yang--Mills theory  that  one obtains at small-$\beta$  by integrating out heavy modes  is composed of the terms like  $|\tr (\Omega^n)|^2, n=1,2, \ldots $  and  are manifestly invariant under this symmetry  \cite{ Gross:1980br}.

    \vspace{3mm}
Let us now describe  how the 0-form center-symmetry is violated in the presence of fundamental fermions. Start with anti-periodic (or any other flavor independent) boundary condition  for fermions,  $\psi (x_4+ \beta) = -\psi (x_4)$. Now, consider a gauge transformation of this condition,  $\psi^G (x_4+ \beta) =- \psi^G (x_4)$, where 
 $G$  is via  a transformation $g(x_4)$ aperiodic up to an element  of center group, $g(x_4+ \beta) = \omega g(x_4)$. 
 Then,  
 $\psi (x_4+ \beta) = -\omega \psi (x_4)$. Therefore,  the center-symmetry transformation   does not respect the original boundary conditions, and leads to explicit breaking of center symmetry. For example, the one-loop potential one obtains at small-$\beta$  by integrating out fermions induce  terms of the form  $\tr (\Omega^n), n=1,2, \ldots $, which   does explicitly break the center symmetry \cite{ Gross:1980br}.
 
 \vspace{3mm}
Surprisingly,  the  0-form center symmetry can actually be ``rescued" even in the presence of fundamental fermions. 
This is appreciated  as a genuine symmetry and called  color-flavor center (CFC)  symmetry  in  \cite{Cherman:2017tey}.   
Also see earlier related work \cite{Iritani:2015ara,  Cherman:2016hcd, Tanizaki:2017qhf}.  
Assume   $N_f=N_c$   momentarily. This assumption will be relaxed.  Impose an $U(N_f)_V$ flavor twisted boundary condition on fermions, where we choose the flavor twist to be  
\begin{align}
 \Omega_F^0 = \textrm{diag}
(1, \omega, \cdots, \omega^{N_f-1}), \qquad \omega = e^{2\pi i /N_f} . 
\label{flavor-hol-1}
\end{align}
 Now, fermion  boundary conditions are
\begin{align}
\psi (x_4+ \beta) = -  \psi (x_4)  \overline \Omega_F^0
\label{tbc-f}
\end{align}
and under a gauge transformation aperiodic up to an element of the center, 
$\psi^G (x_4+ \beta) = -  \psi^G (x_4)  \overline \Omega_F^0  $.  This amounts to changing the 
 boundary conditions   into $\psi (x_4+ \beta) = - \psi (x_4)   \omega^{-1} \overline \Omega_F^0  $. Since the boundary conditions are different, in general, this is again a non-invariance and explicitly breaks center-symmetry.  However, 
  $\omega \Omega_F^0 $ is a cyclic permutation of $\Omega_F^0$ and can be brought to the original boundary conditions by using a flavor rotation generated by the shift matrix $(S)_{a, b} = \delta_{a+1, b}$:
 \begin{align}
 S=  \begin{bmatrix}
  0&1& 0 & \cdots &0  \\
  0&0&1&\cdots &0  \\
  \vdots & \vdots & \ddots &  & \vdots \\
   0&0&0&\cdots &1 \\
  1 & 0 & 0  &   & 0
  \end{bmatrix}
  \end{align}
   obeying the algebra:
  \begin{align}
   S  \Omega_F^0  S^{-1} = \omega  \Omega_F^0,  \qquad S \in  \Gamma_S \subset  SU(N_f)_V 
   \label{algebra}
   \end{align}
   where  $ \Gamma_S \subset  SU(N_f)_V$ is the cyclic permutation subgroup of the vector-symmetry. 
Therefore, the theory with $N_f =N_c$ possesses an exact diagonal  0-form center symmetry,  
\begin{align}
 \Z_{N_c} \; \; \text{CFC-symmetry}:   \qquad  \tr \Omega ({\bf x}) \mapsto   \omega   \tr \Omega ({\bf x}),  \qquad \psi_a \mapsto  \psi_{a+1} 
\label{CFC}
\end{align}
which is  an admixture of  center of gauge group  and   $ \Gamma_S$ subgroup of  flavor rotation.      Since  CFC symmetry intertwines  0-form part of the center  symmetry on $\R^3 \times S^1$, and flavor transformations, it has both local and extended order parameters.   
The order parameters are 
Polyakov loops and 
Fourier transforms of fermion bilinears with respect to flavor index:  
\begin{align} 
 \widehat{  (\psi_{R}  \psi_{L})}_p:
    \equiv
\frac{1}{\sqrt{N_c}}    \sum_{a=1}^{\nf} \omega^{-a p} \,
    \psi_{Ra}  \psi_{L}^{a}:
\end{align}
which transform  non-trivially under CFC-symmetry:
\begin{align}
\label{eq:CFC_action}
 \Z_{N_c}: \qquad &
    \tr \Omega^{p} \; \mapsto \; \omega^{p} \>
    \tr \Omega^{p}\, ,
    \\
    \quad &
 \widehat{  (\psi_{R}  \psi_{L})}_p 
\;\mapsto \; \omega^{p} \>
 \widehat{  (\psi_{R}  \psi_{L})}_p
\end{align} 
Therefore,   $ \tr \Omega^{p}$ is the natural order parameter for CFC symmetry, just like Polyakov loop is the order parameter of center-symmetry in pure Yang--Mills theory.

  \vspace{3mm}
 The CFC  0-form symmetry is present for any  $m_\psi \geq 0 $.  In the limit  $m_\psi \rightarrow  \infty $, the CFC symmetry  is part of genuine 
  $\Z_{N_c}^{[1]}$ center-symmetry  in pure Yang--Mills theory. For general $N_f$,    CFC  0-form symmetry becomes: 
 \begin{align}
\text{CFC-symmetry}:  \Z_{{\rm gcd}(N_f, N_c)}  \qquad {\rm general} \;  N_f
\label{CFC-2}
\end{align}
 
 \vspace{3mm}
 The  exact CFC symmetry will also be important and manifest itself in beautiful ways in  in our discussion of one- and two-loop   potential for gauge holonomy  in QCD(F/adj) where it will be a manifest symmetry as described in Subsection \ref{sec:frustrate}.  More importantly, the realization of 
 of the  CFC symmetry at small $\R^3 \times S^1$   will be important for the idea of adibatic continuity in QCD(F/adj), and  
  in the realization of Nye-Singer index theorem for fermions as described in Section \ref{sec:topology}. 
 
\vspace{3mm}
The price one pays for keeping an exact 0-form center-symmetry appears as explicit reduction of the non-abelian chiral symmetry to its maximal abelian torus. This is  because $\Omega_F^0$ only commutes with the Cartan generators of $SU(N_f)_V$.  
 Therefore, with twisted boundary conditions, the global symmetry is  explicitly broken 
 down to  maximal torus and  the global symmetry of the compactified theory becomes 
 \begin{align}
{\bf G}_{\rm max-ab}={U(1)^{N_f-1}_L  \times U(1)^{N_f-1}_R \times U(1)_V \times U(1)_{A_D}   \times \mathbb{Z}_{2\mathrm{gcd}(N_c,N_f)}\over \mathbb{Z}_{N_c}\times (\mathbb{Z}_{N_f})_L\times (\mathbb{Z}_{N_f})_R\times (\mathbb{Z}_2)_\psi}.
\label{MAG}
\end{align}
 With these boundary conditions, only  
 $N_f$ Nambu--Goldstone   bosons remain gapless at large-$S^1$,  while the off-Cartan  NG-bosons acquire masses  of order $\frac{2 \pi}{ \beta N_f}$.   The chiral Lagrangian reduce to a  non-linear sigma model  on   the maximal torus, as described in Subsection \ref{sec:acont}.

   \section{Frustration, collapse and  a new governance:    Gauge-holonomy  potentials in the presence of  flavor-holonomies}  
 In this section, we describe  some  implications of the grading over the Hilbert space via $(-1)^F   e^{i \pi Q_0}   \prod_{a=1}^{N_f} e^{i \frac{2\pi a}{N_f}  Q_a}  $  in terms of gauge holonomy potentials.  We will show that the grading 
 and consequent quantum distillation maps to an intriguing phenomenon in gauge holonomy potentials.   
 Gauge holonomy potential is  a function of   ${\rm tr}_{\cal R} (\Omega^n) $ for  representations ${\cal R}$ appearing in the microscopic theory. 
 We demonstrate explicitly that some reprepresentations appearing in holonomy potential are prone to frustrations and their effects collapse to zero, while some others are immune and their effect dictates the new  ground states and/or thermal equilibrium states.  This is a fairly entertaining story, as pointed in the section title and depicted in Fig.~\ref{fig:holonomy}.

\vspace{3mm}
 As described in the Introduction,  gauge holonomy $\Omega$ is dynamical and its value at small-$\beta$ is determined by the  potential $V[\Omega]$.  In thermal gauge theory,   this type of potentials are  well understood  at one-loop \cite{Gross:1980br, Weiss:1981ev} and two-loop orders \cite{KorthalsAltes:1993ca,KorthalsAltes:1999cp} and \cite{Guo:2018scp} and in some supersymmetric gauge theories where perturbative-loop  potential vanishes,  it is understood non-perturbatively   \cite{ Poppitz:2012sw, Poppitz:2012nz}.

\vspace{3mm}
It is easy to implement  the non-dynamical $\Omega_F$ flavor backgrounds and examine their  physical effects on the gauge holonomy potentials.   This is the image of quantum distillation of Hilbert space in path integral formalism. We will obtain quite surprising results in QCD(F/adj) as well as QCD(F) for these potentials and their  extrema.
 
 \vspace{3mm}
One  goal is to determine the realization of the color-flavor center symmetry  \eqref{CFC} at small-$\beta$ regime for $SU(N_c )$ gauge theory.  
For $N_c \geq 3$,  we show that 
in the presence of the flavor twist $\Omega_F^0$,  there is an exponentially increasing  number of  degenerate minima   at one-loop order in contradistinction with thermal case, where there is a unique deconfined minimum  at $\Omega=1$. 
 The  degeneracy at one-loop order becomes continuous   in the Veneziano-type  large-$N_c$ limit where the one-loop potential in QCD(F/adj) vanishes. 
This is surprising considering that the theories we are dealing with  are  non-supersymmetric, and  characteristic scaling of the potential must be of order $N_c^2$.  
  At two-loop order, we will show that for the grading that achieves perfect quantum distillation of Hilbert space ($\Omega_F^0$)  leads to a center-symmetric minimum for the holonomy potential!

\begin{figure}[t]
\begin{center}
\includegraphics[width = 1\textwidth]{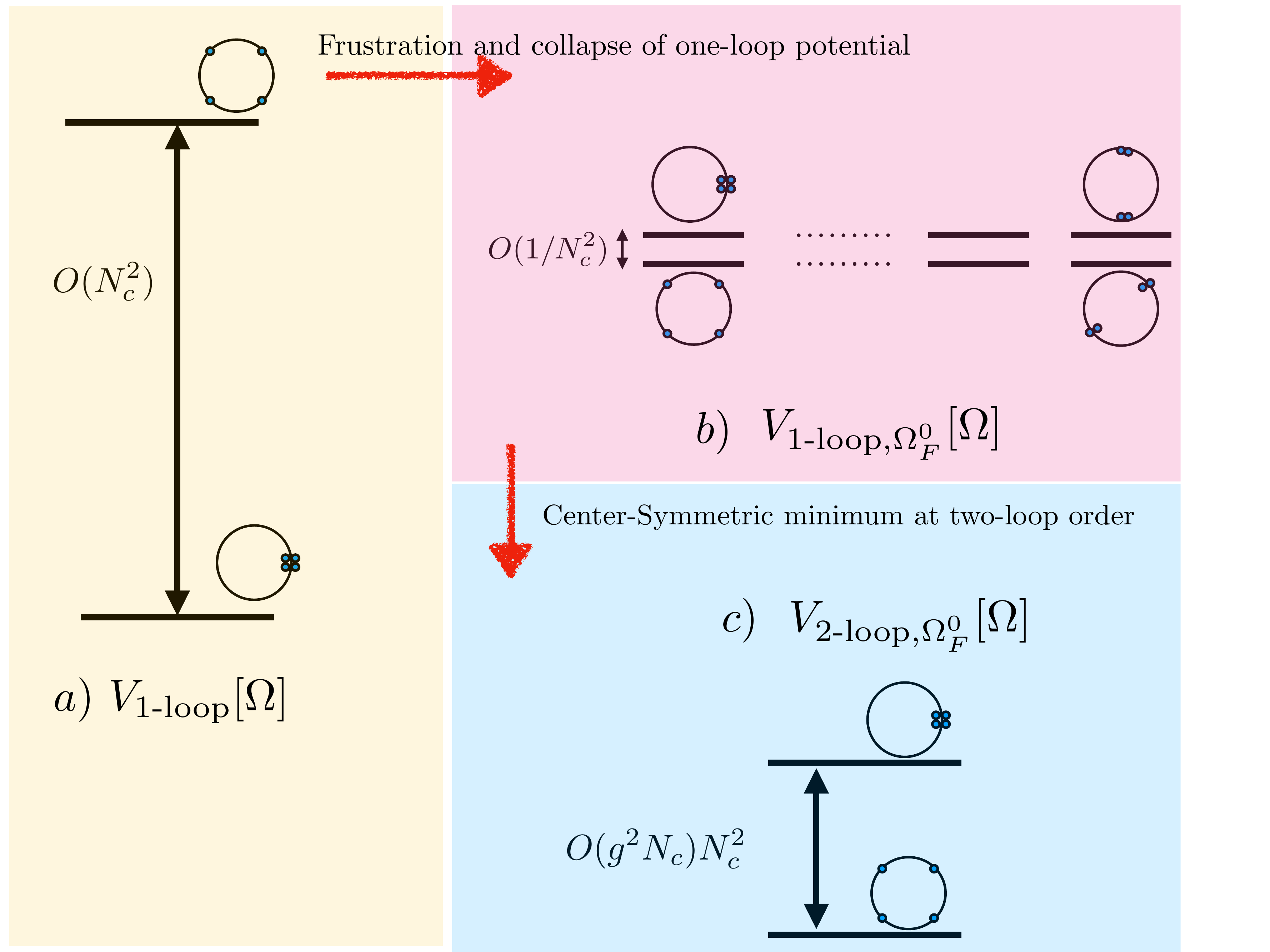}
\caption{The figure describes the image of Hilbert space distillation in path integral and gauge holonomy potential description. 
a) This is the standard implication of Gross-Pisarski-Yaffe analysis. Center-broken configuration is the minimum and governs 
ground state (thermal equilibrium state properties.) In particular, the center-symmetric configuration is extremely disfavored. 
The potential gap between min and max is order $N_c^2$.  
b) The image of distillation of ${\cal H}$ is the frustration of GPY potential at one-loop. Now, there are exponentially increasing number of min and max. In particular, the center-symmetric and center-broken configurations are almost on the same footing as the gap between min and max is order  $1/N_c^2$.  c) At two-loop, there is a center-stabilizing double-trace term in the potential which is resistant to frustration. That term  decides the new ground state of the theory, which is now center-symmetric. This holds to all orders in perturbation theory and non-perturbatively. }
\label{fig:holonomy}
\end{center}
\end{figure}

\subsection{Frustration and  collapse of one-loop  Gross-Pisarski-Yaffe potential}
 \label{sec:frustrate}
 Before turning on the $\Omega_F$ twist, it is useful to remind the thermal one-loop potentials. In the small-circle limit where the theories  are weakly coupled, the one-loop potential for the gauge holonomy  \eqref{vev} receives contributions from the 
 weakly coupled $(A_{\mu}, \lambda, \psi^a)$ fields:
 \begin{align} 
 V_{\rm 1-loop, thermal}(\Omega)   & = \frac{2 }{\beta}  \int \frac{ d^3 p}{(2 \pi)^3}   \left[   + \sum_{i,j} \log (1 - e^{-\beta p + i v_{ij} })  -  \sum_{i,j} 
  \log (1 + e^{-\beta p + i v_{ij} 
    })   \right. 
  \cr 
 &  \left.  \qquad \qquad \qquad  \;\;  -  N_f  \sum_{i} \left( \log (1 + e^{-\beta p + i v_{i}  } )  + \rm c.c. \right) \right]   \cr 
& =  \frac{2}{\pi^2 \beta^4}  \sum_{n=1}^{\infty}  \left[  (-1 + (-1)^n )  \frac{1}{n^4}           |\tr (\Omega^n)|^2   +  
N_f  \frac{(-1)^n}{n^4}         \left(    \tr (\Omega^n)+ {\rm c.c.}  \right)  \right]  \qquad
\label{1-loop-thermal}
 \end{align} 
As the center-symmetry is explicitly broken by the fundamental fermions with anti-periodic boundary conditions, this potential has terms of the form  $\tr (\Omega^n), n=1,2, \ldots$ which  indeed violates center symmetry.     The global minimum of the potential is located at $\Omega=1$. Although this is a result of one-loop calculation, it is true to all orders in perturbation theory and non-perturbatively \cite{Gross:1980br} in small-$\beta$ domain.

\begin{figure}[t]
\begin{center}
\vspace{-1.5cm}
\includegraphics[width = 1.2\textwidth]{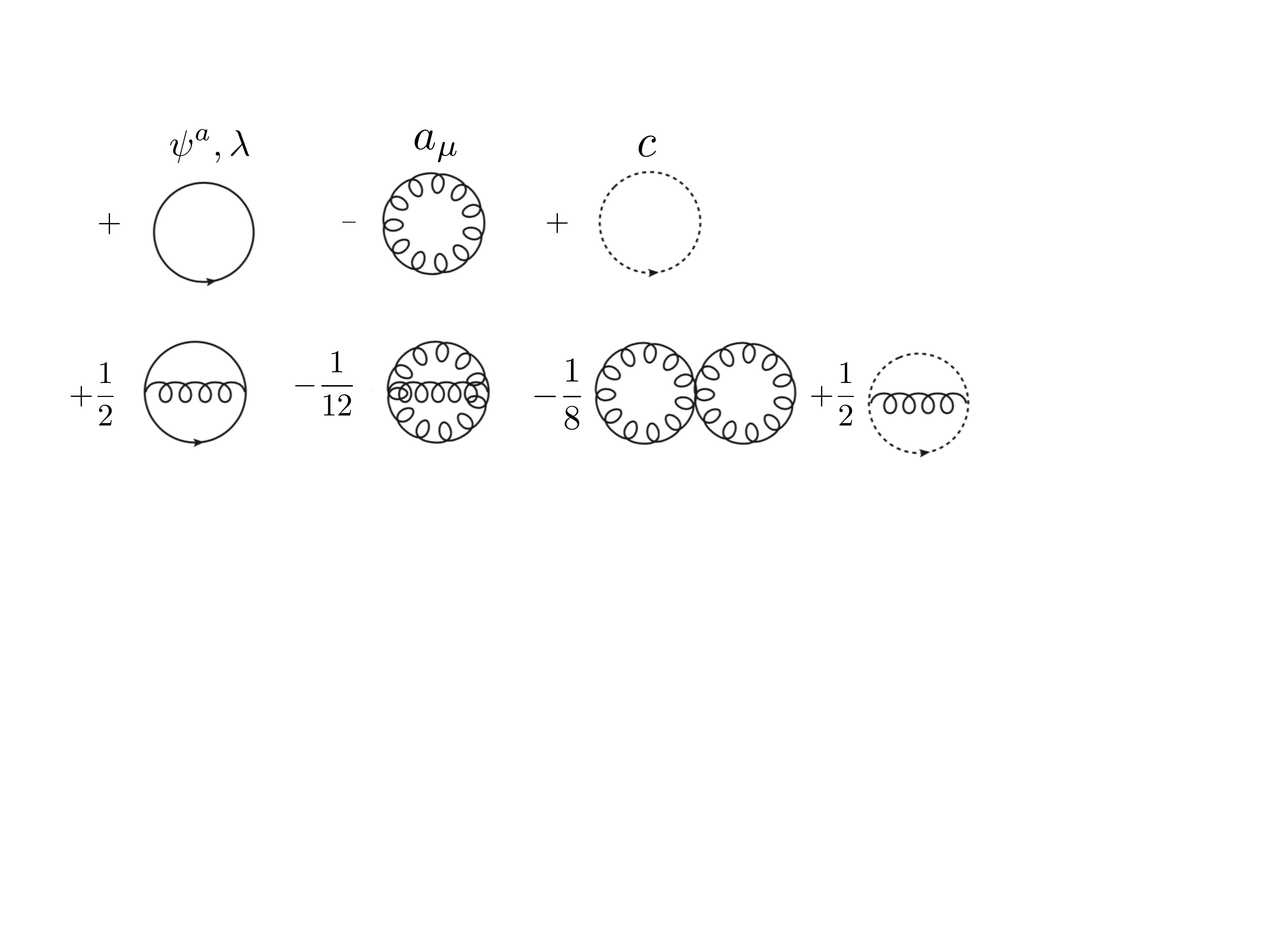}
\vspace{-7.5cm}
\caption{Feynman diagrams for the gauge holonomy potential at one- and two-loop order for $N_f$-Dirac fundamental fermions and one-adjoint Majorana 
($N_{\rm adj}^{\rm Dirac}= \half$) fermion.  Diagrams are  reproduced from  \cite{KorthalsAltes:1993ca, Guo:2018scp}  and slightly generalized according to new matter content.   Up to two-loop order,   the adjoint and fundamental fermions are decoupled in their contributions. In the graded case, where we insert $(-1)^F 
 \prod_{a=1}^{N_f} e^{i \epsilon_a Q_a} $ into the partition function, we determine the  potential for the dynamical gauge holonomy $\Omega$ in the fixed background of $\Omega_F$ flavor holonomy. 
 }
\label{fig:feynman}
\end{center}
\end{figure}

\begin{figure}[t]
\vspace{-0.5cm}
\begin{center}
\includegraphics[width = 0.5\textwidth]{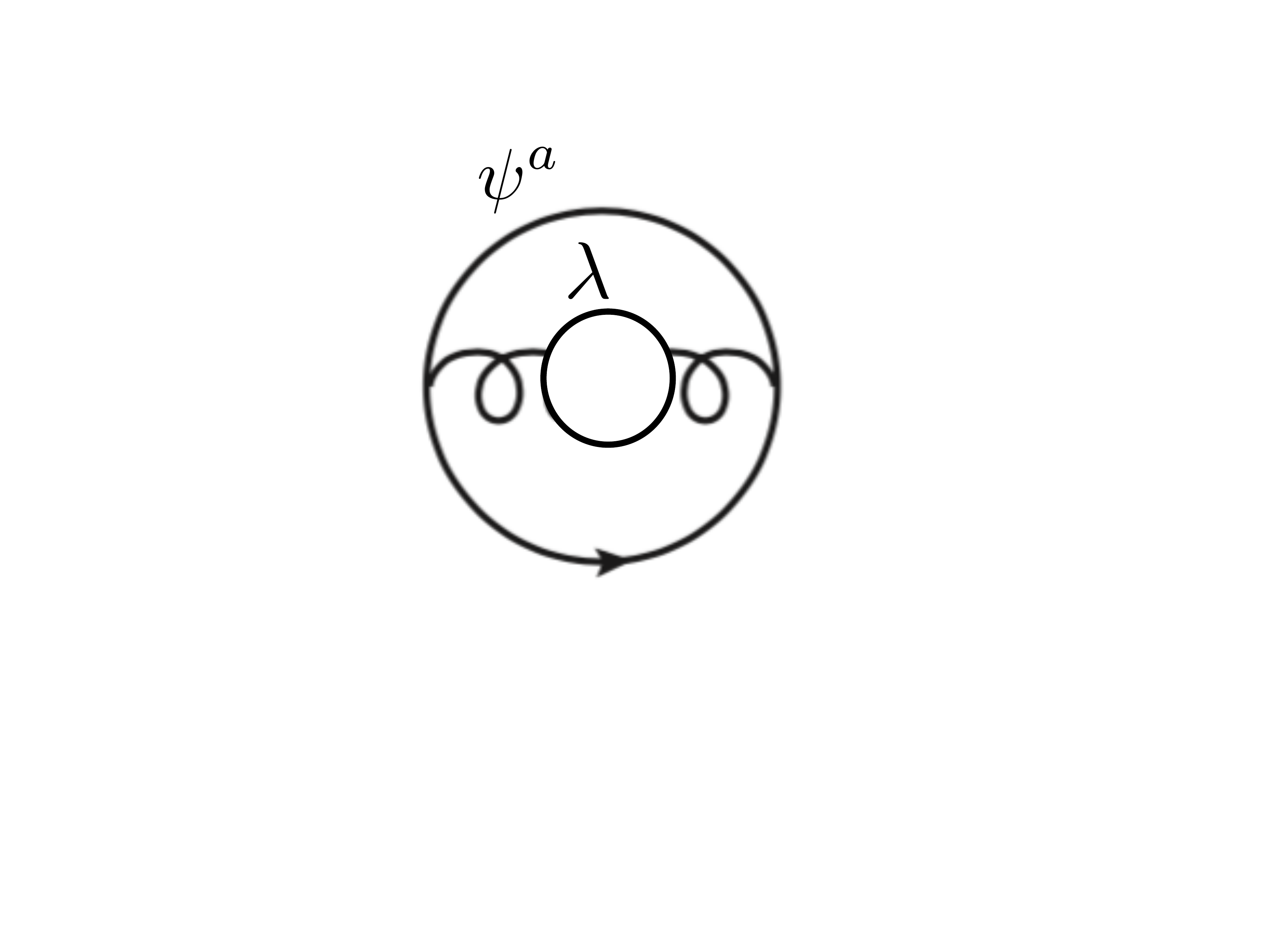}
\vspace{-2.5cm}
\caption{In the determination of gauge holonomy potential, adjoint and fundamental fermions only start to couple at three-loop order in diagrams as above.  However, as explained in the text, two-loop result will suffice for the determination of CFC-symmetry. 
 }
\label{fig:three-loop}
\end{center}
\end{figure}

\vspace{3mm}
\noindent
{\bf A quixotic quest:}  Assume momentarily $N_f=N_c$. 
Clearly, all three contributions in \eqref{1-loop-thermal}  works against what we would like to achieve, and all three terms  prefer $\Omega \propto 1$ as the global minimum. Moreover, all the terms in the fundamental fermion contribution, except the ones of the form $ \tr (\Omega^{N_c  k} )$     explicitly break center-symmetry.  We would like to achieve  two things. 
\begin{itemize}
\item Make the potential CFC-symmetric by using appropriate $\Omega_F$. 
\item Make sure CFC-symmetry is unbroken and   $ \Omega  \propto  \text{diag}(1,\omega,\cdots,\omega^{N_c-1})$ is  the global minimum   of the potential. 
\end{itemize}
Obviously, this goal sounds surreal. But it  is crucial for the  realization of the adiabatic continuity idea. However, the thermal potential for center-symmetric configuration  is of order   $O(N_c^{-2})$,   while the potential for the global minimum is $-O(N_c^2)$, and it is not a priori obvious how to get rid of terms like  $ N_f  \tr (\Omega^n), n \neq N_c k$  in the potential.    It may seem  almost   impossible 
to overcome these difficulties  and achieve   a center-symmetric minimum for QCD(F/adj) or  QCD(F) with heavy adjoint fermion. On the other hand, we know that the holonomy potential at its minimum corresponds to free energy of the QFT, and quantum distillation must alter free energy drastically.   Therefore, if the quantum distillation idea ought to be operative at the path integral level, it must alter the potential in useful ways. In other words, 
in the space of graded partition functions, there may be interesting opportunities and indeed, there are!

 \vspace{3mm}
 In order to capture the graded partition function, we impose somehow exotic looking boundary conditions on fermions. The implication of these boundary conditions on gauge invariant states in the Hilbert space (such as mesons, baryons etc)  will be discussed in  Section \ref{sec:pheno}.   
 In fact, by using the idea of quantum distillation, we will {\it derive} the most efficient  boundary conditions as an extremum of twist free energy.

 \vspace{3mm}
 The  boundary conditions on fermions in the path integral  formulation, which is the   mapping of the \eqref{GPF}  in operator formalism,   are:  
 \begin{align}
\lambda(x_4 + \beta) &= (+)  \lambda(x_4),  \cr
  \psi(x_4 + \beta) &=   (+)    \psi(x_4) e^{i  \pi}  \overline \Omega_F 
  \label{bc}
\end{align}
where  $(+)$  is due to $(-1)^F$  insertion into the state-sum, and  $e^{i  \pi}   \Omega_F $  is a  $U(N_f)_V $ twist. 
  The boundary condition on gauge connection is periodic.

 \vspace{3mm}
 The one-loop potential for gauge holonomy $\Omega$ in the presence of the (non-dynamical)   $\Omega_F$  background  and $(-1)^F$ 
 is given by 
  \begin{align} 
 V_{\rm 1-loop, \Omega_F}(\Omega)   & = \frac{2 }{\beta}  \int \frac{ d^3 p}{(2 \pi)^3}   \left[   + \sum_{i,j} \log (1 - e^{-\beta p + i v_{ij} })  -  \sum_{i,j} 
  \log (1 + e^{-\beta p + i v_{ij} + \color{red}{ i \pi}  })   \right. 
  \cr 
 &  \left.  \qquad \qquad \qquad  \;\;  - \sum_{a=1}^{N_f}  \sum_{i} \left( \log (1 + e^{-\beta p + i v_{i} - {\color{red} i \epsilon_a}  })  + \rm c.c. \right) \right]    
 \end{align} 
 The  first sum is   center-destabilizing gauge boson contribution, the second is center-stabilizing  adjoint fermion contribution  and  the third one is twisted boundary condition fundamental fermion contribution.  
  In terms of gauge-invariant Wilson lines,  the potential can be written as: 
 \begin{align}
 V_{\rm 1-loop,  \Omega_F} & =  V^{\rm  gauge}_{\rm 1-loop} + V^{\rm \lambda}_{\rm 1-loop}  +  V^{\rm \psi}_{\rm 1-loop, \Omega_F}\,,\cr
 V^{\rm  gauge}_{\rm 1-loop} + V^{\rm \lambda}_{\rm 1-loop}  & = (-1 +1) 
 \frac{2}{\pi^2 \beta^4}  \sum_{n=1}^{\infty}   \frac{1}{n^4}   
        |\tr (\Omega^n)|^2  =0 \,, \cr 
V^{\rm \psi}_{\rm 1-loop, \Omega_F} & =
 \frac{2}{\pi^2 \beta^4}  \sum_{n=1}^{\infty}  \frac{(-1)^n}{n^4}    
   \left[\tr (\Omega^n)   \tr (\bar \Omega_F^n)  + \text{c.c.}\right]  .
  \label{one-loop-A}
\end{align}
The  gauge boson  and  one-adjoint fermion  (with periodic boundary condition) contributions cancel each other out 
$V^{\rm \lambda} +  V^{\rm gauge}=0 $ as  in  the  $\N =1$ supersymmetric Yang--Mills theory, not only at one-loop  level but to all-loop orders in perturbation theory.\footnote{More precisely, $V^{\rm \lambda} +  V^{\rm gauge}=0 $ in the $N_f=0$ theory.  For general $N_f$, the subset of diagrams involving  gauge fluctuations and {\it only} adjoint fermion loops  must cancel each other exactly as in $\N=1$ SYM.  At 3-loop order and higher, there are diagrams involving both adjoint and fundamental fermion loops, such as  Fig.~\ref{fig:three-loop}.  These diagrams will not cancel against gauge fluctuations, but they will not cause any difficulty either.}
The center-symmetry breaking effect of gauge fluctuations are undone by the quantum fluctuations of  periodic adjoint fermions. The final decision about the presence/absence of CFC symmetry  and its realization is given by fundamental fermions. 
In \eqref{one-loop-A},   $\tr (\Omega^n) $ is dynamical, determined by extremizing the potential, and  $ \tr (\bar \Omega_F^n)$ is non-dynamical. It is the boundary condition we choose. Clearly, the behavior of the small-$\beta$ theory at one-loop order  crucially depends on our choice of 
  $\Omega_F$.\footnote{The flavor holonomy $\Omega_F$  is indeed a choice at this stage, and we are free to choose any boundary condition. However, not all choices are on equal footing. It is meaningful to ask which choice corresponds to  
  having a CFC-symmetry, or a persistent mixed anomaly or is the 
  most efficient  quantum distillation.   From the last point of view,  there is an extremization problem, despite $\Omega_F$ 
  being non-dynamical!   One can write down 
a twist    free-energy in confined phase  for hadrons and extremize it. Its maximum correspond to $\Omega_F= \Omega_F^0 $.}

\vspace{3mm}
Now comes one of the main  points. For a generic  flavor holonomy backgroud $\Omega_F$, the potential explicitly breaks 
 $\Z_{N_c}^{[0]}$ center-symmetry, which acts on Polyakov loop  $\tr (\Omega^n)  \rightarrow \omega^n  \tr (\Omega^n)$.   For special values of $\Omega_F$,  the one-loop potential is invariant under  various subgroups of  $\Z_{N_c}$. Of course, this is not an accident and in fact, holds to all-loop orders and non-perturbatively.  It is just manifestation of the  relevant subgroup of 
 $ {  \Z_{{\rm gcd}(N_f, N_c)} \;  \rm CFC} $ symmetry as described in Section \ref{CFC-sec} and also 
\eqref{eq:CFC}.  

\vspace{3mm}
The twisted boundary condition fundamental fermion induced potential having  terms of the form $ \tr (\Omega^n) \tr (\bar \Omega_F^n) $ is just the manifestation of the fact that $\psi$ transforms as  fundamental under color and anti-fundamental under flavor: 
\begin{align}
SU(N_c) \times SU(N_f)_V: \;\; \psi (x) \rightarrow g_c(x)  \psi(x)  g_f^{\dagger} 
\end{align}
 where the first one is gauge structure and latter is vector-like flavor symmetry. If flavor symmetry were to be gauged, this would indeed be the holonomy potential for fermions in QCD with bi-fundamental fermions, see e.g.  \cite{Tong:2002vp,  Shifman:2008ja}. 
  But in our current discussion,   $SU(N_f)_V$ is not gauged, only a background is turned on. 
 
 \vspace{3mm}
 For $\Omega_F= \Omega_F^0$ given in   \eqref{flavor-hol-0}, the potential becomes invariant under the full  $ {  \Z_{{\rm gcd}(N_f, N_c)}} $   CFC-symmetry.  Indeed, 
    in the $V^{\rm \psi}_{\rm 1-loop, \Omega_F^0} $,  all terms which transform non-trivially under the CFC
vanish identically:  
\begin{align}
&\tr (\Omega_F^0)^n =0 \qquad {\rm for}  \; \;  n\neq 0 \mod N_f \cr
&\tr (\Omega_F^{0})^{N_f k} = N_f  
\end{align}
Consequently, $V^{\rm \psi}_{\rm 1-loop, \Omega_F^0} $ becomes  a sum of the terms of the type  $\tr (\Omega^{N_fk})$  and the  full one-loop potential 
takes the simple form: 
 \begin{align}
 &V_{\text{1-loop}, \Omega_F^0}   =  \frac{2}{\pi^2  N_f^3 \beta^4}  \sum_{k=1}^{\infty}   \frac{(-1)^{ N_f k}}{ k^4}      \left[  \tr (\Omega^{N_fk}) + \text{c.c.}\right]  
 \label{one-loop}
\end{align}
This is a quite peculiar potential whose number of degenerate  minima (and maxima) is exponentially increasing for $N_f\sim N_c$. \footnote{The vacuum structure for $N_c=N_f=3$ was analyzed numerically in \cite{Kouno:2013mma}.}  
The characteristic size of the potential   \eqref{one-loop} is of order  $O \big( \frac{N_c}{N_f^3} \big)$, much  smaller compared to thermal gauge holonomy  potential,  which is   of order $O(N_f N_c)$ induced by thermal fundamental fermion induced term in  \eqref{1-loop-thermal}.

\subsection{Counting the min-max  and exponential degeneracy at one-loop} 
\label{sec:minimum}
 The one-loop potential \eqref{one-loop} is valid for general $N_f$ and $N_c$. 
Below, we describe the minima and maxima  of this potential for the  case $N_f= N_c = N$.\footnote{We do determine both minima and maxima, because  in the graded partition function, by turning on a $U(1)_V$-twist,  we can  switch the minima with maxima. }
We dropped the subscript to lessen the clutter. 
The main contribution to the one-loop potential \eqref{one-loop} comes from the $k=1$ term.  This amounts to finding the
extremum of the potential 
\begin{align}
V(v_i)=  \sum_{i=1}^{N}   (-1)^N \cos (N v_i), \qquad \sum_{i=1}^{N}  v_i   = 0  \mod 2\pi
\label{toy}
\end{align} 
within the Weyl-cell of the $SU(N)$. The Weyl group of $SU(N)$ is the permutation group $S_N$. Therefore, configurations  $(v_1, \ldots v_N) $ related to one another by Weyl permutations are gauge equivalent and should not be counted independently. 
This is the process of removing gauge redundancy by using the quotient with $S_N$. 
The minima  of the potential 
is thus given by $\Omega$ such that $\Omega^N=\1$   for odd $N$ and $\Omega^N=-\1$   for even $N$. 
  Furthermore, $\det \Omega=1$. There are four distinct cases.

\vspace{3mm}
\noindent 
{\bf Minima set conditions, $N$-odd}
\begin{enumerate}
\item $\Omega =  \text{diag}(e^{ i  \frac{2 \pi}{N} a_1}, \ldots, e^{ i  \frac{2 \pi}{N} a_N})$
  \item Each $a_i$ takes a number from $0$ to $N-1$.
  \item $a_1\leq a_2\leq \cdots \leq a_N$.
  \item $ \sum_{i=1}^{N}a_i=0$ (mod $N$). 
\end{enumerate}

\vspace{3mm}
\noindent 
{\bf Maxima  set conditions, $N$-odd}
\begin{enumerate}
\item $\Omega = \text{diag}(e^{ i  \frac{2 \pi}{N^2} a_1}, \ldots, e^{ i  \frac{2 \pi}{N^2} a_N}) $
  \item Each $a_i$ takes a number from the set of $N$ positive-integers   $ \frac{N-1}{2} + 2N k, \;  k=0, N-1$ in the range   $0$ to $N^2$.
   \item  Choose $N$ $a_i$'s such that  $a_1\leq a_2\leq \cdots \leq a_N$.
  \item $\sum_{i=1}^{N}a_i=0$ (mod $N^2$). 
  \item Duplicate the same process with the set   $ \frac{N+1}{2} + 2N k,  \; k=0, N-1$.
 \end{enumerate}
 
\vspace{3mm}
\noindent 
{\bf Minima set conditions, $N$-even}
\begin{enumerate}
\item $\Omega =  \text{diag}(e^{ i  \frac{2 \pi}{2N} a_1}, \ldots, e^{ i  \frac{2 \pi}{2N} a_N})$
  \item Each $a_i$ takes an odd  number from $0$ to $2N-1$. 
  \item $a_1\leq a_2\leq \cdots \leq a_N$.
  \item $\sum_{i=1}^{N}a_i=0$ (mod $2N$). 
\end{enumerate}

\vspace{3mm}
\noindent 
{\bf Maxima set conditions, $N$-even}
\begin{enumerate}
\item $\Omega =  \text{diag}(e^{ i  \frac{2 \pi}{N} a_1}, \ldots, e^{ i  \frac{2 \pi}{N} a_N})$
  \item Each $a_i$ takes a  number from $0$ to $N-1$. 
  \item $a_1\leq a_2\leq \cdots \leq a_N$.
  \item $\sum_{i=1}^{N}a_i=0$ (mod $N$). 
\end{enumerate}

\vspace{3mm}
\noindent 
For $N=2$, there is a unique minimum given by $(a_1, a_2)=(1,3) \mod 4 $, which corresponds to $\Omega=\text{diag}(e^{i \pi/2},e^{i 3\pi/2})$, and center-symmetry is already stable at one-loop level. This is an exceptional case and with increasing $N$, vacuum degeneracy increases quickly. In Table~\ref{tab:list},  we list the   Minima-Set and Maxima-Set  for $N=2,3,4,5$.  Note that the value of the toy  potential \eqref{toy} for both $N$-even and  $N$-odd is $V_{\rm min}=-N$, the maximum  for $N$-even is at 
$V_{\rm max}=+N$, while for $N$-odd, it is located at $V_{\rm max}=N \cos \frac{\pi ( N \pm 1)}{N} = N \cos \frac{\pi} 
{N}$. This is the reason that the number of maxima for $N$-odd being roughly twice as many as the number of minima.

\begin{table}[htp]
{ \scriptsize
\begin{center}
\begin{tabular}{cc}
\begin{tabular}{| c|c|}  
	\hline
Min-SU(2) & Max-SU(2)\\ 
Mod  4& Mod  2\\
		\hline
		{\bf 13}   &  00  \\ 
		  &  11  \\ 
	\hline
	\hline
Min-SU(3) & Max-SU(3)\\ 
Mod  3& Mod  9\\
		\hline
		000   &  117  \\ 
		 {\bf 012}  &  144  \\ 
		  111 &    477  \\ 
		222&  	 225   \\ 
		 &  288 \\ 
		 & 588 \\
	\hline
	\hline 
Min-SU(4) & Max-SU(4)\\ 
Mod  8& Mod  4\\
		\hline
		1115   &  0000  \\ 
		 1133  & 0013  \\ 
		  1177 &    0022  \\ 
		{\bf 1357} &  	 0112   \\ 
		1555 &   0233   \\ 
	3337 &   1111 \\ 
	3355 & 	 1133    \\ 
	3777 & 	 1223   \\
	 5577  & 2222 \\  
		  & 3333  \\
		\hline 
	\end{tabular} &
	\begin{tabular}{| c|c|} 
	\hline
Min-SU(5) & Max-SU(5)\\ 
  Mod 5 & Mod  25\\
		\hline
00000 & (2, 2, 2, 2, 17) \\ 
 00014 & (2, 2, 2, 7, 12) \\ 
 00023  & (2, 2, 2, 22, 22) \\ 
  00113 & (2, 2, 7, 7, 7) \\ 
 00122 & (2, 2, 7, 17, 22) \\ 
 00244 & (2, 2, 12, 12, 22) \\ 
 00334 & (2, 2, 12, 17, 17) \\ 
 01112 & (2, 7, 7, 12, 22) \\ 
  01144 & (2, 7, 7, 17, 17) \\ 
 {\bf 01234} & (2, 7, 12, 12, 17) \\ 
 01333 & (2, 7, 22, 22, 22) \\ 
 02224& (2, 12, 12, 12, 12) \\ 
 02233 & (2, 12, 17, 22, 22) \\ 
 03444 & (2, 17, 17, 17, 22) \\ 
 11111 & (7, 7, 7, 7, 22) \\ 
 11134 & (7, 7, 7, 12, 17) \\ 
 11224 & (7, 7, 12, 12, 12) \\
 11233 & (7, 7, 17, 22, 22) \\ 
 12223 & (7, 12, 12, 22, 22) \\ 
 12444 & (7, 12, 17, 17, 22) \\ 
13344  & (7, 17, 17, 17, 17) \\ 
  22222 & (12, 12, 12, 17, 22) \\ 
 22344& (12, 12, 17, 17, 17) \\ 
  23334& (12, 22, 22, 22, 22) \\ 
  33333& (17, 17, 22, 22, 22) \\ 
 44444 & (3, 3, 3, 3, 13) \\ 
 & (3, 3, 3, 8, 8) \\ 
 & (3, 3, 3, 18, 23) \\ & (3, 3, 8, 13, 23) \\ & (3, 3, 8, 18, 18) \\ & (3, 3, 13, 13, 18) \\ & (3, 3, 23, 23, 23) \\ & (3, 8, 8, 8, 23) \\ & (3, 8, 8, 13, 18) \\ & (3, 8, 13, 13, 13) \\ & (3, 8, 18, 23, 23) \\ & (3, 13, 13, 23, 23) \\ & (3, 13, 18, 18, 23) \\ & (3, 18, 18, 18, 18) \\ & (8, 8, 8, 8, 18) \\ & (8, 8, 8, 13, 13) \\ & (8, 8, 13, 23, 23) \\ & (8, 8, 18, 18, 23) \\ & (8, 13, 13, 18, 23) \\ & (8, 13, 18, 18, 18) \\ & (8, 23, 23, 23, 23) \\ & (13, 13, 13, 13, 23) \\ & (13, 13, 13, 18, 18) \\ & (13, 18, 23, 23, 23) \\ & (18, 18, 18, 23, 23) \\
		\hline 	
	\end{tabular}
	\end{tabular}	
	\end{center}
}
\caption{The min and max set for one-loop  $V_{\text{1-loop}, \Omega_F^0}$. 
In the thermal case, there is a unique minimum. 
}
\label{tab:list}
\end{table}%

\vspace{3mm}
The exact formula for the number of  gauge inequivalent   minima  and maxima of the one-loop potential  as described above maps to a problem 
in  additive number theory.\footnote{All of the above problems fall  into a set of  combinatorial problems studied
 by Erd\H{o}s,  Ginzburg and Ziv  \cite{erdos}. The minima set follows the   OEIS entry https://oeis.org/A145855}
The number of degenerate minima  are given by :
\begin{align}
{\frak {N}}_{\rm min} (N)= \frac{1 }{2N} \sum_{d|N} (-1)^{N+d} \;  \phi\left( \frac{N}{d}\right) \:
 {2d \choose d}
\label{nummin}
\end{align} 
where the sum runs over all positive divisors of $N$, $  \phi\left( \cdot \right) $ is the Euler totient function, and the last term is binomial coefficient.  The 
number of degenerate maxima   are given by :
\begin{align}
{\frak {N}}_{\rm max} (N)=  
\left\{ \begin{array}{ll} 
\frac{1 }{2N}  \displaystyle \sum_{d|N} \;  \phi\left( \frac{N}{d}\right) \: {2d \choose d}
 & \qquad N \; {\rm even}  \cr   \cr
\frac{1}{N} \displaystyle  \sum _{ d |N}   \;  \mu (d)  \; 
 {{2 N/d} \choose {N/d}}  & \qquad   N \; {\rm odd} 
 \end{array}
 \right.
 \label{nummax}
\end{align} 
where   $\mu (d)$ is   M\"obius function.

\vspace{3mm}
Let us derive an explicit growth in the number of minima within the asymptotic approximation. 
First, the number of sequences obeying the conditions 2 and 3 for minima set is given by combinatorial expression  
$ 2N-1 \choose N$.
  Statistically, the condition  4 is met with probability $\sim 1/N$, hence the rough number of minima  of  one-loop potential is 
   $\sim \frac{1}{N}  {2N-1 \choose N}$. This is nothing but the dominant  $d=N$ term in the summation \eqref{nummin}. 
  Asymptotically we have an exponential growth in the number of gauge inequivalent minima  of the one-loop potential:
  \begin{align}
{\frak {N}}_{\rm min} (N) \approx 	\frac{(2N-1)!}{(N!)^2} \underbrace{\longrightarrow}_{{\rm large-}N}  \frac{2^{2N-1}N^{-3/2}}{\sqrt{\pi}}\;. 
\end{align}
  Comparison of this formula with the exact number (see  \cite{Sloane_theencyclopedia} sequence  A145855)
 below shows that the approximate  formula is rather accurate.
 
{\scriptsize
\begin{align}
	\begin{tabular}{|c|ccccccccccc|}
		\hline
		$N$ & 2 & 3 & 4 & 5 & 6 & 7 & 8 & 9 &10 &11  &12
		\\\hline
		Minima  Exact & 1 & 4 & 9 & 26 & 76 & 246 & 809 & 2704 &9226 & 32066  &  112716
		\\
		Approx. & 1.5 & 3.33 & 8.75 & 25.2 & 77 & 245.14 & 804.38 & 2701.11 &9237.8 & 32065.1 &  112673.16
		\\\hline 
		Maxima  Exact & 2& 6 &  10 & 50  &  80 & 490 &  810 & 5400   & 9252 & 64130 & 112720   \\ 			\hline 
	\end{tabular}
\end{align}
}

\noindent
\vspace{3mm}
There are two important physical points to be made: 

\vspace{3mm}
\noindent
{\bf 1) Exponential degeneracy:} Recall that typical one-loop gauge holonomy potentials \cite{Gross:1982at}  for  non-supersymmetric $SU(N)$ gauge theory  has 1, 2  or  at most $N$ minima or maxima.  
For the $\Omega_F^0$ twist, which seems to lead to the most efficient distillation on the Hilbert space,  
the corresponding one-loop potential has  exponentially increasing number of minima  and maxima.  

\vspace{3mm}
\noindent
{\bf 2) Collapse of the one-loop potential:}
In the thermal case where there is  no twist, the gap between the minima and maxima of the one-loop potential 
\begin{align}
\Delta V_{ \rm 1-loop} \equiv  V_{ \rm 1-loop, max} -  V_{ \rm 1-loop, min}  \sim O(N^2) 
\end{align}
Since the one-loop potential of $\Omega_F^0$-twisted fermions  is suppressed by a factor of $\frac{1}{N^4}$ compared to the case with no twist,  the gap  between the maximum and minimum  of the potential becomes 
 \begin{align}
 \Delta V_{ \rm 1-loop, \Omega_F^0} \sim O(N^{-2}).
 \end{align}  
 In fact, if we take large-$N_c$ limit with $N_f=N_c$ or more general Veneziano large-$N_c$ limit,   the one-loop potential 
\eqref{one-loop} vanishes as $\frac{1}{N_c^2} $ and we obtain a extremely dramatic result for a non-supersymmetric QFT: 
 \begin{align}
 V_{\text{1-loop}, \Omega_F^0} [\Omega]   =  0,  \qquad N_c =\infty. 
  \label{one-loop-largeN}
\end{align}
Namely, in large-$N_c$ limit, there is a moduli space at one-loop order in perturbation theory. 
In supersymmetric theories,  the moduli space for gauge holonomy persists to  all orders in perturbation theory (assuming supersymmetry preserving boundary conditions)
 both at finite and large $N_c$,  and it may only be  lifted non-perturbatively. 
Below, we show that,   in QCD(F/adj) the  degeneracy is  lifted at two-loop order, and 
center-symmetry is stabilized completely.

\subsection{Frustration tolerant  operators and center-stability  at two-loop order } 
\begin{figure}[h]
	\begin{center}
		\includegraphics[width=7cm]{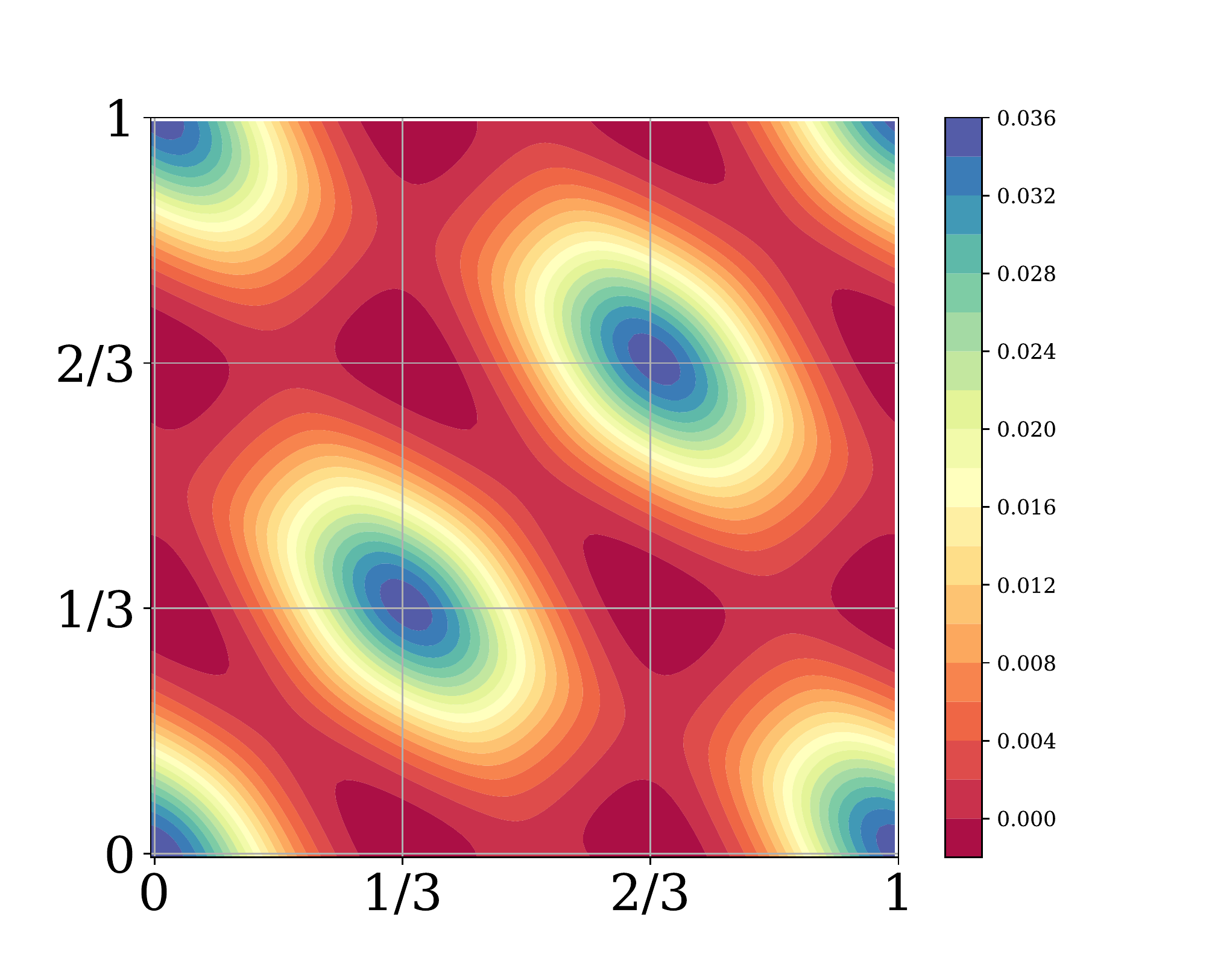}
		\put(-34,-5){\large $\displaystyle \frac{q_1}{2\pi}$}
		\put(-67,25){\large $\displaystyle \frac{q_2}{2\pi}$}
	\end{center}
	\vspace{-\baselineskip}
	\caption{\label{fig_2loop}The two-loop potential of twisted fundamental fermions $g^{-2} \beta^4 V_{ \rm 2-loop, \Omega_F^0}^\psi $ for $N_c=3$ as a function of the eigenvalues of the holonomy 
	$\Omega=\text{diag}(e^{iq_1}, e^{iq_2}, e^{iq_3})$.}
\end{figure}

The presence of a plethora of degenerate minima  of one-loop potential  for $N>2$ prompts us to ask if they are stable or not with the inclusion of higher order contributions. In the usual  thermal QCD, the two-loop potential for the holonomy has been computed in \cite{KorthalsAltes:1993ca,KorthalsAltes:1999cp,Guo:2018scp}.  See  formula (17) in  
\cite{KorthalsAltes:1999cp} and  formula (5.11) in \cite{Guo:2018scp}. These two formulas at first sight looks quite different, the fist one is in  terms of products of Bernoulli polynomials $B_1, B_2, B_3$  whose arguments are either 
eigenvalues of gauge holonomy or eigenvalue differences, 
   while the second one is only 
in terms of $B_4$ polynomial.  Due to non-trivial  Bernoulli polynomial identities, these two formulas are actually identical!

\vspace{3mm} 
Remarkably, the full  two-loop potential of the whole QCD(F/adj)  can be calculated by just using 
 the two-loop result based on the fundamental fermions.   This is due to the fact that the adjoint  and fundamental fermion loops  are decoupled from each other at one- and two-loop order level.  The one-loop order diagrams are individual fermion bubble diagrams, and at two-loop, we have fermion bubble diagrams with the gluon propagator insertion,  see Fig.~\ref{fig:feynman}.
At three loop order, there will be be mixing effects between adjoint and fundamental fermions, such as fermion bubble diagram with gluon propagator  with adjoint fermion bubble insertions, see Fig.~\ref{fig:three-loop},   or vice versa. But these effects are of order 
$O((g^2N_c)^3)$ compared to classical action and in order to determine CFC realization in the full theory, we will need only $O((g^2N_c)^2)$. 

\vspace{3mm}
Re-expressing    the formula (5.11) from \cite{Guo:2018scp} in terms of gauge invariant Polyakov loops, the two-loop potential induced by $N_f$ flavors of massless fundamental fermions obeying the thermal (anti-periodic) boundary condition is given by 
\begin{align}
	V_{ \text{2-loop}, \rm thermal}^\psi & = \frac{g^2 N_f}{\beta^4}
	\frac{3}{\pi^4}\left\{-\frac{N_c^2-1}{8N_c}\sum_{n=1}^{\infty}\frac{(-1)^n}{n^4} [\text{Tr}(\Omega^n)+\text{c.c.}]    + \frac{1}{24}\sum_{n=1}^{\infty}\frac{\big| \text{Tr} (\Omega^n) \big|^2}{n^4} \right\} 
\end{align}  
What we need is  the two-loop potential in the presence of the  $\Omega_F$  flavor-twisted boundary conditions \eqref{bc}. 
 With the appropriate insertion of $\Omega_F$ flavor-holonomy eigenvalues into  the analysis of  \cite{Guo:2018scp}, we find:  
\begin{align}
	V_{ \text{2-loop}, \Omega_F}^\psi & = \frac{g^2}{\beta^4}
	\frac{3}{\pi^4}\left\{-\frac{N^2_c-1}{8N_c}\sum_{n=1}^{\infty}\frac{(-1)^n}{n^4}[\text{Tr}(\bar{\Omega}_F^n)\text{Tr}(\Omega^n)+\text{c.c.}] + \frac{N_f}{24}\sum_{n=1}^{\infty}\frac{\big| \text{Tr} (\Omega^n) \big|^2}{n^4} \right\} 
	\label{V2psi}
\end{align} 
 Employing the flavor-twist  $\Omega_F= \Omega_F^0$
given in \eqref{flavor-hol-0}, the two-loop potential takes the form: 
\begin{align}
	V_{\text{2-loop}, \Omega_F^0}^\psi & \rightarrow  \frac{g^2}{\beta^4}
	\frac{3}{\pi^4}\left\{-\frac{N_c^2-1}{8N_cN_f^3} \sum_{n=1}^{\infty}\frac{(-1)^{N_fk}}{k^4}[\text{Tr}(\Omega^{N_fk})+\text{c.c.}] + \frac{N_f}{24}\sum_{n=1}^{\infty}\frac{\big| \text{Tr} (\Omega^n) \big|^2}{n^4} \right\} 
	\label{V2psi-min}
\end{align} 
The first term is proportional to the one-loop potential. For $N_f=N_c$,  it  represents an $\mathcal{O}(g^2N_c)$ correction to \eqref{one-loop} 
 by the $\Z_{N_c}$ symmetric single-trace operators   $\text{Tr}(\Omega^{N_ck})$. 
As such, it does not alter the exponentially large degeneracy of the one-loop potential. 
By  contrast,  the second term is quite different.   In particular, this double-trace operator  is independent of the flavor-twisted boundary condition, and we refer to it as {\it  frustration tolerant}.  
 Much like adjoint fermions endowed with periodic boundary conditions,  \eqref{V2psi-min} is minimized at   the center-symmetric  minimum: 
 \begin{align}
	\Omega & = \text{diag}(1,\omega,\cdots,\omega^{N_c-1}), \qquad  \qquad N_c~\text{odd}, \cr
	\Omega & = \omega^{1/2}\text{diag}(1,\omega,\cdots,\omega^{N_c-1}), \qquad N_c~\text{even}.
	\label{centersym}%
\end{align}
As a result, the exponential  degeneracy of the one-loop potential is lifted  and the total two-loop potential picks out the center-symmetric vacuum \eqref{centersym} as a global minimum. 

\vspace{3mm}
This result cannot be altered by three- or higher- loop orders  or non-perturbatively  due to weakness of the coupling constant  at the scale of compactification,  thanks to  asymptotic freedom.  Therefore, the center-symmetry is stable in QCD(F/adj) with $N_f=N_c$ at small-$\beta$.

\vspace{3mm}
In Figure~\ref{fig_2loop} the potential at two loop order   $	V_{\text{2-loop}, \Omega_F^0}^\psi $ is depicted for $N_c=3$. The minima are located at $\Omega=(1,\omega,\omega^2)$ and its gauge-equivalent copies. On the other hand,  the CFC-breaking holonomies  $\Omega=\1, \omega \1$ and $\omega^2 \1$ are maxima of the potential and hence unstable. 
This   structure  holds for all $N_c$. Since the color-flavor center symmetry is stable for both small $\beta$ and large $\beta$ we have a  good chance to have unbroken CFC symmetry for all $\beta$, 
at least for sufficiently light $m_\lambda$.

\begin{figure}[t]
\vspace{-3cm}
\begin{center}
\includegraphics[width = 1\textwidth]{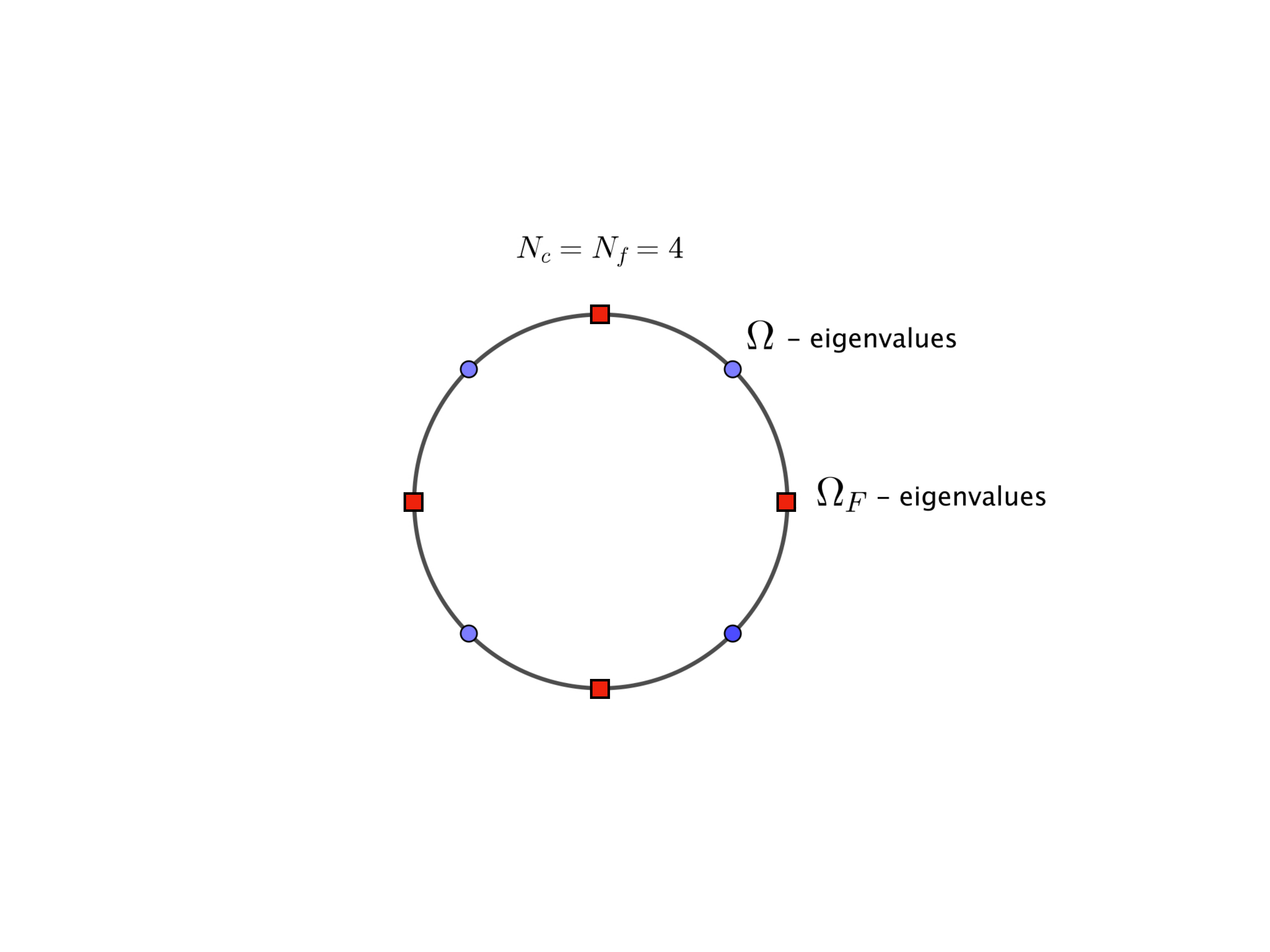}
\vspace{-3cm}
\caption{{
 The flavor-holonomy is non-dynamical, and gauge holonomy is dynamical. But the value of flavor-holonomy  determines  the  gauge holonomy potential  and its  minimum.  In  quantum distillation discussion over the Hilbert space, we  establish $\Omega_F^0$   \eqref{flavor-hol-0} as an extremum of the  twist-free energy  for flavor holonomy. 
 Despite  $\Omega_F$ being non-dynamical, the extremization of flavor-holonomy  potentials are physically meaningful. For $\Omega_F=1$, the pressure is maximized, for  $\Omega_F$ shown in figure, the  pressure 
  is minimized. This  means  as if less  degrees of freedom are contributing to the state sum. In this way, we will be able to circumnavigate   all 
  phase transitions in certain  4d QFTs. 
  } }
\label{fig:holonomy}
\end{center}
\end{figure}

\subsection{(Non)decoupling  of massive adjoint fermion}
Our ultimate goal in this work is to find a method to study ground state structure, and low-lying states in QCD(F).  In 
\eqref{V2psi-min}, we showed that the ground state of the holonomy potential for the  $N_f=N_c$ QCD(F/adj) theory is at a center-symmetric point \eqref{centersym}. Ultimately, we would like to turn on a mass term for adjoint fermion $m_\lambda$  and take  $m_\lambda \gg \Lambda$ so  that adjoint fermion decouples from the IR physics. In this way,  IR-physics reduce to the one of QCD(F). At the same time, for center-symmetry stability, sufficiently light adjoint fermions are necessary.  Can both of these requests be satisfied simultaneously? 

\vspace{3mm}
  Turning on a soft  adjoint fermion mass $m_\lambda$, the balance between the gauge fluctuations and adjoint fermion fluctuations  shown in penultimate line in \eqref{one-loop-A} breaks down. However,  the massless fundamental fermions still generate a CFC-stabilizing holononomy potential at two-loop order. The one-loop gauge plus adjoint fermion contribution is 
  soft, and is of  order $m_\lambda^2 \beta^2 O((g^2N_c)^0)$, and the second order contribution is further suppressed,  of order $- m_\lambda^2 \beta^2 (g^2N_c) $, and can be neglected below. 
 The combined potential which determines the center-symmetry realization (which picks the minimum among the states described in  Section. \ref{sec:minimum}) is given by 
 \begin{align}
 V^{\rm  gauge}_{\rm 1-loop} + V^{\rm \lambda}_{\rm 1-loop} + V^{\rm \psi}_{\rm 2-loop, \Omega_F}
&= 
 \frac{2}{\pi^2 \beta^4}  \sum_{n=1}^{\infty}  \frac{ \overbrace{ \left[ -1+ \frac{1}{2} (n\beta m_\lambda)^2 K_2 (n \beta m_\lambda)  + \frac{g^2N_c}{16 \pi^2} \right]  }^{a_n} }{n^4}  |\tr (\Omega^n)|^2     \cr 
 & \approx   \frac{2}{\pi^2 \beta^4}  \sum_{n=1}^{\infty}  \frac{ \left[ -  \frac{1}{4} (n\beta m_\lambda)^2   + \frac{g^2N_c}{16 \pi^2} \right] }{n^4}  |\tr (\Omega^n)|^2     
\end{align}
In the second line, we used small $z$ asymptotic of the Bessel function, $K_2(z) \sim \frac{2}{z^2}  - \frac{1}{2}  + O(z^2) $, at fixed $m_\lambda$ and sufficiently small $\beta$.  
The first line of the  potential is reliable provided $\beta \Lambda \ll 1$.  
The  $ \Z_{N_c}$ CFC-symmetry will remain intact provided the coefficients of first $\lfloor \frac{N}{2} \rfloor$ 
Polyakov loop operators remains positive \cite{Unsal:2008ch}, giving the  condition $ (\beta m_\lambda)   <  \frac{(g^2N_c)^{1/2} }{  N_c \pi} $. 
Therefore, in the window, 
\begin {align}
\Lambda  \ll  m_\lambda   <  \frac{(g^2N_c)^{1/2} }{\beta   N_c \pi} 
\label{dnd}
\end{align}
adjoint fermion exhibits the following striking properties:
\begin{itemize} 
\item It decouples from the long distance physics (chiral Lagrangian etc.) because $\Lambda  \ll    m_\lambda$    so that the IR theory is essentially QCD(F) both on $\R^4$ as well as on $\R^3 \times S^1$. 
\item It does not decouple from the holonomy potential, rather, for that purpose, it  acts in the same way as  light  (or even  massless)  adjoint fermion because $ m_\lambda   <  \frac{(g^2N_c)^{1/2} }{\beta   N_c \pi} $. 
\end{itemize}
These conditions accommodate adjoint fermions {\it heavy} compared to strong scale, but {\it light} compared to compactification scale provided $N_c \sim 3, 4, \ldots.$. When $N_c \rightarrow \infty$, the above condition cannot be satisfied, and 
$m_\lambda$ must be light compared to strong scale to keep the full $\Z_{N_c}$ center-symmetry intact.

\section{Quantum distillation as reduction of effective density of states}  
\label{sec:pheno}
\noindent 
{\bf Motivation:} Consider thermal partition function: 
\begin{align} 
 {\cal Z} (\beta ) = \tr \Big[ e^{-\beta  H} \Big]  = \sum_{n \in \B} e^{-\beta  E_n} {\rm deg} (n)  + \sum_{n \in \F } e^{-\beta  E_n}  {\rm deg} (n) 
 \label{partition}
\end{align} 
which correspond to a state sum over the Hilbert space of the QFT.   In QCD(F), the states are mesons, baryons, glueballs, resonances and alike. In the low-temperature limit, $\beta  \rightarrow \infty$, the thermal partition function is dominated by low-lying states in the Hilbert space.
In the high-temperature limit, $\beta  \rightarrow 0$, all the states contribute to the state sum on a similar footing.  

\vspace{3mm}
The phase transitions are associated with the singularities in the free-energy or other thermodynamic observables. We will consider any non-analytic point of $\F_{\rm thermal} (\beta ) $ taking place for real positive $\beta  \in [0, \infty)$ as a phase transition point.  
Since $H$ is Hamiltonian of QFT on $\R^3$, we are already in thermodynamic limit, hence, according to Lee--Yang analysis, one expects singularities in  $\F_{\rm thermal} (\beta ) $  as $\beta $ varies. It is in general very hard to determine the Lee--Yang singularities analytically. 

\vspace{3mm}
The idea of quantum distillation of Hilbert space is ultimately tied with the manipulation of the singularities of partition function.  
Clearly, each term in \eqref{partition} is positive definite and analytic in $\beta $. However, in thermodynamic limit,
the sum need not be analytic, it generically  has singularities.  On the other hand, it is also clear that the insertion of phase factors 
associated with global symmetries ${\bf G}$  
into the  thermal partition function, hence obtaining a generalization     \eqref{GPF},    necessarily reduces its magnitude: 
\begin{align} 
| {\cal Z}(\beta , \epsilon_1, \ldots, \epsilon_N)|  \leq {\cal Z}(\beta )
\label{ineq}
\end{align} 
Clearly, this manipulation does not alter the Hilbert space ${\cal H}$ of $H$, because ${\cal Z}(\beta , \epsilon_1, \ldots, \epsilon_N) $ is a symmetry graded partition function.    The intention is to show that 
${\cal Z}(\beta ,  \epsilon_1, \ldots, \epsilon_N)$  can be tame enough such that it does not  possess 
singularities in  $\beta  \in [0, \infty)$ while ${\cal Z}(\beta )$ does.  
In order to provide one perspective on the quantum distillation idea, we need to first explain the density of states in a generic QFT, especially for asymptotically high-energy states.

\vspace{3mm}
\noindent 
{\bf Density of states:}  The density of states is the inverse Laplace transform to the partition function. 
Although we cannot calculate the partition function at arbitrary temperatures, we can easily  calculate its leading order behavior  at  high temperature for asymptotically free theories.  From there, we can   infer the density of  hadronic states, by applying  an inverse Laplace transform. 

\vspace{3mm}
In small-$\beta $ limit where the theory becomes weakly coupled at the scale $\beta $, a way to calculate the asymptotic form of the partition function is to calculate the  free energy of the system.   At leading order, the free energy of the QCD(F/adj) is given by the Stefan-Boltzmann law, see standard  texts  \cite{Philipsen:2012nu, Kapusta:2006pm, Laine:2016hma}, which is equal to 
one-loop potential \eqref{1-loop-thermal} evaluated at its minimum:
\begin{align}
\F_{\rm thermal} (\beta )&= -  \frac{\pi^2}{90}   \frac{V_3}{\beta ^4}  \left[  \underbrace{2(N_c^2-1)}_{ \rm gluons} +  \frac{7}{8} \times    \underbrace{  2 (N_c^2-1)}_{\rm adj.  Weyl \; ferm.}  + 
\frac{7}{8} \times  \underbrace{ 4 N_f  N_c }_{\rm fund. D.  ferm.} \right]  \qquad \beta  \rightarrow 0 
\end{align} 
This is the sum of  $(A_\mu,  \lambda,  \psi^a)$ contributions. The $\frac{7}{8}$ factor arises due to  Dirac-Fermi distribution corresponding to  fermions with thermal boundary conditions.

\vspace{3mm}
\noindent
{\bf Quark-Hadron duality:}
The free energy  has at least  two interpretations,  and the two  are related  via the quark-hadron duality \cite{Shifman:2000jv,
Shifman:2005zn}
\begin{itemize} 
\item {\bf Micro interpretation:} One interpretation is in terms of microscopic constituents, gluons and quarks. 
The numbers appearing in free energy such as   $ (N_c^2-1)$, $ (N_c^2-1)$,  $4 N_f N_c $  count respectively,  the number of microscopic  bosonic and fermionic degrees of freedom in the QFT. 
\item    {\bf Macro  interpretation:}  
The other interpretation is in terms of macroscopic states, the hadrons  in physical Hilbert space ${\cal H}$. 
The inverse Laplace transform of the partition function is the density of states of hadrons: 
\begin{align}
{\cal Z}(\beta ) \sim e^{- \beta  \F_{\rm thermal}}  \sim e^{+a  N_c^2   V_3  T^3 }   \qquad  \Longleftrightarrow  \qquad 
\rho_{SB}(E) \sim  e^{E^{3/4} N_c^{1/2} { (a V_3)}^{1/4}}   
\label{SB}
\end{align}
where $ \rho_{SB} (E)$ is the growth in correspondence with  Stefan-Boltzmann(SB)  law, and $a$ is a pure order one number.  SB growth is special in the sense that it is the largest asymptotic  growth in a local finite-$N_c$ QFT.  Only at $ N_c=\infty$ and string theory, one can obtain a Hagedorn-growth.  
\end{itemize} 
 
\vspace{3mm}
Let us now consider the graded partition function and associated  graded free energy and density of states. 
The translation  of the center-symmetry preserving boundary conditions \eqref{bc} to operator formalism is:  
\begin{align} 
{\cal Z}(\beta ) = 
\tr \Big[ e^{-\beta H} (-1)^F   e^{i \pi Q_0}   \prod_{a=1}^{N_f} e^{i \frac{2\pi a}{N_f}  Q_a}  \Big], 
\label{graded}
\end{align} 
where $Q_0 \equiv  \int d^3 x  \, \sum_a   \psi_a^{\dagger}   \psi^a$ and    $Q_a \equiv  \int d^3 x  \;   \psi_a^{\dagger}   \psi^a$. 

\vspace{3mm}
The twist free energy associated with this partition function can be found by calculating the holonomy  potential at its minimum. 
We have the holonomy potential at two loop order at our disposal, and the global minimum of it is stable to all orders in perturbation theory and non-perturbatively. Plugging the global minimum  \eqref{centersym} to one-loop potential (to make comparison with thermal result clearer), 
yields the   twist free energy $\F_{\Omega_F^0}(\beta )$:
 \begin{align}
\F_{\Omega_F^0}(\beta ) &= -  \frac{\pi^2}{90}   \frac{V_3}{\beta\beta^4}  \left[  \underbrace{2 (N_c^2-1)}_{ \rm gluons}   \underbrace{  {  \color{red} - }2 (N_c^2-1)}_{\rm adj. Weyl \; ferm. pbc.}  + \frac{7}{2}
\underbrace{ \color {red} \frac{1}{N_c^2}  }_{\rm fund. D. ferm. tbc.} \right]  \cr
&=  -  \frac{\pi^2}{90}   \frac{V_3}{\beta^4}    \left[   \frac{7}{2} \frac{1}{N_c^2}   \right]   \xlongrightarrow{ N_c \rightarrow \infty }   {\color{red} 0}
\end{align} 
This is a quite striking result. It   is as if there is merely  $\frac{1}{N_c^2}$ quark degree of freedom in the system instead of $\sim N_c^2$ bosonic and $\sim N_c^2$  fermionic degrees of freedom, similar to the description in  \cite{Cherman:2018mya} in large-$N_c$ QCD(adj). 
The twist free energy behaves as if   the QFT at hand does not even have a single particle worthy degree of freedom in 4d.  Yet we did not touch  the Hamiltonian and the Hilbert space.
The corresponding density of states  for the growth of the hadronic states  (after quantum distillation) takes the form 
\begin{align}
{\cal Z}_{\Omega_F^0} (\beta) \sim e^{-\beta  \F_{\Omega_F^0}(\beta)  }  \sim e^{+ a   \frac {1}{N_c^2}   V_3/\beta^3   }    \qquad  \Longleftrightarrow  \qquad 
\rho_{\Omega_F^0} (E) \sim  e^{ \frac{1}{N_c^{1/2}} E^{3/4}  { (a V_3)}^{1/4}}  
\label{redux}
\end{align}
in sharp contrast with the Stefan-Boltzmann growth \eqref{SB}. This implies 
that a   dramatic amount of cancellation ought to occur in the graded state sum over the Hilbert space $\cal H$ of QFT 
among  high energy physical states in ${\cal H}$.  
\begin{align}
\text {distillation at high-$E$ spectrum}: \;\;\;  \rho_{\Omega_F^0} (E)  \ll \rho_{SB}(E)  \qquad  \qquad 
 \qquad 
\end{align}
The mapping of the quantum distillation of the Hilbert space ${\cal H}$ to thermodynamics is following: 
\begin{quote}
Despite the fact that  the state sum using  $\rho_{SB}(E)$ leads to a  CFC broken and chirally restored phase at 
 at small-$\beta$ and   phase transitions in ${\cal Z}(\beta)$, the state sum with    $\rho_{\Omega_F^0} (E)$   leads to  CFC unbroken,  and chiral symmetry broken phase at small-$\beta$ adiabatically connected to large-$\beta$.  In particular, possible phase transitions can  be avoided as the theory interpolates from large to small 
$ \beta$ by using  ${\cal Z}_{\Omega_F^0}(\beta)$. 
\end{quote}

\vspace{3mm} 
A remark is on the  large-$N$ limit following the rationale of  \cite{Cherman:2018mya}.  If we consider our theory on a three dimensional spatial manifold ${\cal M}_3$  with a characteristic curvature scale $\ell$ and volume $V$,  it is possible to show that 
\begin{align}
 \rho_{\Omega_F^0} (E) \sim e^{\sqrt{\ell E}},  
 \label{2d}
 \end{align}
exhibiting  a two-dimensional scaling. In particular, Hagedorn growth $e^{\beta_H E}$   expected in large-$N$ theories  and string theory, as well as Stefan-Boltzmann growth  $e^{V^{1/4} E^{3/4}}$  which is natural in  four-dimensional theory in spatial volume $V$ completely disappears.  In the sense of graded density of states,  our non-supersymmetric theory acts in a similar way to supersymmetric theories on curved spaces \cite{DiPietro:2014bca}, similar to   \cite{Cherman:2018mya}.

\vspace{3mm}
Similar effects are also present in \cite{Cherman:2018mya, Basar:2014jua, Kutasov:1990sv, Dienes:1995pm,Dienes:1994np}. In string theory, this  has parallels  to \cite{Kutasov:1990sv}  where despite the fact that bosonic and fermionic sector of the theory has tacyonic instabilities, the $(-1)^F$ graded sum is tachyon-free and stable.


\subsection{Distillation of  mesons: Cancelling bosons against bosons}
 Below, we describe the quantum distillation in the Hilbert space explicitly in terms of low lying hadronic states.  
 This discussion is aimed to provide  simple  pedagogical insights  into the process.
 
Consider $m_{\psi_1}  = m_{\psi_2}  = \ldots =m_{\psi_{N_f}} \geq 0$  in QCD(F/adj), a positive semidefinite equal mass for the $N_f$-flavors. If $m_\psi >0$, then the global symmetry of the theory on  $\R^4$ that acts on fundamental fermions is  
given in \eqref{eq:non_ab2}, roughly  $U(N_f)_V$.    In QCD(F) phenomenology corresponding to $N_f=3$, this is  called  flavor-$SU(3)$ limit, 
see eg. \cite{Manohar:1998xv}.

\begin{figure}[h]
\begin{center}
\includegraphics[width = .4\textwidth]{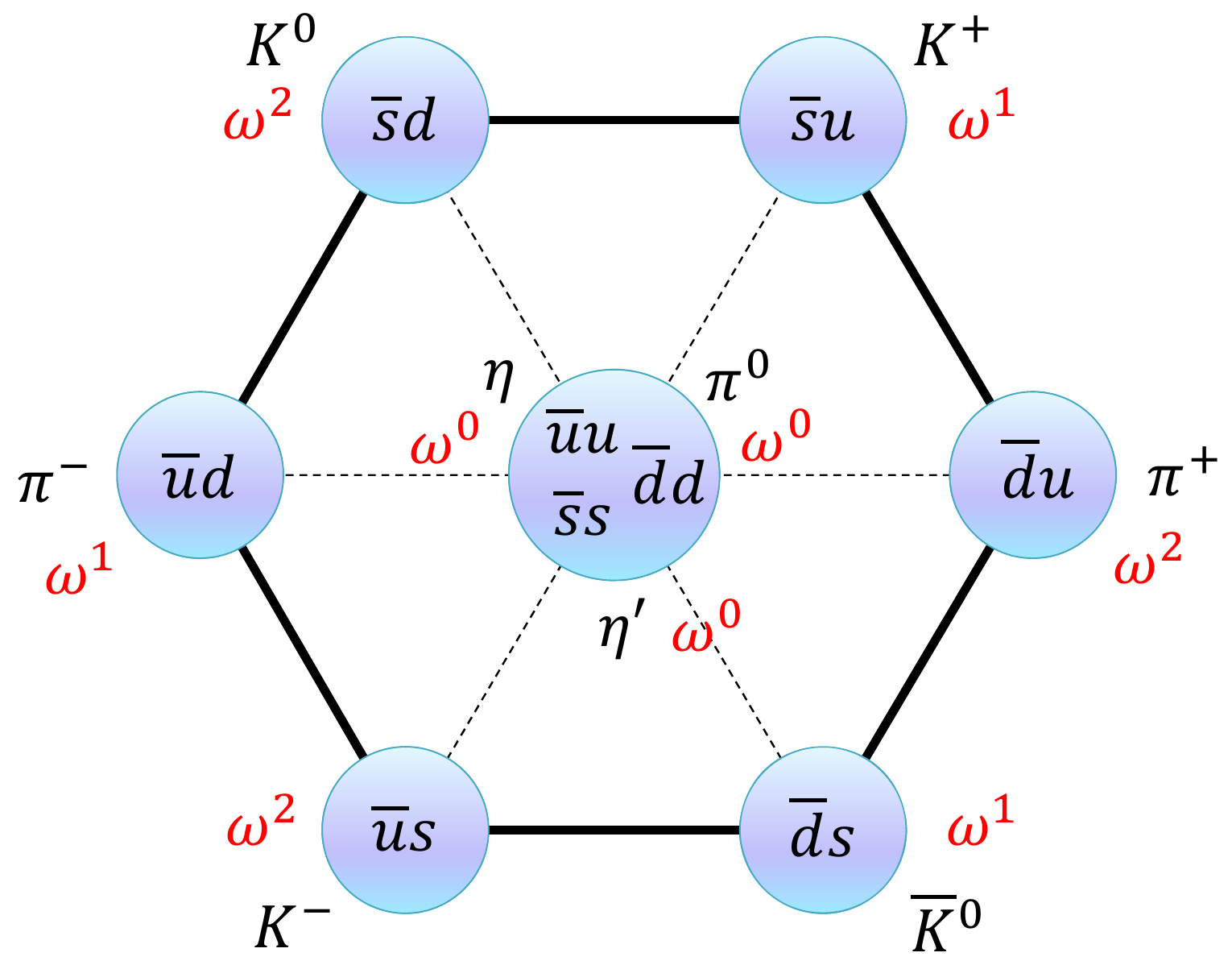}
\vspace{-2mm}
\caption{ Grading and quantum distillation  in the spin-zero meson octet  sector.  Grading assigns phases $\omega^k = e^{\frac{i 2 \pi k}{3}}$ to different mesons in such a way that  the graded sum is zero. This means spin-zero meson contribution disappear in the graded sum.  }
\label{fig:meson}
\end{center}
\end{figure}

\vspace{3mm}
In QCD(F/adj), as discussed  in Section \ref{sec:expectations}, there are two possible patterns of chiral symmetry breaking, 
down  to a vector-like subgroup \eqref{pattern1} or  to a subgroup which has a chiral $U(1)_{A_D}$ factor\eqref{pattern2}. If the first is true behavior of the theory on  $\R^4$,   the IR theory  is described in terms of    $N_f^2$ Nambu--Goldstone bosons, whereas  if the second  is true behavior,  it is described in terms of    $N_f^2-1$ Nambu--Goldstone bosons, $(N_f^2 -1)$ massless fermionic  mesons and two massless baryons,  tabulated in \eqref{composite}.  These IR-degrees of freedom satisfy rather non-trivial collection of  't Hooft anomalies, and hence,  we view both  
as logical possibilities.

\vspace{3mm}
As an example, consider the first pattern  \eqref{pattern1}, in which  IR theory  has   $N_f^2$ Nambu--Goldstone bosons, with quantum numbers
\begin{align} 
M_{a}^{b}= \overline\psi_a \psi^b,  \qquad a, b=1, \ldots, N_f 
\end{align}
In thermal case, these states would contribute to the thermal free energy as $ {\cal F}_{\rm mesons}  (\beta) \sim  -  \frac{\pi^2}{90}   \frac{V_3}{\beta^4}  N_f^2$. 
However,  in the state sum, the presence of the   $ \prod_{a=1}^{N_f} e^{i \epsilon_a Q_a} $ operator amounts to grading the states with phases, which are roots of unity. See Fig. \ref{fig:meson} for the assignment of phases to mesons in theory with $N_f=3$ flavors. The state $M_{a}^{b}$  is assigned a phase
\begin{align} 
M_{a}^{b} \mapsto   e^{ - 2 \pi i (b-a)/N_f} M_{a}^{b}
\label{meson-g}
\end{align} 
modifying the terms in the state sum into 
\begin{align} 
 N_f^2 e^{-\beta E_{\pi}}  \mapsto  \sum_{a, b=1}^{N_f}    e^{ - 2 \pi i (b-a)/N_f}   e^{-\beta E_{\pi}} =0
\end{align} 
Namely, in the graded partition function, the contribution of the scalar mesons to the  state sum  \eqref{graded} vanishes 
and corresponding free energy for this sector is mapped to zero  $ {\cal F}_{{\rm mesons}, \Omega_F^0}  (\beta)=0$ as if there are no mesons in the spectrum at all!

\vspace{3mm}
If we turn on a mass term for adjoint fermion $m_\lambda$, then, as described in  Section~\ref{sec:general}, $U(1)_{A_D}$ reduces to $\Z_{2N_f}$ and $\eta'$ is no longer degenerate with the rest of the pions. In this limit, the state sum is modified as: 
\begin{align} 
 (N_f^2-1) e^{-\beta E_{\pi}}  +  1  e^{-\beta E_{\eta'}} \mapsto (-1) e^{-\beta E_{\pi}}  +  1  e^{-\beta E_{\eta'}} 
 \label{meson-cancel}
\end{align} 
which means  that the contribution of this sector   can even be negative.

\vspace{3mm}
In QCD(F/adj), there are   also fermionic mesons such as $\bar\psi_a \lambda \psi^b  \equiv \psi_{Ma}^{\;\;\;\;\; b}$, see 
\eqref{mesino}. If  \eqref{pattern2} of chiral symmetry breaking is realized, these are gapless modes.  
Even if a mass term is turned on for  quarks,  the masses of $(N_f^2-1)$ of these fermionic mesons are equal due to flavor symmetry. 
This is so because  $\psi_{Ma}^{\;\;\;\;\; b}$ decompose as ${\bf adj} \oplus {\bf 1}$ under $SU(N_f)_V$. 
Their contribution is essentially diminished by  grading in the same way as mesons \eqref{meson-cancel}.

\vspace{3mm}
The meson  states in the spectrum transform  in the  singlet, adjoint or product of adjoint  representations under flavor symmetry. 
This  comes from the fact that  in   the global symmetry, we must mode out the gauge redundancies.  For $N_f=N_c$,  the global symmetry is not 
 $SU(N_f)_V$,   but  $PSU(N_f) = SU(N_f)/\Z_{N_f}$.  
 This aspect is exactly parallel to $\mathbb {CP}^{N-1}$  model where a perfect quantum distillation   takes place  in the large-$N$ limit,  as explained in 
 \cite{Sulejmanpasic:2016llc,    Dunne:2018hog}
   in order to explain the path integral with twisted boundary conditions \cite{Dunne:2012ae}. 
 We refer  the distillation in bosonic large-$N$  $\mathbb {CP}^{N-1}$ as {\it perfect distillation}, because the graded sum cancels all states 
 in the Hilbert space,  but the ground state(s). 
 In this respect, despite the theory being purely  bosonic, the graded partition function can be engineered to emulate the  supersymmetric  index in a supersymmetric gauge  theory, which counts just the ground states \cite{Witten:1982df}.

\subsection{Distillation of baryons: Fermions against fermions}
Baryons in $SU(N_c)$  QCD(F) are fermions for  odd $N_c$ and  bosons for even $N_c$.  Below, we assume $N_c$ is odd, where we will describe cancellation of fermions against fermions.  The discussion can straightforwardly  be generalized to $N_c$ even. 

\vspace{3mm}
For simplicity, consider $SU(3)$ QCD with  $N_f=3$ fermions. Lightest baryons decompose into $S= \frac{1}{2}$ octet and   
$S= \frac{3}{2}$ decuplet.  In the $SU(3)$-flavor limit where all quark masses are equal, the decuplet masses are equal to each other and 
the octet masses are equal to each other, but $m_{\rm decouplet } \neq  m_{\rm octet}$.

\begin{figure}[h]
\begin{center}
\raisebox{5mm}{\includegraphics[width = .45\textwidth]{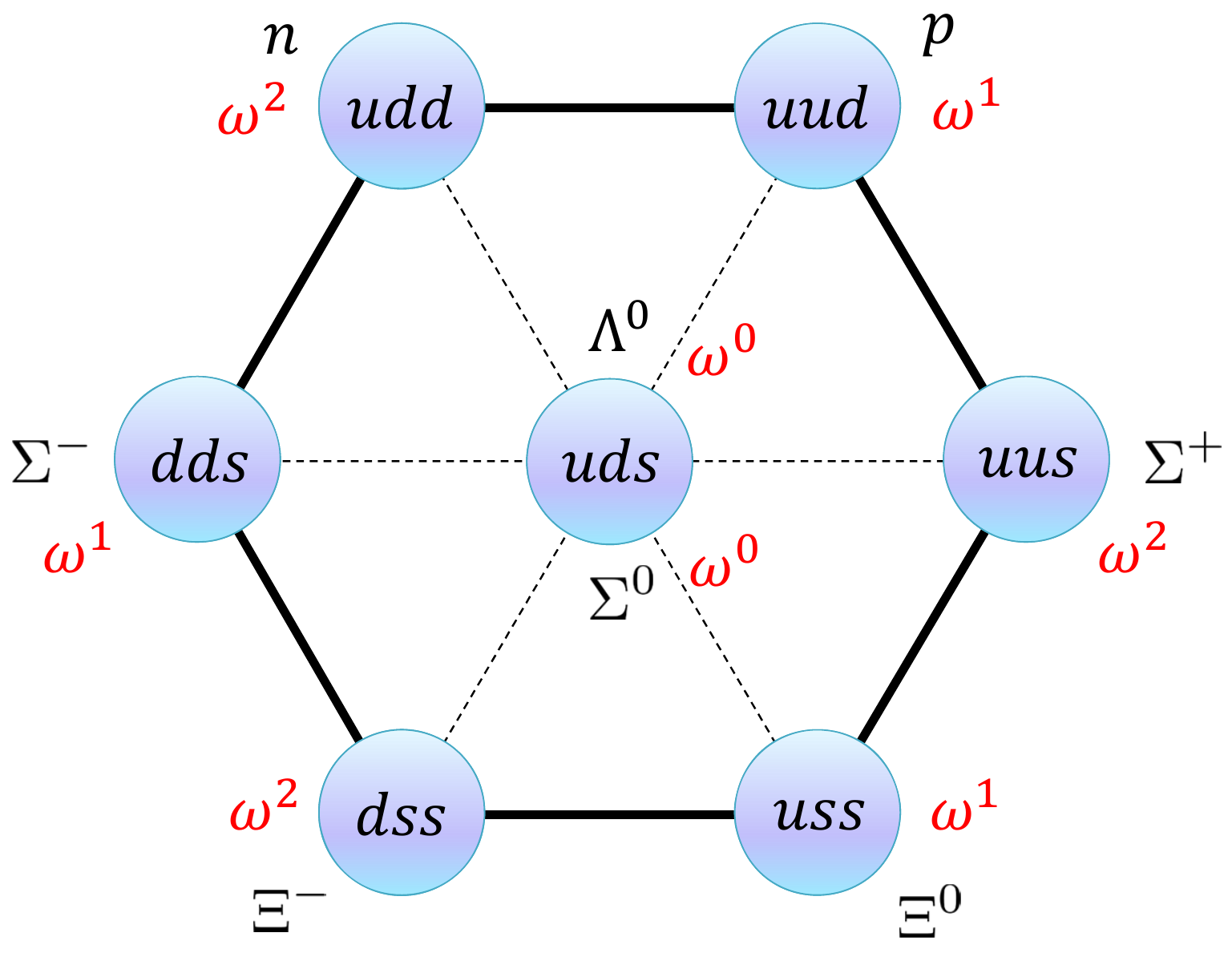}}
\includegraphics[width = .48\textwidth]{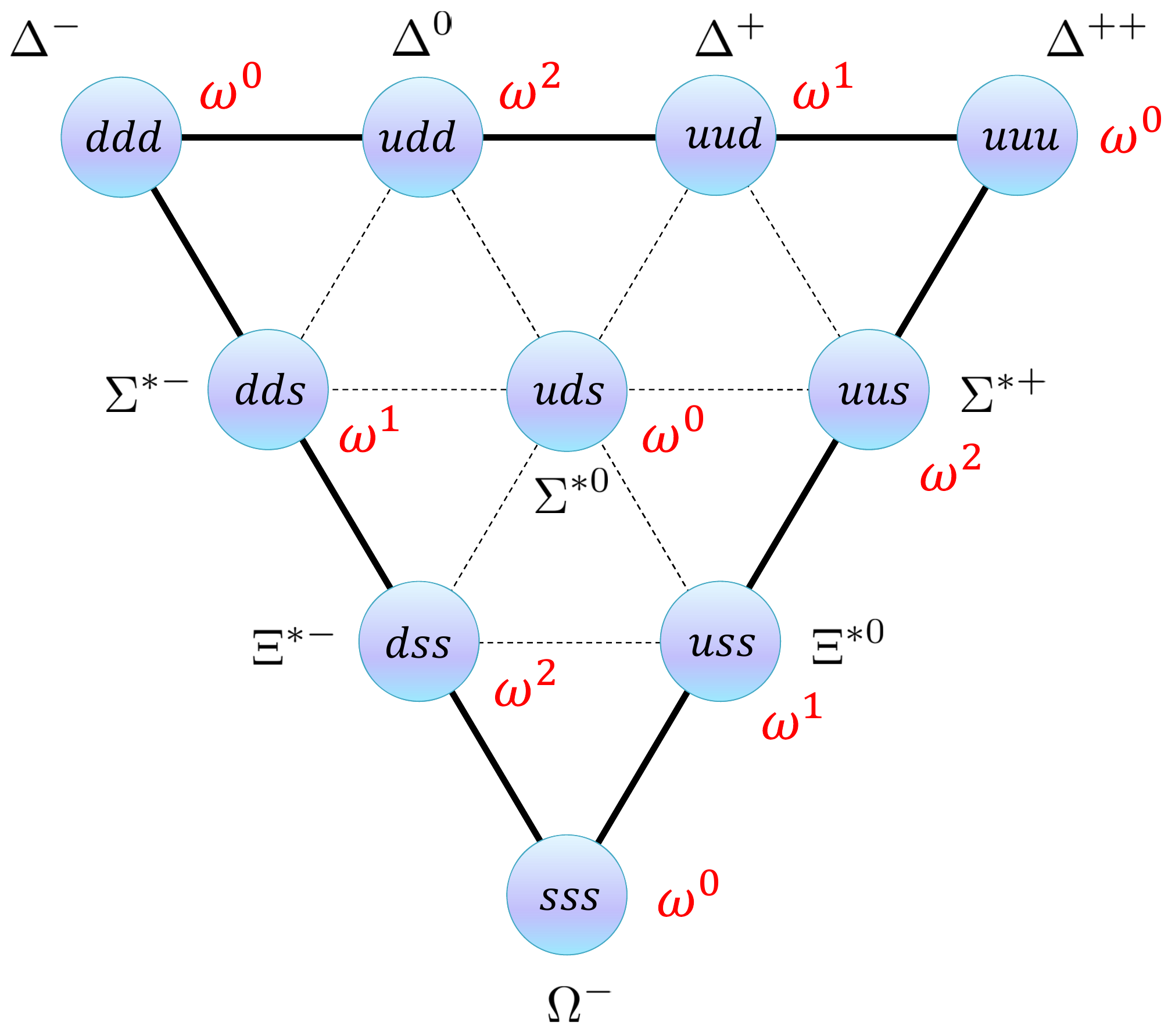}
\vspace{-3mm}
\caption{ Grading and quantum distillation  in the baryon  spin-$\half$ octet and spin-$\frac{3}{2}$ decouplet sectors }
\label{fig:baryon}
\end{center}
\end{figure}

\vspace{3mm}
The action of grading on the decouplet baryons is as follows. The spin-3/2 wave-function in the $m = 3/2$ state is  just $| \uparrow   \uparrow  \uparrow \rangle$  and is completely symmetric, and so are the other $m$ states.  The flavor wave function must also be completely symmetric. (We suppress completely anti-symmetric color structure.) Then, a baryon in decouplet is 
\begin{align} 
B^{abc}= \psi^{\{ a}    \psi^b  \psi^{c \}}  ,  \qquad a,b,c =1, \ldots, 3
\end{align} 
In thermal state sum, the decouplet baryons would contribute as  $40  e^{-\beta E_{ \rm decouplet}}$. However, 
in the graded  state sum, similar to the discussion of mesons,  
the presence of the   $ \prod_{a=1}^{N_f} e^{i \epsilon_a Q_a} $ operator amounts to phases for  $B_{abc}$  
of the form: 
\begin{align} 
B_{abc} \rightarrow   e^{ - 2 \pi i (a+b+c)/3} B_{abc}
\end{align} 
The action of the grading into the state sum for the  $S= \frac{1}{2}$ octet  and $S= \frac{3}{2}$ decuplet are: 
\begin{align}
&16   e^{-\beta E_{\rm octet}} \underbrace{\rightarrow }_{\Omega_F^0}
(-2)  e^{-\beta E_{\rm octet}} \cr 
& 40  e^{-\beta E_{\rm decouplet}} \underbrace{\rightarrow }_{\Omega_F^0}
(+4)  e^{-\beta E_{\rm decouplet}}
\end{align}
Clearly, the contribution of baryons  to graded partition function and graded  thermodynamics will be rather small. 
We will momentarily quantify how small this contribution is.

\vspace{3mm}
If we take masses  of quarks larger than strong scale, we can assume quark 
are  sufficiently non-relativistic. The non-relativistic quark model has an  enhanced $SU(6)$  spin- flavor symmetry, under which spin up and down  and 3-flavor indices combine to $\bf 6$ dimensional fundamental representation of  $SU(6)$, 
  see \cite{Manohar:1998xv} for a review. 
The lightest  baryons transform as the symmetric product of three $\bf 6$'s  of $SU(6)$, which is the {\bf 56} dimensional representation. In this case, the effect of the quantum grading is the reduction
\begin{align}
56  e^{-\beta E} \underbrace{\rightarrow }_{\Omega_F^0}
(+2)  e^{-\beta E }
\end{align}
The contribution of 56 lightest baryonic states  reduce to the one of 2 states only.  

\vspace{3mm}
In the case of general $N$ and heavy quark limit,  the spin-flavor symmetry group is $SU(2N)$ and the baryons transform in the $N$-index symmetric representation. Its dimension   ${3N-1} \choose N$
 grows exponentially as $\sim (27/4)^N/\sqrt{3\pi N}$. For $N=2,3,4,5,6,7,8,9$ we verified numerically that the quantum grading reduces this to just 2 states only.\footnote{Thanks to Takuya Kanazawa for demonstrating  this.} Hence, we have the reduction
\begin{align}
{{3N-1} \choose N}   e^{-\beta E_{\rm baryon}} \rightarrow 
(+2)  e^{-\beta E_{\rm baryon} }.
\end{align}
Also note that at large-$N$ limit,   the masses of baryons scale as $N$. As such, at finite-$\beta$,  
their contribution vanishes either in thermal state sum or in graded 
state sum. Therefore, at least for the large-$N$ theory,  the baryon effects  are doubly suppressed, both due to quantum distillation and due to the fact that they gradually decouple from the spectrum.

\vspace{3mm}
\subsection{Graded Thermodynamics of  QCD(F)  $\approx$ Thermodynamics of  YM}
\label{graded-thermo}

Consider graded thermodynamics of the QCD(F) defined through the generalized   partition function:
\begin{align} 
{\cal Z}(\beta) =  
\tr \Big[ e^{-\beta H}  \prod_{a=1}^{N_f} e^{i \frac{2\pi a}{N_f}  Q_a}  \Big],    \label{graded2}
\end{align} 
where for $N_f$ odd, the twist lives in $SU(N_F)_V$ and for $N_f$ even, it  lives in $U(N_F)_V$.

\vspace{3mm}
We claim  that the graded thermodynamics of QCD(F) should be approximately  equivalent to regular thermodynamics of the pure Yang--Mills theory {\it both} in the confined and deconfined phases  provided a physical scale (deconfinement temperature or strong scale) are matched between the two theories.  
We  provide some  evidence for this claim and use it to explain  lattice simulation results of Ref.~\cite{ Iritani:2015ara}.

\vspace{3mm}
First consider the  thermal  and graded  free energy of QCD(F) in comparison with thermal free energy in YM in the very high temperature regime,  $\beta \rightarrow 0$.  
Assume $N_f=N_c$. 
 \begin{align}
\F_{\rm thermal} ^{\rm QCD(F)}  &= -  \frac{\pi^2}{90}   \frac{V_3}{\beta^4}  \left[  \underbrace{2(N_c^2-1)}_{ \rm gluons} + 
\frac{7}{8} \times  \underbrace{ 4 N_f  N_c }_{\rm fund. D.  ferm.} \right]  \cr
{\cal F}_{\Omega_F^0}^{\rm QCD(F)} &= -  \frac{\pi^2}{90}   \frac{V_3}{\beta^4}  \left[  \underbrace{2 (N_c^2-1)}_{ \rm gluons}      +  \frac{7}{2}
\underbrace{\frac{1}{N_c^2}  }_{\rm fund. D. ferm. tbc.} \right]     \cr 
&= {\cal F}^{\rm YM} \left(1+ O(1/N_c^4) \right)  
\end{align} 
The thermal free energy of QCD(F)  in the $\Omega_F^{0}$ background is essentially same as pure YM theory. It only differs from it 
by  an  $O(1/N_c^4)$ correction.  

\vspace{3mm}
As discussed in the context of quark-hadron duality, this result has two related interpretations. In the microscopic picture, it implies that the contribution of quarks is suppressed by a factor $1/N_c^4$. 
Instead of having $N_fN_c\sim N_c^2   $ quark degree of freedom, one has effectively  $N_fN_c/N_f^3 \sim 1/ N_c^2   $ of them contributing to the graded-thermodynamics of QCD(F).
 In the language of hadrons, this result means that the asymptotic growth of the non-glue-ball   hadrons  in Hilbert space at high energies  is  reduced from the Stefan-Boltzmann growth  $\rho_{SB}(E) \sim  e^{E^{3/4} N_c^{1/2} {(aV_3)}^{1/4}}   $ down to $\rho_{\Omega_F^0}(E) \sim  e^{E^{3/4} N_c^{-1/2} { (aV_3)}^{1/4}}   $.   

\vspace{3mm}
The distillation of Hilbert space does not mean that the quarks are projected out from the theory, despite the fact that their effect in ${\cal F}_{\Omega_F^0}^{\rm QCD(F)}$ disappear in the large-$N_c$ limit. 
 The quarks  are  present  in the Lagrangian, and similarly, the baryonic or mesonic hadrons they build are present  in the physical spectrum of the theory. In particular, the renormalization group $\beta$ function of the graded QCD(F)   is not changed at all.  The first coefficient of $\beta$ function  is  still $\beta_0 =  \frac{11}{3}N_c - \frac{2}{3}N_f$, and the theory runs in the same way as the regular QCD,  with a strong-scale $\Lambda_{\rm QCD}  = \mu \exp[- \frac{8 \pi^2}{g^2(\mu) \beta_0}]$.  However, the graded thermodynamics associated with the QCD(F)  approximates the thermodynamics of pure Yang--Mills theory at very high temperatures. Below, we argue that  the matching of the thermodynamics must also be valid  at low temperature strong coupling regime.  

 \vspace{3mm}
 \noindent
{ \bf Quantum distillation of flavor channels:}  Quantum distillation acts on degrees of freedom which transform under flavor (such as 
vector mesons, other resonances and  baryons), but not on the glueball sector. 
Using the rationale described around \eqref{redux}, the density of states coming from the mesons and baryons reduce so much so that 
no single 4d degrees of freedom coming from the flavor sector contribute to the graded-thermodynamics of QCD(F).    This is  completely analogous to the  discussion in 2d large-$N$  bosonic $\mathbb {CP}^{N-1}$ model  where all  states except the ground state cancel among themselves  \cite{Dunne:2012ae} in the $\Omega_F$ twisted partition function.  

 \vspace{3mm}
 Therefore, we expect the  graded free energy of QCD(F) to be approximately    the  thermal free energy of YM
\begin{align}
{\cal F}^{\rm QCD(F)}_{\Omega_F^0} (\beta)  \approx  {\cal F}^{\rm YM} (\beta) , \qquad \forall \beta \in [0, \infty]
\end{align}
{\it both}  in the confined and deconfined phases.  In deconfined phase, using the one and two-loop potentials  of the fundamental 
fermion given in  \eqref{one-loop} and \eqref{V2psi-min} for $N_f \sim N_c$ theory,     and the one of the pure gauge sector,   it is straightforward to show that 
\begin{align}
{\cal F}^{\rm QCD(F)}_{\Omega_F^0} (\beta)  =   {\cal F}^{\rm YM} (\beta) + O(N_c^{-2}) + O(g^2N_c)
\label{YM-QCD-Twisted}
\end{align}
i.e. the difference in free energy is either suppressed with the weak coupling at small $\beta$ and by   factors of $N_c$.  

As emphasized earlier around \eqref{SB},  this has implications for the   asymptotic growth of the density of states of hadrons in the Hilbert space ${\cal H}$.  In particular, the  \eqref{YM-QCD-Twisted} imply that leading order effective density of states in QCD(F) with $\Omega_F^0$-graded Hilbert space and the usual density of states in Yang--Mills theory are equal,  
\begin{align}
\rho_{\Omega_F^0}^{\rm QCD(F)} (E)  =  \rho_{SB}^{\rm YM}(E), \qquad {\rm leading \;  order,} \;  E{\rm-large}
\label{largeE-spec}
\end{align}
Furthermore, at the low-energy end of QCD(F) with $\Omega_F^0$-grading, there are tremendous cancellations. 
For example, as described in  \eqref{meson-cancel}, the contribution of $N_f^2-1$ NG bosons and one 
$\eta'$ to the partition function becomes 
$ (-1) e^{-\beta E_{\pi}}  +  1  e^{-\beta E_{\eta'}} $, which is $ O(N_f^0)$ instead of $O(N_f^2)$.  It is highly conceivable that the contribution of low-lying hadrons (which are not glueballs and alike) cancel to a great extend.   In the low energy limit, $O(N_c^2)$ spectral density of the  hadrons just reduce to  $O(N_c^0)$, just like glueball spectral density. 
\begin{align}
\rho_{\Omega_F^0}^{\rm QCD(F)} (E)   \sim   \rho_{SB}^{\rm YM}(E) \sim O(N_c^0)  \qquad    E{\rm-small}
\label{largeE-spec-2}
\end{align}
  Provided a  physical scale matching is done, 
for example the  CFC breaking temperature   $T_d[ \Omega_F^0] $   in QCD(F) and  deconfinement  temperature $T_d^{\rm YM}$ at which the $\Z_{N_c}$ center symmetry realization changes in YM are set equal, 
 $T_d[\Omega_F^0]  \equiv T_d^{\rm YM}$, we expect that  all common observables related to thermodynamic properties, such as free energies, pressure, entropy  to behave  in a similar  manner  in these two theories. 
 Quantum distillation of Hilbert space, and in particular    \eqref{largeE-spec} and  \eqref{largeE-spec}  is the 
physical  explanation  for the intriguing simulation results that appeared  in Ref.~\cite{Iritani:2015ara}. 


\section{Topology of field space and chiral symmetry }  
\label{sec:topology}
So far, we have shown that with the $N_f = N_c$ QCD(F/adj), the CFC symmetry can be made stable   at small-$\beta$ 
with a judicious choice of grading over Hilbert space. This regime (or its small perturbations) is expected to be 
 adiabatically  connected to large-$\beta$ strongly coupled regime. 
 In this section, we examine the non-perturbative aspects of the small-$S^1$  regime with unbroken CFC symmetry.

\subsection{Monopole-operators and  index theorem} 
At small $\beta$,  $a_4$ acts as an compact adjoint Higgs field. The two-loop holonomy potential \eqref{V2psi-min}  leads to a 
CFC-stabilizing $\Z_{N_c}$ symmetric minimum  \eqref{centersym}, see Fig.~\ref{fig:holonomy} as well.  The gauge structure of the theory abelianizes 
 at the length scales larger than the inverse  lightest $W$-boson mass $(m_W)^{-1}=  \frac{\beta N_c}{2 \pi}$.
\begin{align}
    SU(N_c) \rightarrow  U(1)^{N_c-1} \,.
    \label{eq:higgsing}
\end{align}
Small  $S^1 \times \R^3$   regime can be described by an abelian 
 3D effective field theory. Note that the microscopic theory is always four dimensional;  the symmetry and ABJ anomaly of the theory are the ones of the 4d theory.   
 
  \vspace{3mm}
The long distance theory is  different in different parts of parameter space $(m_\lambda, m_\psi)$.  In the vicinity of 
$(m_\lambda, m_\psi)=(0,0)$, Cartan subalgebra gluons
$a^{i}_{\mu}$  ($i = 1, \ldots, N_c-1$,  $\mu = 1,2,3$) remain gapless to all orders in perturbation theory.  
$a^{i}_{4}$ modes acquire  masses  due to two-loop potential  covering  the range 
\begin{align} 
m_{a^{i}_{4}} \in \left[ \frac{(g^2 N_c) } {(\beta N_c)},    \frac{(g^2 N_c)}{\beta} \right] \; . 
\end{align} 
   Therefore,\footnote{Note that this is a mass induced  by two-loop center-stabilizing potential. This scale is different from the electric  mass scale $m_{\rm el.} \sim g T  $ induced by one-loop center-destabilizing potential and 
  magnetic mass  scale  $m_{\rm mag.} \sim g^2  T\equiv g_3^2  $ which is  the strong scale  of long-distance 3d YM theory. 
 It is  similar in structure but still  different from the gap of the holonomy field in QCD(adj) which is induced at one-loop level by a center-stabilizing potential 
 and taking values in the range $m_{a^{i}_{4}} \in \left[ \frac{ \sqrt{(g^2 N_c)}} {(\beta N_c)},    \frac{\sqrt{(g^2 N_c)}}{\beta } \right]$. See  
 \cite{Argyres:2012ka} for the latter and \cite{Laine:2016hma} for thermal case. }    
we can ignore $a^{i}_{4}$  fluctuations in the description of physics at distances larger than $({\rm min}(m_{a_4^i}))^{-1}$ and not include at all in the long-distance effective Lagrangian. However, in the vicinity of  $m_\lambda = 0,   m_\psi =\infty$,  QCD(F/adj) reduces to  supersymmetric $\N=1$ SYM. There, $a^{i}_{4}$ field remains gapless to all orders in perturbation theory with the rest of supersymmetric multiplet and must be included in the long-distance Lagrangian.  Therefore, we will include $a^{i}_{4}$ in the long-distance EFT despite the fact that it is not needed everywhere  of the phase diagram for capturing  long-distance EFT. 

 \vspace{3mm}
In order to describe topological defects, it is more convenient to use 
 Abelian duality transformation \cite{Polyakov:1976fu}: 
 \begin{align}
 F^{i}_{\mu \nu} 
=  g^2/(2\pi \beta ) \epsilon_{\mu\nu\alpha} \partial^{\alpha} \sigma^{i}, 
\end{align}
 relating 
the Cartan gluons to 3d scalars $\sigma^{i}$, which is called ``dual photons".   Although there are $N_c-1$ dual photons, 
we will use a sligthly redundant description, and take $ i=1, \ldots N_c$. One extra mode decouples from 
dynamics.\footnote{This would be trivially true if all matter was in adjoint representation. Showing this in the presence of fundamental matter requires some care. We will return to this issue at the end of this section.}
 Define the dimensionless combination 
\begin{align} 
a^{i}_{4} \beta  \equiv v^i +  \phi^i,  \qquad i=1, \ldots N_c
\end{align} 
where  the   $v^i$ is    minimum of the holonomy potential, and  $\phi^i $  indicates  the fluctuations of the fields around it. 
 It is also convenient to combine the two-scalars into a complex scalar:\footnote{This definition is sufficient for our purpose. If desired, $\theta$-angle can be incorporated to this definition as in \cite{Davies:2000nw, Poppitz:2012sw}.}
\begin{align}
z^i =  -  \frac{4\pi}{g^2} v^i    -  \frac{4\pi}{g^2} \phi^i +  i \sigma^i, \qquad 
 z\equiv -  \frac{4\pi}{g^2} v    -  \frac{4\pi}{g^2} \phi  +  i \sigma,
 \label{z-def}
\end{align} 
With these conventions,  the kinetic term for the scalars can be written as
\begin{equation}
\label{kinetic1}
{\cal L}_{kin.} =  {g^2 \over 16 \pi^2 \beta  } |\partial_\mu z|^2  = {g^2 \over 16 \pi^2 \beta  }
    \left( (\partial_\mu \phi)^2 
         + (\partial_\mu\sigma)^2 \right)\, , 
\end{equation}
where $g^2= g^2(m_W)$.

\vspace{3mm}
In the weak coupling regime where the dynamics abelianize, there are saddle-points of the classical action, solution to classical Euclidean monopole-instanton equation, which are just  the dimensional reduction of the self-duality equation: 
\begin{align}
Da_4 = *_3 F
\label{mon}
\end{align}
The  solutions are fully classified. There are  $N_c$ types of monopole-instantons
\cite{Lee:1997vp,Kraan:1998sn} associated with the affine root system of Lie algebra $\frak{su}(N_c)$. The action of these configurations is given by the distance between the consecutive eigenvalues of Wilson line $\Omega$. 
\begin{align}
S_i =  \frac{4\pi}{g^2} v  \cdot \alpha_i  =    \frac{4\pi}{g^2}  ( v_{i+1} - v_{i} ), \qquad  i=1, \ldots, N_c
\label{action}
\end{align}
In the 
center-symmetric minimum of two-loop potential \eqref{centersym}, $( v_{i+1} - v_{i} ) =   \frac{2 \pi}{N_c}$ and 
  all of the $N_c$ monopole-instantons  have identical Euclidean actions  given by 
\begin{align}
S_0 = 
  \frac{8\pi^2}{g^2N_c} \equiv 
 \frac{S_{\cal I}}{N_c}
\end{align}
where $S_{\cal I} = 
\frac{8 \pi^2}{g^2}$ is the 4d  instanton action.

\vspace{3mm}
 In the absence of the fermionic zero modes, the leading monopole amplitudes would be given by 
 \begin{align}
 {\cal M}_{i} =    e^{-S_0}  e^{ -\frac{4\pi} {g^2} \alpha_i \cdot \phi +   i \alpha_i \cdot \sigma },  \; i=1, \ldots N_c 
 \end{align}
  similar to   Polyakov model  \cite{Polyakov:1976fu}.   However, in the presence of massless fermions, the monopole-instantons also pick up a number of fermionic zero modes as dictated by  
  index theorem for Dirac operator on $\R^3 \times S^1$  \cite{Nye:2000eg, Poppitz:2008hr}.    
  
  \vspace{3mm}
  With periodic boundary conditions for adjoint fermions, each monopole has two adjoint zero modes \cite{ Affleck:1982as, Seiberg:1996nz, Katz:1996th,  Aharony:1997bx, Davies:2000nw}. For the fundamental fermions, the story is quite different as explored in \cite{Cherman:2016hcd}. For $N_f=N_c$  theory,   if no flavor twisting is used  i.e. $\Omega_F =1$, all $2N_c$ zero modes would be localized to one type of monopole. If one imposes a $U(1)_V$ twist into the 
  fermions, $\psi (x_4+\beta ) = e^{i \delta}  \psi (x_4)$, the zero mode is 
  localized to the monopole with charge $\alpha_i$ if $\delta \in   (v_i, v_{i+1})$. As $\delta$ crosses an eigenvalue of gauge holonomy, the fundamental fermion zero mode 
   jumps from one-type of monopole to the neighboring one  \cite{GarciaPerez:1999ux, Bruckmann:2003ag}, see 
   \cite{Poppitz:2008hr} for a  discussion of the jumping phenomenon for general representations, and \cite{Moore:2014gua} for a brane interpretation of the jumping phenomena. 
 Ref. \cite{GarciaPerez:1999ux} shows this explicitly by using ADHM construction. 
  The jumping phenomenon is dictated by the structure of index theorem, and interplay between gauge holonomy and boundary conditions on fermions  \cite{ Poppitz:2008hr, Misumi:2014raa, GarciaPerez:2009mg}.  
With the center-symmetric flavor holonomy $\Omega_F^0$, the $2N_c$ zero modes fractionalize into $N_c$ groups of two and  
 each monopole acquires  two fundamental zero modes, mimicking  exactly adjoint fermions (modulo some minor caveats described below).  
 
   \vspace{3mm}
 For a general (abelianizing)  gauge holonomy  and  arbitrary flavor holonomy of the form 
 \begin{align}
 \Omega &= {\rm Diag} \left( e^{i v_1}, e^{i v_2},  \ldots, e^{i v_{N_c}}  \right) \cr 
 \Omega_F &= {\rm Diag} \left( e^{i \epsilon_1 }, e^{i \epsilon_2},   
   \ldots, e^{i \epsilon_{N_f} } \right)
   \label{back-fg}
 \end{align} 
 the number of fermionic zero modes localized at monopole-instanton $\alpha_i$  for periodic adjoint fermion $\lambda$ and the  $N_f$  fundamental fermions  $\psi^{a}, a=1, \ldots, N_f $ are  given by:
 \begin{align}
&{ \cal I}_{\alpha_i} =2 \qquad {\rm adjoint}  \cr
 & {\cal I}_{\alpha_i}  = \sum_{a=1}^{N_f} \left( {\rm sign}[ \epsilon_a -v_i] -  {\rm sign}[ \epsilon_a -v_{i+1}] \right)
\label{index}
 \end{align}
 
   \vspace{3mm}
We have few remarks.  Index jumps each time $\epsilon_a$ crosses a $v_i$. For $ v_{i-1} < \epsilon_a < v_i$, the zero mode is localized to monopole with charge $\alpha_{i-1}$. The fermion zero mode wave functions decays exponentially as $e^{- |\epsilon_a - v_i| r} $
and decays algebraically for $|\epsilon_a - v_i| =0$. For $ v_i < \epsilon_a < v_{i+1}$, the zero mode do get   localized to monopole $\alpha_{i}$. For $ |\epsilon_a - v_i|  =0 $ i.e., crossing the boundary, the index is not  well-defined.

 \vspace{3mm}
For 
$N_f=N_c$ =even, $\Omega_F^0$ eigenvalues given in \eqref{flavor-hol-0} are interspersed between the gauge holonomy eigenvalues $\Omega$ \eqref{centersym}, see Fig. \ref{fig:holonomy}.  However, for $N_f=N_c$ =odd, $\Omega_F^0$ eigenvalues coincide precisely with the  gauge holonomy eigenvalues $\Omega$. One can actually generate a gap between the two set of eigenvalues by imposing slightly more 
general boundary condition, twisted by $ e^{i \alpha}   \Omega_F^0  \in U(N_f)_V$.   From now on, we will treat the  $N_f=N_c$ odd case on the same footing with 
$N_f=N_c$ even case, and declare that the flavor holonomy eigenvalues are interspersed between gauge holonomy eigenvalues as evenly as possible.

   \vspace{3mm}
  For $N_f=N_c$ QCD(F/adj), the $N_c$-tuple of the indices  is given by:
   \begin{align}
\left[ { \cal I}_{\alpha_1},   { \cal I}_{\alpha_2}, \ldots, { \cal I}_{\alpha_{N_c}}\right]= \underbrace{[2, 2, \ldots, 2]}_{\text{ adj. fermion}} +  \underbrace{  [2, 2, \ldots, 2]}_{\text{ fund.  fermion}}  
\label{index-2}
 \end{align}
Consequently, the monopole-instanton amplitudes  in the $(m_\lambda, m_\psi)=(0,0)$ theory are given by:  
\begin{align}
\label{monopoles-2}
{\cal M}_{i} &=    
e^{  \alpha_i \cdot z } 
      ( \psi_{Ri} \psi_{L}^{i}) (\alpha_i \cdot  \lambda)^2 ,  \qquad i=1, \ldots, N_c.  
\end{align}
and possess a total of 4 fermi zero modes.   In the $N_f=N_c$ theory in the   \eqref{back-fg} background, adjoint and fundamental fermion zero modes appear exactly on  the same footing, and this will help us when we decouple one or the other. 

   \vspace{3mm}
 Note the usual relation between the 4d  instanton amplitude   \eqref{eq:instanton_fermions} and monopole-instanton amplitudes: 
 \begin{align}
{\cal I}_{4d} & \sim \prod_{i=1}^{N_c}{\cal M}_{i}   \sim 
 e^{- \frac{8 \pi^2}{g^2}} \prod_{i=1}^{N_c}   ( \psi_{Ri} \psi_{L}^{i}) (\alpha_i \cdot  \lambda)^2 ,  
\end{align}  
following from the fact$ \sum_{i=1}^{N_c} \alpha_i = 0$. 
The 4d instanton does  couple neither to holonomy field $\phi$    nor to dual photons $\sigma$. It is also  exponentially  suppressed with respect to  monopole-instantons. However,  they are topological configurations which comply with the ABJ anomaly  and    determine the anomaly free quantum symmetry of the theory  given in \eqref{diagonal} 
or anomaly free discrete remnants once a mass term is turned   either for $\lambda$  or $\psi$.  Also note that the $4N_c$ bosonic zero modes of the 4d instanton  and its 
$2N_c + 2N_c$ fermionic zero modes are matched by the $N_c$  types of monopole-instantons.

 \vspace{3mm}
 Introducing a soft mass term either for adjoint fermion or fundamental fermion, some or all fermi zero modes of the monopoles can be soaked up.    As such, the dynamical role of  monopole operators, as well as bion operators,  does  change in different   regions in the $(m_\lambda, \; m_\psi)$ plane. 
 Below we list the possible forms of the monopole operators in different regime in the   parameter space $(m_\lambda, \; m_\psi)$.
 \begin{align}
{\cal M}_{i}  = 
  \left\{
  \begin{array}{ll}
e^{-S_i}  e^{ - \frac{4 \pi}{g^2} \alpha_i \cdot \phi +   i \alpha_i \cdot \sigma } 
      ( \psi_{Ri} \psi_{L}^{i}) (\alpha_i \cdot  \lambda)^2,      & \qquad  m_\lambda=0, \; m_\psi=0 \cr \cr
    e^{-S_i} f_\lambda   \;  e^{ -  \frac{4 \pi}{g^2}  \alpha_i \cdot \phi +   i \alpha_i \cdot \sigma } 
      ( \psi_{Ri} \psi_{L}^{i}),      & \qquad  m_\lambda>0, \; m_\psi=0 \cr \cr   
      e^{-S_i}  f_\psi  e^{ -  \frac{4 \pi}{g^2}  \alpha_i \cdot \phi +   i \alpha_i \cdot \sigma } 
    (\alpha_i \cdot  \lambda)^2,      & \qquad  m_\lambda=0, \; m_\psi>0 \cr \cr
          e^{-S_i}   f_{\lambda \psi}  e^{ -  \frac{4 \pi}{g^2}  \alpha_i \cdot \phi +   i \alpha_i \cdot \sigma },      & \qquad  m_\lambda>0, \; m_\psi>0 
  \end{array} \right.
  \label{mon-op}
\end{align}
where 
 $f_\psi = \frac{\det(\gamma_\mu D_{\mu}\big|_{\rm{monopole}} +m_\psi)}{\lim_{m_\psi \to\infty} \det( 
\gamma_\mu D_{\mu}
\big|_{\rm{monopole}} +m_\psi)} $ etc. 
For small parameters,  the functions $ f_\lambda \sim m_\lambda $ and  $ f_\psi  \sim m_\psi $ and 
 $ f_{\lambda \psi} \sim  m_\lambda m_\psi $ as can be inferred  by soaking up the  fermion zero modes to soft mass term. 
 In the limit where the masses decouple, e.g. $m_\psi / m_W \gg 1$,   the corresponding functions approach to a constant.  See Section 2.1.2 of \cite{Cherman:2017dwt}. We will drop $f_\lambda, \;  f_\psi, \; 
f_{\lambda \psi}$ unless it is not strictly necessary to keep them,  in order  to lessen the clutter.

\begin{figure}[t]
\begin{center}
\vspace{-2.5cm}
\includegraphics[width = 1.4\textwidth]{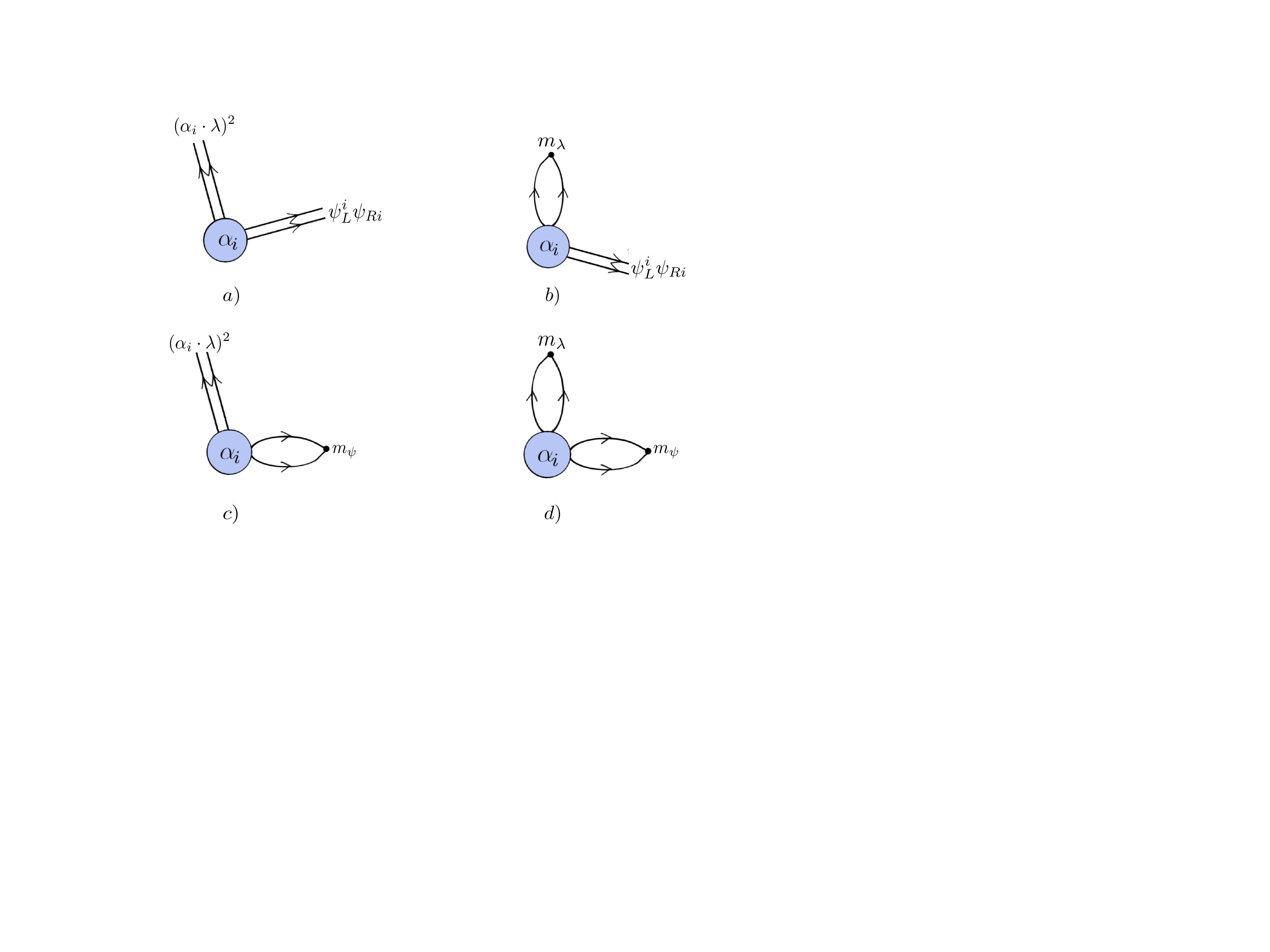}
\vspace{-8cm}
\caption{a)Monopole operators  in $N_f= N_c$ QCD(F/adj)  with $\Omega_F^0$ twisted boundary conditions
 b) with soft $m_\lambda$. c)with soft $m_\psi$ d) with soft $m_\lambda$ and  $m_\psi$.  
  In the limits where  
 either of the  quark masses is taken large, the effect of the corresponding mass on the monopole operator disappears.
  }
\label{fig:monop}
\end{center}
\end{figure}

\vspace{3mm}
  There are other correlated topological configurations, but the first two-orders in semi-classics (for this particular problem) suffice to capture  the interesting features of  non-perturbative dynamics.

\subsection{Grand canonical ensemble  and 
EFT}
\vspace{3mm}
In the semi-classical regime,  the long-distance theory at scales larger than $m_W^{-1}$  
is described   in terms of proliferation of 
topological 1-defects,   2-defects, 3-defects  etc.  We may view the Euclidean vacuum as a grand canonical ensemble  of these configurations. 
These events are: 
\begin{itemize}
\item{Monopole-instantons  ${\cal M}_i, \;   i=1. \ldots, N_c$ which are 
saddle points (critical points)  of the classical action with characteristic size $r_{\cal M}\sim m_W^{-1} $ and density  $e^{-S_i}$. }
\item{ Neutral bions   ${\cal B}_{ii} = [{\cal M}_i \overline {\cal M}_i] $ and magnetic bions  ${\cal B}_{i, i\pm1} = [{\cal M}_i\overline {\cal M}_{i\pm1}]  $ \cite{Unsal:2007jx, Shifman:2008ja} with characteristic size  $r_{\cal B} \sim  
r_{\cal M}/g^2 $ and densities  $e^{-2S_i}$ and $e^{-S_i-S_{i+1}}$. 
  These  contributions should be interpreted as exact critical points at infinity in the context of Picard-Lefschetz theory \cite{Behtash:2018voa}.  The characteristic size $r_{\cal B} $ arises because 
 Lefschetz thimbles associated with such critical points are non-Gaussian due to  quasi-zero mode (QZM) directions. 
 The characteristic size emerges as the point on the thimble where the integration is dominated. There is also a crucial hidden topological angle associated with neutral bion, such that $\arg {\cal B}_{ii} =  \arg {\cal B}_{i,i+1}  + \pi$, due to the fact that the QZM thimble of  ${\cal B}_{ii}$ makes a journey to complex domain. \footnote{In ${\cal N}=1$ SYM, for example, the relative phase is crucial to explain the vanishing of the gluon condensate, $\langle \tr F_{\mu \nu}^2\rangle$, 
which receives positive contribution from  ${\cal B}_{i,i+1}$ and negative contribution from  ${\cal B}_{ii}$. This is  because in vacuum   $
\langle {\cal B}_{i,i+1} \rangle = e^{i \pi}  \langle {\cal B}_{ii} \rangle$. This observation resolved an old puzzle why the gluon condensate can ever vanish in a vector-like gauge theory with a positive-definite path integral measure \cite{Behtash:2015kna}.}}
 \item{The higher order terms in the semi-classical expansion can be incorporated via the  cluster (virial) expansion for an interacting gas of  monopole-instantons. But  the qualitatively important behavior of the theory is  dominated by first two-order in semi-classics, while higher orders are needed for resurgent cancellations.  
 }
\end{itemize}

\vspace{3mm}
Let ${\cal T}$ denote the set of all topological defects, 1-defects, 2-defects, etc as described above. 
 \begin{align}
{\cal T} =  \left\{ {\cal M}_i,  \;\  \bar {\cal M}_i, \;\;   [{\cal M}_i  \bar {\cal M}_{i\pm1}],  \;\;
 [{\cal M}_i  \bar {\cal M}_{i}], \;\;  [{\cal M}_i  \bar {\cal M}_{j}{\cal M}_k ], \ldots \right\} 
\end{align}
 The ordering is according to the action $(S)$  or density ($e^{-S}$) of the corresponding topological event.  The leading ones are rare, their density is proportional is $e^{-S_i} $  and subleading ones are rarer, their density is  proportional is $e^{-S_i- S_{i+1}} $,    but nevertheless all are present.  It is also crucial that not all the terms come with the same sign due to 
 hidden topological angles \cite{Behtash:2015kna}, in other words, the  densities are real, but fugacities may be complex. 

\vspace{3mm}
The sum over all  events   is equivalent to adding  all operators induced by these topological configurations to the 
 action \eqref{kinetic1}: 
\begin{align}
& \prod_{\cal T}   \sum_{n_{\cal T}=0}^{\infty}    \frac{1}{n_{\cal T}! } \left[ \int d^3x \; {\cal T}   \right]^{n_{\cal T}} \cr
&= 
 \prod_{i=1}^{N_c}  \left( \sum_{   n_{{\cal M}_i} = 0}^{\infty}   \frac{1}{n_{{\cal M}_i!} }   \left[ \int d^3x \; {\cal M}_i  \right]^{n_{{\cal M}_i}}  \right)  \left( \sum_{   n_{ \bar {\cal M}_i} = 0}^{\infty}   \frac{1}{n_{ \bar {\cal M}_i!} }   \left[ \int d^3x \; \bar {\cal M}_i  \right]^{n_{ \bar {\cal M}_i}}  \right)  \ldots  \cr
&= \exp\left[ \sum_{i=1}^{N_c}  \int d^3x \;  \left( {\cal M}_i   + \bar {\cal M}_i  + {\cal B}_{ii} +{\cal B}_{i,i+1}  \ldots  \right) \right]
\label{EFT}
\end{align}
resulting in an EFT which is valid provided the theory is at sufficiently weak coupling. 
 The range of physical phenomena that this effective EFT explains is  quite diverse in various limits of QCD(F/adj),  
 and  the rest of the paper is devoted to explanation of these effects.  This is hardly  surprising  because the EFT is derived within the domain of applicability of semi-classical methods.

  \vspace{3mm}
Before proceeding, we remind that the existence of monopole operator in the effective Lagrangian does not imply   the existence of mass gap for gauge fluctuations,  despite the fact that monopole operator has both   holonomy and dual photon dependence.  
 When 
at least    one of the fermion type, F or adj,   is exactly massless,  the proliferation of monopoles do  not induce a  mass gap for gauge fluctuations, 
as evident from the monopole operators \eqref{mon-op}. 
 Only when both fermions are massive, the monopole operators themselves  may  induce a mass gap for gauge fluctuations as in \cite{Polyakov:1976fu} and   \cite{Unsal:2008ch}.

\subsection{Can gluons  acquire  a chiral charge?}
{\bf Topological shift symmetry:} 
In the absence of monopole instantons, the dual formulation has an $[U(1)_J]^{N_c -1}$  topological shift symmetry, which protects the gaplessness of the  dual photon to all orders in perturbation theory. 
The symmetry and its Noether current are: 
\begin{align}
[U(1)_J]^{N_c -1}:  \sigma  \rightarrow  \sigma   +   \varepsilon,  \qquad  {\cal J}_{\mu} = \partial_{\mu}  \sigma  
\label{shift-mix}
\end{align}
By abelian duality relation, the current is  the (euclidean) magnetic field, $ B_{\mu}$, and current conservation is the statement of the absence of monopoles: 
\begin{align}
\partial_{\mu}  {\cal J}_{\mu} =\partial_{\mu}   B_{\mu}   = 0. 
\end{align}
If there are  monopoles in the theory,  (and no massless fermions), then, of course, current is no longer conserved:
\begin{align}
\partial_{\mu}  {\cal J}_{\mu} =\partial_{\mu}   B_{\mu}   =  \rho_m(x)  \neq 0
\end{align}
where $\rho_m$ is the magnetic charge density. This implies that the  topological shift symmetry   \eqref{shift-mix} is  not present at non-perturbative level. 
Indeed, absence of the shift symmetry is synonymous with the proliferation of the monopole-instantons, and  non-perturbatively  $[U(1)_J]^{N_c -1}$ violating monopole-operators  are induced. 
The effective long distance Lagrangian exhibits both confinement of electric charge with finite string tensions and a non-perturbative mass gap for the gauge fluctuations, see   \cite{Polyakov:1976fu} on  $ \R^3$ and \cite{Unsal:2008ch} for  its generalization to locally four-dimension on $ \R^3 \times S^1 $.

\vspace{3mm}
\noindent
{\bf Mixing of topological shift symmetry with  chiral symmetry:} 
With the inclusion of massless fermions, the story is different and  not less    interesting, as it captures other non-perturbative phenomena. 
 What happens precisely depends on the theory. 
However, overall story develops as follows. 
\begin{itemize} 
\item{
Unlike the 4d instanton amplitude which is singlet under the  chiral symmetries of the theory  
$ {\bf G}_{\rm max-ab} \subset  {\bf G}$ given in \eqref{MAG},  the fermion zero mode structure of the monopole operator is generically not  a singlet under the  chiral symmetry.  } 

\item{Since the corresponding symmetry is a genuine  {\it non-anomalous} symmetry of the QFT,  the monopole operator as a whole {\it must be} singlet under it. }

\item{What happens is that a subgroup (or whole) of  $[U(1)_J]^{N_c -1}$    intertwines  with $ {\bf G}_{\rm max-ab}$. 
This mixing guarantee that  $ {\bf G}_{\rm max-ab}$ is respected by monopole-operators. }

\item {This is a  strange phenomenon, however.   It is usually believed that  in theories with massless fermions, only microscopic fermions and their composites  are charged  under chiral symmetry. A  priori, gauge field has ``nothing" to do with  chiral charges associated with \eqref{MAG}.  Yet,  the intertwining of emergent   shift symmetry  and microscopic chiral symmetry  tells us that the gauge fluctuations {\bf must}  acquire a chiral charge, and there is  no other option!  This type  effect first appeared  in a gauge theory on $\R^3$  in the work of Affleck, Harvey and Witten \cite{Affleck:1982as}. Its generality and usefulness in gauge theories on $\R^3 \times S^1$ is newer  \cite{Unsal:2007jx,  Cherman:2016hcd}.  
It actually provides an example of analytically calculable  chiral symmetry breaking in both QCD(F/adj), QCD(F) and $\N=1$ SYM, a phenomenon that is believed to be an incalculable  strong coupling effect is realized in weak coupling calculable domain. }
\end{itemize}
Below, we explain this qualitative picture quantitatively.

\vspace{3mm}
\noindent 
{\bf What is conserved  and what is violated at the monopole-instanton event?}  
The charge associated with the topological current is magnetic charge  ${\bf Q}_m = \int d^2x \;  {\cal J}_0 =    \int d^2x  \; \partial_0 \sigma $ and at the monopole-event, it  is clearly 
violated. Consider a collection of $n_i$ monopoles of type $\alpha_i$  for $i= 1, \ldots, N_c$ sprinkled  in between two asymptotic time slice.  Then, 
the magnetic charge non-conservation is 
\begin{align}
\Delta {\bf Q}_m &= {\bf Q}_m (t= \infty) - {\bf Q}_m (t= - \infty) = \int d^2 x  F_{12}  \Big |_{t=-\infty}^{t=+\infty}  \cr 
& =  \int_{S^2_{\infty}}  F_{12}  \cr
&= \frac{4 \pi }{g} \sum_{i=1}^{N_c} n_i \alpha_i \cr
&  = \frac{4 \pi }{g}  \left( n_1 - n_{N_c}, n_2 -n_1, n_3-n_2, \ldots,  n_{N_c -1} -  n_{N_c} \right)
\end{align} 
These charges violate {\it emergent} $ [U(1)_J]^{N_c -1}$ explicitly and completely. 
However, the non-conservation of magnetic charge is  not whole story in theories with  dynamical fermions.

\vspace{3mm}
%
In the presence of massless fermions,  each monopole-event carries certain number of fermionic zero modes \cite{Nye:2000eg, Poppitz:2008hr}. 
The zero modes are  charged under maximal torus of {\it microscopic} {\it non-abelian} axial symmetry, $ {\bf G}_{\rm max-ab}$ given in \eqref{MAG}.    However, unlike the $U(1)_A $ which is anomalous in QCD(F), the non-abelian chiral symmetry   \eqref{symmetry}  is non-anomalous, and  
any of its subgroups  cannot be violated  by the non-perturbative events! The resolution of this puzzle is the key piece to understand the chiral symmetry breaking and the fact that gauge fields in gauge theories may and do acquire a chiral charge.

\vspace{3mm}
The axial current associated with non-abelian chiral symmetry in 4d can be written as  
$  J^{5A}_{\mu} =\bar  \psi  \gamma_{\mu} \gamma_5 T^A  \psi $  where $T^A, A=1, \ldots, N_f^2-1$ are generators of $SU(N_f)$. 
This  current is conserved,  and the corresponding charges are: ${\bf Q}^{5A}= \int d^3 x  \; \psi^{\dagger} \gamma_5 T^A \psi$. Namely, the charge commutes with the Hamiltonian $ [H, {\bf Q}^{5A}]=0$ for all $A$. However, in the 
the graded partition function  \eqref{GPF}, we have $SU(N_f)_V$ charges along Cartan sub-algebra,  and 
the operator  
$H' = H- i   \sum_{a \in {\rm Cartan}} \frac{\epsilon_a}{\beta }   Q_a$ only commutes with the Cartan generators of the axial charges,    $ [H', {\bf Q}^{5A}]=0 $ for all $A \in {\rm Cartan} $. 
Let us use an explicit basis for Cartan generators:
\begin{align}
(H^{A=a})= \half {\rm Diag} ( 0, \ldots, 0\underbrace{1}_a, \underbrace{-1}_{a+1}, 0, \ldots, 0  ) 
\end{align}
In this case, we can express the chiral charge as: 
\begin{align} 
{\bf Q}^{5,A=a}=  \half ( Q^{5,a} - Q^{5,a+1} ) = \alpha_a \cdot Q^5, \qquad  A= 1, \ldots N_f-1  
\label{charge-1}
\end{align}
where  ${Q}^{5,a}= \int d^3 x  \; \psi^{\dagger, a} \gamma_5 
\psi_a$ and 
all ${\bf Q}^{5,A=a}$ commutes with Hamiltonian and  $H' $.  We can augment this list with an affine-charge, which is conserved because it is a linear combination of the other conserved charges: 
\begin{align} 
{\bf Q}^{5,A=N_f}  \equiv - \sum_{a=1}^{N_f-1} \alpha_a \cdot Q^5 =       \half ( Q^{5,N_f} - Q^{5,1} )  =  \alpha_{N_f}  \cdot Q^5
\end{align}

\vspace{3mm}
Here is the main point: Despite the fact that  ${\bf Q}^{5,A=a}$ commutes with the Hamiltonian as well as grading, 
it appears  to be violated by the fermionic zero mode structure of the  monopole-instantons: 
\begin{align}
{\bf Q}^{5,A=a}:  ( \psi_{Ri} \psi_{L}^{i})   \rightarrow  (  \delta_{ai}- \delta_{a+1i})  ( \psi_{Ri} \psi_{L}^{i})
\end{align} 
  How can a symmetry be  simultaneously non-anomalous and also    ``apparently"  violated by monopole-instanton amplitude? Of course, this is not possible. A non-anomalous symmetry of the theory must be respected by all topological configurations. 

\vspace{3mm}
Consider again  collection of $n_i$ monopoles of type $\alpha_i$  for $i= 1, \ldots, N_c$ sprinkled  in between two asymptotic time slice.  The background will have  $n_1$ many   $ (  \psi_{R1} \psi_{L}^{1})$ zero modes,  $n_2$ many   $ ( \psi_{R2 }\psi_{L}^{2} )$  zero modes and  $n_{N_f}$ many   $ ( \psi_{RN_f} \psi_{L}^{N_f} )$  zero modes. Then, the apparent axial  charge non-conservation will be 
\begin{align}
\Delta {\bf Q}^5  &= {\bf Q}^5 (t= \infty) - {\bf Q}^5  (t= - \infty) \cr 
&= \sum_{A=1}^{N_f} n_A  \alpha_A \cr
&  =2   \left( n_1 - n_{N_f}, n_2 -n_1, n_3-n_2, \ldots,  n_{N_f -1} -  n_{N_f} \right)
\end{align} 

  \vspace{3mm}
Assume momentarily $N_f=N_c$. In this case, it is clear that magnetic charge non-conservation and chiral charge non-conservation are exactly proportional to each other for any background. In fact, we can construct a linear combination of these two-charges which is respected by all non-perturbative and topological configurations:
\begin{align} 
{\bf \tilde Q} =  \frac{g}{4\pi} {\bf  Q}_m  -   {\bf Q}^5, \qquad   {\rm such \;  that} \qquad \Delta  {\bf \tilde Q} =0 
\label{conserved}
\end{align}
What does this mean?  Here,  $ {\bf Q}^5 $ microscopic chiral  charge is associated with  Cartan subgroup $[U(1)_A]^{N_f -1}$  of the full non-abelian chiral symmetry, while 
${\bf  Q}_m  $ is an emergent symmetry  in EFT, but it  is only  valid to all orders in perturbation theory.  So, the genuine microscopic symmetry here is only chiral symmetry, and this whole mechanism is present  so that the chiral charge of the fermion bilinear can be transferred to gauge fluctuations!  We can also equivalently say that  the diagonal of the  
 ${\bf G}_{\rm max-ab}$ and topological shift symmetry 
\begin{align}
[U(1)^{N_c-1}]_{AJ} = { \rm Diag} \left( [U(1)^{N_c-1}]_A \times [U(1)^{N_c-1}]_J \right)
\label{shift}
\end{align}
  is the symmetry of the long-distance effective field theory.   

\vspace{3mm}
The conserved charge \eqref{conserved} is the global content of the local  
 current conservation, $ \partial_{\mu}  \left( \frac{g}{4\pi}  {\cal J}_{\mu}  -J^5_{\mu} \right) =0.$ Naively, the fermionic zero mode part of the monopole operator would induce an anomaly, because it is proportional to
$ \psi_{Ri} \psi_{L}^{ i}$.  But this is always accompanied by a magnetic  flux event, and a correlated change in the magnetic flux compensates the change in the fermion number. 
The combination in \eqref{conserved} is conserved  in any perturbative or non-perturbative process. As described below, the choice of the vacuum breaks this symmetry spontaneously. 

\vspace{3mm}
The mechanism described above can be generalized to all QCD(F) with $ N_f \leq N_c$ and 
 $1\leq N_f \leq N_c-1$  QCD(F/adj) in abelianizing gauge holonomy background.  We will show that in those cases, it accounts for all Nambu--Goldstone bosons in the theory with twisted boundary condition.  

 \vspace{3mm} 
 \noindent
{\bf Chiral symmetry order parameters:}
 Because of the topological shift and chiral symmetry mixing,  in gauge theories in general there are two  types of chiral order parameters: 
\begin{align}
&{\rm  Monopole  \; (magnetic \;  flux) \;  operators:}   \qquad   \qquad   e^{   \alpha_i \cdot z 
}   \cr
&{\rm Fermion  \; bilinears, \;  multilinears:}  \qquad \qquad \qquad  \psi_{Ra} \psi_{L}^{b},    \;\;  \tr \lambda \lambda,  \;\;   
(\psi_{Ra} \psi_{L}^{b}  \tr \lambda \lambda)  \qquad \qquad \qquad \qquad \qquad \qquad 
\end{align}
In all calculable examples in semi-classical domain on $\R^3 \times S^1$, $\chi$SB occurs due to condensation of the magnetic flux operators. 
In  most interesting cases, $(m_\lambda=0, m_\psi =0), (m_\lambda>0, m_\psi =0), (m_\lambda=0, m_\psi >0)$, the realization of this scenario differs in crucial ways. Below, we describe each in some detail.

\subsection{QCD(F/adj)}
\noindent
$\bf m_\lambda=0, m_\psi =0 :$
What happens in the presence of  massless fermions and twisted boundary conditions?  As described in Section \ref{sec:general},  
 although the microscopic theory has non-abelian chiral symmetry 
${\bf G}$ given in  \eqref{symmetry}, the twisted boundary conditions explicitly reduces the global symmetry down to its maximal abelian subgroup  ${\bf G}_{\rm max-ab} \subset  {\bf G} $ \eqref{MAG}. 

\vspace{3mm}
The action of the axial subgroup of  ${\bf G}_{\rm max-ab}$ on   fermion bi-linears is, for $N_f=N_c$:
 \begin{align}
[U(1)_A]^{N_f-1}:  &\qquad  ( \psi_{Ri} \psi_{L}^{i})  \rightarrow e^{i \varepsilon_i } ( \psi_{Ri} \psi_{L}^{i}),   \qquad   \qquad \lambda \lambda \rightarrow \lambda \lambda  \cr \cr 
U(1)_{A_D}:  &\qquad  ( \psi_{Ri} \psi_{L}^{i})  \rightarrow e^{ -2i \gamma  } ( \psi_{Ri} \psi_{L}^{i}),    
\qquad   \qquad \lambda \lambda \rightarrow  e^{ 2i \gamma  }  \lambda \lambda  
\label{transform}
\end{align} 
such that there is one constraint among $N_c$ continuous variables $\varepsilon_i$. 
\begin{align}
 \sum_{i=1}^{N_c} \varepsilon_i=0, 
 \label{constraint}
\end{align}
This means the fermion zero mode part of the monopole operator \eqref{monopoles-2} transform non-trivially under $[U(1)_A]^{N_c-1}$.
In order for it to be invariant under the continuous chiral symmetries, 
the pure flux part of the monopole operator must transform as 
 \begin{align}
 e^{  \alpha_i \cdot z}  \rightarrow  e^{-i \varepsilon_i }   e^{    \alpha_i \cdot z},  \qquad  \varepsilon_i= \alpha_i\cdot \varepsilon \qquad 
\end{align} 
which is nothing but the shift symmetry \eqref{shift-mix} described above and $\varepsilon_i$ satisfy the constraint \eqref{constraint}. 
The axial-emergent  topological symmetry mixing is the mechanism that the monopole operators respect invariance under the non-anomalous chiral symmetry and this is an explicit realization of the $ [U(1)^{N_c-1}]_{AJ}$ 
given in \eqref{shift}. 
 Note that both the flux operator as well as four-fermi operator in \eqref{monopoles-2} are singlet under $U(1)_{A_D}$.

\vspace{3mm}
Unlike the discrete shift symmetry, the continuous shift symmetry  \eqref{shift} forbids a mass term for the dual photon at  any non-perturbative order. To appreciate this contrast, 
recall that  in $\N=1$ SYM,  we only have topological shift  symmetry intertwining with the discrete  chiral $\Z_{2N_c}$ symmetry.    Although mass term for gauge fluctuations cannot be induced at first order in semi-classics, at second order,  semi-classical magnetic bion effects induce a mass term for gauge fluctuations \cite{Unsal:2007jx}.  
But the magnetic bions in QCD(F/adj) has  fermionic zero modes which cannot be contracted. The magnetic bion amplitudes in these two cases 
are: 
\begin{align} 
&{\cal B}_{i, i+1}  \sim e^{-2S_0}   e^{-  \frac{4 \pi}{g^2} ( \alpha_i + \vec\alpha_{i+1}) \cdot \phi } 
        e^{ i (\alpha_i - \vec\alpha_{i+1}) \cdot \sigma }, \qquad  \qquad   \qquad  \qquad \qquad   \;\; \N=1 \; {\rm  SYM}, \cr 
  &      {\cal B}_{i, i+1}   \sim e^{-2S_0}   e^{-  \frac{4 \pi}{g^2} ( \alpha_i + \vec\alpha_{i+1}) \cdot \phi } 
        e^{ i (\alpha_i - \vec\alpha_{i+1}) \cdot \sigma }  ( \psi_{Ri} \psi_{L}^{i})  (\bar \psi_{R}^{i+1} \bar \psi_{L,i+1})  \qquad  
        \rm QCD(F/adj)
        \label{m-bions}
\end{align}
Therefore, in $\N=1$ SYM,  magnetic bions induce a mass term for $\sigma$ fluctuations and in QCD(F/adj),  
 the continuous  shift symmetry forbids formation of any potential for the dual photon.  

\vspace{3mm}
The chiral symmetry is spontaneously broken with a magnetic flux order parameter  acquiring a vev.  In center-symmetric minimum, 
\begin{align} 
 &\langle {\rm VAC}|  e^{  -  \alpha_i  \cdot z }  | {\rm VAC} \rangle = e^{- \frac{4\pi}{g^2}  ( v_{i+1} - v_{i} )}  \langle e^{  -   \frac{4 \pi}{g^2} \alpha_i  \cdot \phi + i  
  \alpha_i  \cdot \sigma }  \rangle = e^{-S_0} e^{i \delta_i},  \cr \cr
& {\rm Diag} [ e^{i \delta_1}, \ldots, e^{i \delta_{N_f}}] \in  {\bf T}^{N_f-1}
  \label{flux-condensation}
\end{align}
leading to spontaneous breaking of chiral symmetry: 
 \begin{align}
[U(1)_A]^{N_f-1} \times U(1)_{A_D} \longrightarrow U(1)_{A_D}
\end{align} 
Therefore, the $N_c -1$ dual photons (which can be viewed as fluctuations around constant phases $e^{i \delta_i}$)  are identified with  the $N_c -1$  Nambu--Goldstone bosons of the spontaneously broken maximal abelian chiral symmetry.  These NG-bosons   live in the maximal torus  ${\bf T}^{N_f-1}$  of the $SU(N_f)_A$, where $N_f=N_c$, and   constitute the fields that make the chiral Lagrangian.

\vspace{3mm}
\noindent
{\bf Fermion multi-linear condensate in  statistical interpretation:} 
 In the vacuum, we can set the flux part of the monopole-operator to its vev 
\eqref{flux-condensation} following a similar rationale in \cite{Davies:1999uw, Davies:2000nw}. 
 Therefore, 
at leading order in semi-classics,  the vacuum is  a dilute gas of $N$ types of monopole-instantons  
 each with complex fugacity 
\begin{align}
\zeta_i = e^{-S_0} e^{i \delta_i},  \qquad i= 1, \ldots, N_c 
\label{f-condense-2}
\end{align}
Physically, the magnitude of fugacity $|\zeta_i|$    is the density of monopole of type-$i$: 
${\cal N}_{{\cal M}_i} / V_{\R^3 \times S^1}  $ where ${\cal N}_{{\cal M}_i} $ is the number of type-$i$ monopoles in volume 
$V_{\R^3 \times S^1}$.  
In the statistical interpretation, we can think of fermion multi-linear condensate 
$ \langle   \psi_{Ri} \psi_{L}^{ j}     \tr \lambda \lambda
\rangle$ as follows. The multi-linear order parameter will pick up contributions only from the support of the monopole-cores. 
The zero modes can be saturated by the zero modes of the monopole. Depending on the monopole type, each species has a different complex phase. In this statistical interpretation, 
$ \langle     \psi_{Ri} \psi_{L}^{ j}     \tr \lambda \lambda
\rangle \sim \delta^i_j e^{-S_0} \beta^{-6}e^{i \delta_i} $ in the semi-classical domain where $\beta N_c\Lambda \lesssim 1$, and $\Lambda$ is the strong scale of QCD(F/adj).   In this regime,   there is no reason for the remaining  $U(1)_{A_D}$ to break as  
\begin{align}
\Delta {\cal L} \sim  e^{-S_0}  \langle e^{  i  \alpha_i  \cdot \sigma } \rangle    \sum_{i=1}^N    \psi_{Ri} \psi_{L}^{ i}     (\alpha_i \cdots \lambda)^2 + {\rm h.c.}
\end{align}
 is quite  weak, and  the fermion bilinears are not capable of 
producing a vev. As a result, 
\begin{align}
   \langle  \psi_{Ri} \psi_{L}^{ j} \rangle 
   =  \langle    \tr \lambda \lambda \rangle =0,   \qquad   \langle   \psi_{Ri} \psi_{L}^{ j}     \tr \lambda \lambda
\rangle  =   \delta_i^j  \Lambda^6 (\Lambda \beta N_c)^{-11/3}  e^{i \delta_i},  \qquad  \beta N_c\Lambda \lesssim 1
\label{four-fermi}
   \end{align}

%
%
%

\vspace{3mm}
\noindent
{\bf Two possibilities at large $S^1 \times \R^3$ and adiabatic continuity:} 
In the semiclassical regime, $U(1)_{A_D}$ axial chiral symmetry remains unbroken, because the flux part of the monopole-operator is not charged under this symmetry and the four-fermi operator obtained upon condensation of flux operator is very weak. 
 On the other hand, in the strong coupling domain, as we described earlier, two patterns of chiral symmetry breaking are plausible, \eqref{pattern1} and \eqref{pattern2}.

\vspace{3mm}
If \eqref{pattern2}  and \eqref{condensate2} is 
realized in $\R^4$, then, QCD(F/adj) exhibits adiabatic continuity, i.e. there are no singularities of the graded partition function \eqref{graded} for any value of $\beta \in (0, \infty)$. Then, $\Z_{N_c}$ CFC-symmetry  remains unbroken, and  $[U(1)_A]^{N_c-1}$ is broken  and   $U(1)_{A_D}$ is unbroken  for any value of $\beta \in (0, \infty)$. This option is not completely unreasonable because there is a large-amount of non-trivial anomaly matching  that  works out on $\R^4$ \eqref{0-anomaly}. 

\vspace{3mm}
If \eqref{pattern1}  and  \eqref{condensate1} is 
realized in $\R^4$,   and chiral symmetry is broken to vector-like subgroup, 
as the radius is varied from small-$\beta $ to large-$\beta $, there must exist  a chiral transition associated with a change in the realization of 
 $U(1)_{A_D}$.  Note that at small-$\beta $, we have shown that $\Z_{N_c}$ remains unbroken, and  $[U(1)_A]^{N_c-1}$ is broken, and we expect these symmetry realizations to be exactly the same at  large-$\beta $ strong coupling. Therefore, the only symmetry that can clash with analyticity is the  $U(1)_{A_D}$. 
 
  \vspace{3mm}
 On dynamical grounds, the four-fermi operators induced by condensation of flux operators can in fact generate fermion bilinear condensates if its  coefficient becomes sufficiently strong.  However,  strong coupling, assuming that it sets in (which is reasonable),  will take place at  $\beta N_c\Lambda \gtrsim  1 $ and as such,  we may expect a phase transition at the non-'t Hooftian scale  $\beta_c \sim \Lambda^{-1}/ N_c   $.  
 However, this  is parametrically  at the boundary of the region of validity of EFT. As such, it is not completely reliable, yet, we find more reasonable than the first option above.

 \vspace{3mm}
 At this stage,  
despite the fact that   $\Z_{N_c} \times  [U(1)_A]^{N_c-1}$ symmetry realizations are same in small- and large-$\beta $, $U(1)_{A_D}$ realization may change.    As a result, 
 adiabatic continuity in QCD(F/adj)  may or may not work as a function of $S^1$ radius just because of a single $U(1)_{A_D}$  factor.   See Fig.  \ref{fig:continuity}.

%

\subsection{SQCD with $N_f=N_c$  on $\R^4$  and QCD(F/adj) on $\R^3 \times S^1$: Adiabatic continuity}Seiberg showed that  the ground states of $N_f=N_c$  (quantum moduli space) SQCD is parametrized   by chiral meson and baryon superfields obeying $\det M - B\bar B= \Lambda^{2N_c}$ \cite{Seiberg:1994bz}.  At the point $B=\bar B =0$,  we have    $\det M = \Lambda^{2N_c}$,  and the chiral symmetry is broken  as in \eqref{pattern2}, $ \bf G \rightarrow   G_V \times U(1)_{A_D} $ to a subgroup which has a  chiral component. As described around \eqref{condensate2}, at this point in moduli space, fermion bilinear condensate cannot form, as they are charged under the 
$U(1)_{A_D}$. Instead  the operator the scalar component of the meson superfield acquires a vev: 
$  M^{a}_{b}=Q_{Ra}Q_{L}^{b} = \Lambda^2 \delta_{a}^{b} $. This is singlet und

   
\vspace{3mm}
Aharony et.al. \cite{Aharony:1995zh} showed that  this patterns continues to hold with a supersymmetry breaking soft mass for the 
 scalar  quark  field  (squark). When the squark mass infinite, and the theory reduces to QCD(F/adj), as  explained in  Ref.~\cite{Aharony:1995zh}   and reviewed   in Section \ref{sec:expectations}, 
 there are two physically well-motivated  possibilities,    
 \begin{align}
 { \bf G \rightarrow   G_{\rm V}}  \qquad  {\rm  or}  \qquad { \bf G \rightarrow   G_{\rm V} \times U(1)_{\rm A_D}  } 
 \label{options} 
 \end{align}
 As we explained in   Section \ref{sec:expectations}, the second option is also  reasonable because it  large amount of zero-form mixed anomalies can be matched by massless composite fermions  \eqref{0-anomaly}.  Ref.  \cite{Aharony:1995zh}  argued, as the squark mass is increased, 
 the theory may or may not have a phase transition, i.e. SQCD may or may not be adiabatically connected to QCD(F/adj) on $\R^4$.  This question is still unsettled and we do not know the answer either.  However, we will provide an adiabatic continuity between SQCD  on  arbitrarily large $\R^3 \times S^1$ and QCD(F/adj) on small $\R^3 \times S^1$.
 
 \vspace{3mm}
 First, consider SQCD with $\Omega_F^0$  twisted boundary condition on large $S^1 \times \R^3 $: 
 \begin{align} 
 W_{\alpha} (\beta) &=  W_{\alpha} (0),  \cr
  Q_{R/L} (\beta) &=   Q_{R/L} (0)  \overline \Omega_F^{0} e^{i \pi} 
  \end{align} 
  where  $W_{\alpha}$ is field strength multiplet and $Q_{R/L}$  are $N_f=N_c$  chiral multiplets.    Since these  boundary conditions are  
  implemented at the level of superfields, they respect ${\cal N}=1$ supersymmetry.  
 
 \vspace{3mm}
 With the use of $\Omega_F^0$ twisted boundary condition on large $S^1 \times \R^3 $,  the chiral symmetry explicitly reduces to its maximal torus at the scale of compactification.  By twisted compactification, the pattern of the chiral symmetry breaking becomes  
 $ \bf G_{\rm max-ab} \rightarrow   G_{\rm V, max-ab} \times U(1)_{\rm A_D} $  in the compactified supersymmetric theory at  $B=\bar B =0$.  This also holds   for sufficiently soft squark mass. Note  
 that    with this explicit breaking,  the $N_f^2 -1$ Nambu--Goldstone bosons  reduce  to $N_f-1$ NG bosons, and 
 $N_f^2 -N_f$ modes acquire masses are in the range $[\frac{2\pi}{\beta N_f}, \frac{2\pi}{\beta}]$.  These massive modes become gapless in the $\beta\rightarrow \infty$ limit.  Turning on a soft mass for squark field, $m_{q_a}  \neq 0$, this  pattern holds for sufficiently small mass perturbation.  

 \vspace{3mm} 
The moral behind this boundary condition is that, just like our non-supersymmetric example,  the Polyakov loop wrapping the  $S^1$ cicrle  becomes a genuine order parameter under $\Z_N$ color-flavor center symmetry, see section \ref{CFC-sec}.  In the non-supersymmetric QCD(F/adj) case, these boundary condition make sure that the gauge holonomy potential is minimized at the center-symmetric point \eqref{centersym}.   In supersymmetric theory, the gauge holonomy potential is zero to all orders in perturbation theory even in the presence of  $\Omega_F^{0}$  background, as it respects supersymmetry.  The contributions of component fields in   $W_{\alpha}$  cancel exactly, and the contributions of scalar and fermionic contribution in   $Q_{R/L}$ cancel exactly.  Therefore, since the effect of fermionic component of $Q_{R/L}$ is in favor of  stabilization of CFC symmetry, \eqref{V2psi-min},  the effect of the scalar must be other way around.   
Lifting the scalars (whose effect is to undo center-stabilizing contribution of fundamental fermions with  $\Omega_F^{0}$ boundary conditions), the 
${\cal N}=0$ deformation of the susy   theory  will land on the CFC symmetric phase on $S^1 \times \R^3 $.

 \begin{figure}[h]
\vspace{-0cm}
\begin{center}
\includegraphics[width = 0.7\textwidth]{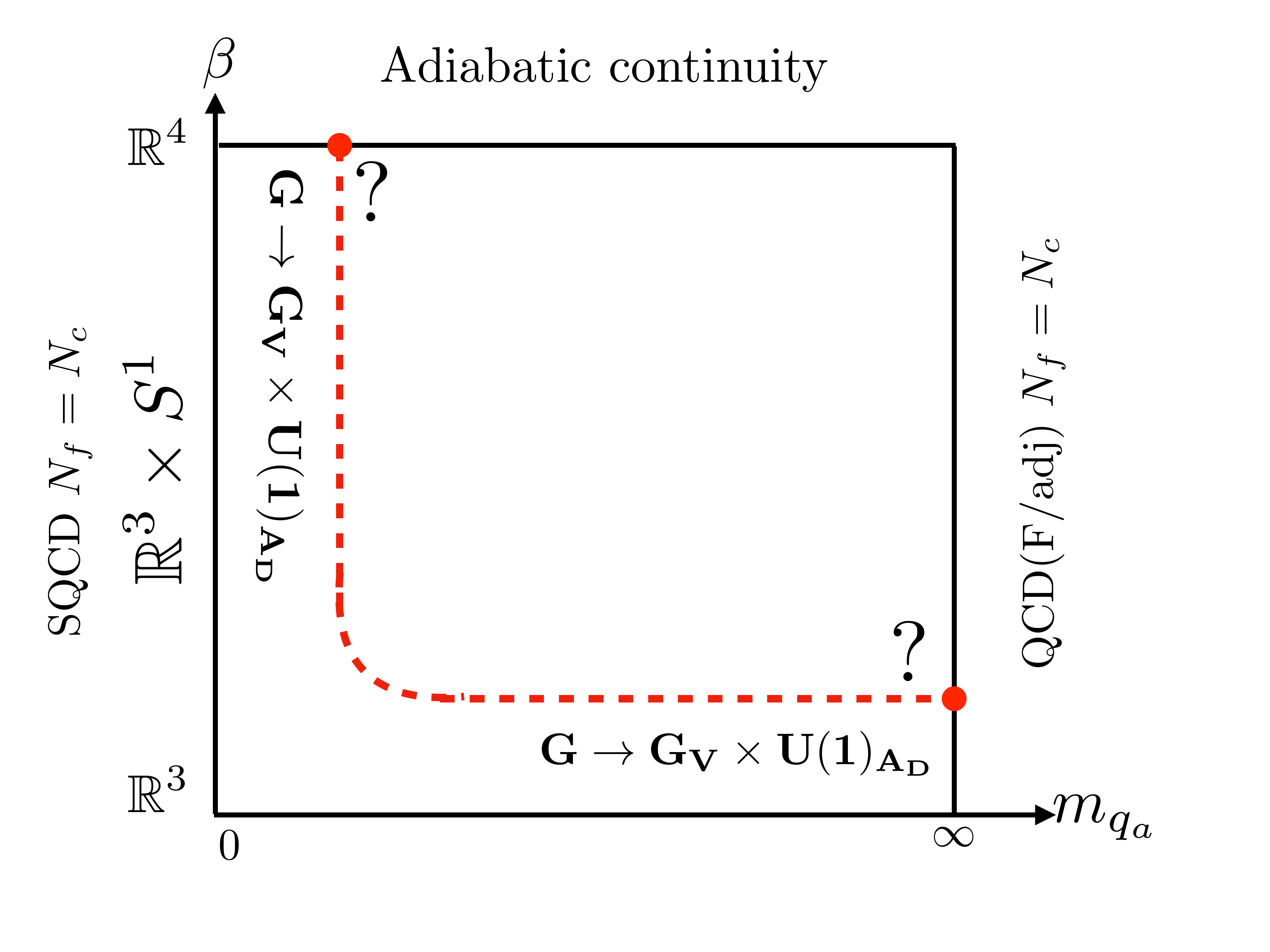}
\vspace{-.5cm}
\caption{There exists an adiabatic continuity between $N_f=N_c$ SQCD and its soft supersymmetry breaking scalar mass deformation $ m_{q_a}  $  at strong coupling on   large  $S^1 \times \R^3$  and non-supersymmetric QCD(F/adj) at weak coupling on small $S^1 \times \R^3$.  
There  is very likely a phase transition associated with $U(1)_{A_D}$ crossing red-dotted line, but absence of phase transition is also consistent with mixed anomalies.  }
\label{fig:continuity}
\end{center}
\end{figure}

\vspace{3mm}
It is intriguing to note that in QCD(F/adj) on small $\R^3 \times S^1$, we find ourself in exactly the same situation.  In the calculable weak coupling domain, the chiral symmetry is broken as $ \bf G_{\rm max-ab} \rightarrow   G_{ \rm V, max-ab} \times U(1)_{\rm A_D} $ by the condensation of monopole flux operators \eqref{flux-condensation}, but  there may be  a $U(1)_{A_D}$ changing phase transition as $\beta$ is dialed.  

\vspace{3mm}
On the other hand, there seems to be an adiabatic continuity between the strongly coupled SQCD with scalar mass deformation and QCD(F/adj) in the weak coupling domain (both endowed with flavor twisted boundary conditions with $\Omega_F^0$)  provided 
\begin{align}
{\rm max} ( m_{q_a} \Lambda^{-1}, \beta \Lambda) \lesssim 1
\end{align}
in the phase diagram in the $(m_{q_a}, \beta) $ plane. (We assumed $N_c$ is fixed and small here.) 
\footnote{
A related  interesting problem is the  $\N=2$ SYM with scalar mass  $m_\Phi $   breaking it to $\N=0$ \cite{Konishi:1996iz, Cordova:2018acb},  and its  compactication to $\R^3 \times S^1$.  The  $(m_\Phi, \beta)$ plane admits a calculable decoupling limit  $m_\Phi \rightarrow \infty$ provided 
$\beta \Lambda N \lesssim 1$.  This  work is  ongoing in collaboration with M. Anber.}

\subsection{QCD(F) and  adiabatic continuity}
\label{sec:acont}
Let us first remind some aspects of the chiral Lagrangian in QCD(F). 
 At $\R^4$,   the  continuous chiral symmetry  $SU(N_f)_L \times SU(N_f)_R $  is spontaneously broken  to vector-like subgroup  
 and the infrared physics is described by chiral 
  Lagrangian 
 $S= \int_{\R^4}  \left[  \frac{f_\pi^2}{4} \tr |\partial_{\mu} \Sigma|^2  + \ldots \right]  $ where chiral 
  field $\Sigma(x)$ captures $N_f^2-1$ NG bosons.  
Turning on a background vector-like flavor field amounts to $\partial_{\mu} \rightarrow D_{\mu} = 
  \partial_{\mu}  + i [ A_{\mu}, \; ]$. Hence, the effect of the flavor homonomy in the compactified direction is 
  to turn on 
  \begin{align}
  A_{\mu}= \delta_{\mu 4} A_4 \equiv   \delta_{\mu 4}  \frac{1}{\beta } {\rm Diag} \left( \epsilon_1,  \epsilon_2, \ldots,  \epsilon_{N_f} \right), 
  \end{align}
   The chiral Lagrangian on $\R^3 \times S^1$  in the flavor holonomy background can be written as 
  \begin{align}
  S_{\Omega_F} = \int_{\R^3 \times S^1}  \left[  \frac{f_\pi^2}{4} \tr |D_{\mu} \Sigma|^2 \right] 
  \end{align}
 The background flavor holonomy  gives a mass to the  off-Cartan components of the meson field  of the order of $\frac{2\pi}{\beta N_f}$. Therefore, only $N_f-1$ mesons remain exactly massless in the $\Omega_F^0$-twisted  background.

\begin{figure}[t]
\begin{center}
\includegraphics[width = 0.9 \textwidth]{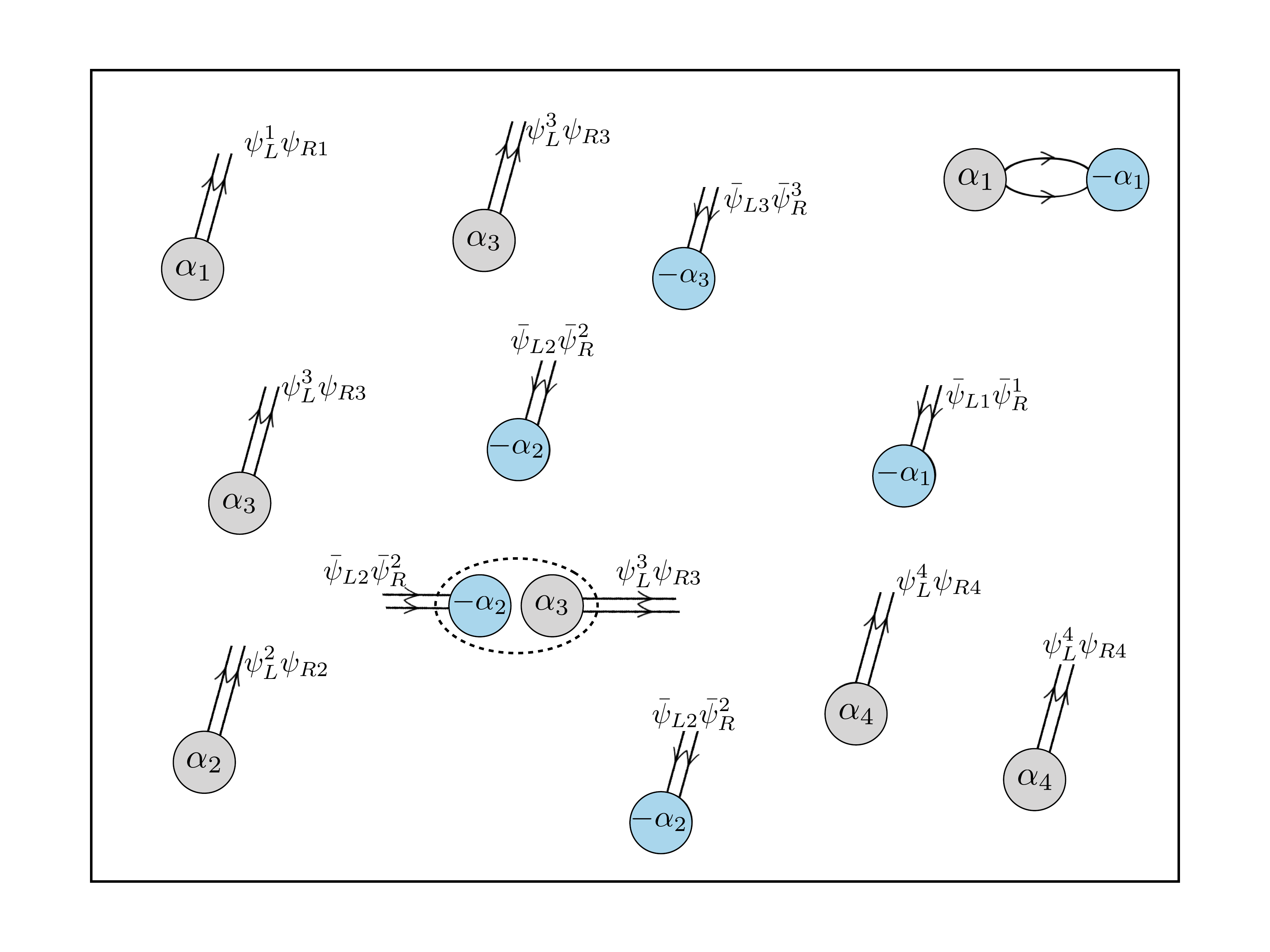}
\caption{Dilute gas of topological configurations in the weak coupling regime obtained by  using graded partition function.  Due to $\Omega_F^0$ background, each species of monopole-instanton has exactly two fermionic zero modes.  Naively, each monopole-event violates the maximal abelian chiral symmetry $U(1)^{N_f-1}$. However, this is not the case because 
of the  intertwining of chiral symmetry with topological shift symmetry, i.e. gauge fluctuations carry a chiral charge. Once a chiral order parameter  acquires a   vacuum expectation value  $\langle {\rm VAC}|  e^{  -  \alpha_i  \cdot z }  | {\rm VAC} \rangle  = e^{-S_0} e^{i \delta_i}$, the vacuum looks as if  it is populated by    $ ( \psi_{Ri} \psi_{L}^{i})$ pairs. In a given vacuum,   $ \langle \psi_{Ri} \psi_{L}^{j}\rangle = \delta_i^j \Lambda^3$ where $\Lambda$ is strong scale.  
 This is the chiral symmetry breaking vacuum of QCD. Note that this is a mechanism taking place at weak coupling.  Historically, it  is  believed that 
the chiral symmetry breaking in QCD-like theories is inherently a strong coupling phenomenon and incalculable. 
 }
\label{fig:mongas}
\end{center}
\end{figure}

\vspace{3mm}
\noindent
$\bf m_\lambda>0, m_\psi =0 :$ Once a  mass term for adjoint fermion is added, the theory resembles more closely to the flavor-limit of QCD. In this case, one looses the $U(1)_{A_\lambda}$  to begin with and ABJ anomaly reduces  $U(1)_{A_\psi}$ down to $\Z_{2N_f}$.    With the $\Omega_F^0$-twisted boundary condition, 
the continuous global symmetry of the  QCD(F)  is: 
 \begin{align}
{\bf G}_{\rm max-ab}={U(1)^{N_f-1}_L  \times U(1)^{N_f-1}_R \times U(1)_V \times  \mathbb{Z}_{2N_f}\over \mathbb{Z}_{N_c}\times (\mathbb{Z}_{N_f})_L\times (\mathbb{Z}_{N_f})_R\times (\mathbb{Z}_2)_\psi}.
\label{max-ab-2}
\end{align}
The action of the  axial chiral symmetry  on fundamental  fermion bilinears is given in \eqref{transform}. Again,  the invariance of the monopole operator leads to intertwining of topological shift symmetry with the $[U(1)_A]^{N_f-1}$ symmetry, and the gauge fluctuations acquire a chiral charge as described after \eqref{transform}.

\vspace{3mm}
In QCD(F/adj) as well as QCD(F) obtained by decoupling adjoint fermion (while keeping the CFC-symmetry intact),  the chiral symmetry breaking occurs at weak coupling. To see this and its consequences, note that 
the proliferation of the monopoles   induces the  terms in the effective long-distance Lagrangian: 
\begin{align}
\label{monopoles-Lag}
{\cal L}^{m} \supset  - \sum_{i=1}^{N_f=N_c} 
 e^{   \alpha_i \cdot z} 
   ( \psi_{Ri} \psi_{L}^{i}) 
+ {\rm h.c.} 
\end{align}

\vspace{3mm}
Similar to QCD(F/adj), $[U(1)_A]^{N_c-1} $ intertwining with the topological shift symmetry forbids a mass generation for the 
dual photon field $\sigma$.  \eqref{monopoles-Lag} is responsible for the transmutation of the chiral charge of fermion bilinear into gauge fluctuations $\sigma$. 
The spontaneous symmetry breaking occurs by the condensation of the pure flux part of the monopole operator 
\eqref{flux-condensation}, and 
 the NG-field is valued in the maximal torus ${\bf T}^{N_c-1}$.  In this case, since there are only two fermion zero mode per 
 monopole instanton,  
the   chiral symmetry breaking  yields  a chirally non-invariant constituent mass for the microscopic quarks.  Indeed, setting the vev for the flux part of the monopole operator \eqref{flux-condensation}, 
 we find: 
\begin{align}
\label{monopoles-Lag-2}
{\cal L}^{m}  
&\sim  -\sum_{i=1}^{N_f=N_c}  e^{-S_0}  e^{ i \delta_i}     (\psi_{L,i} \psi_{R,i}) 
+ {\rm h.c.} 
\end{align}
 All fermion species acquire a chirally non-invariant dynamical mass apart from the real mass which appears due to combined effect of gauge holonomy $\Omega$ and flavor holonomy $\Omega_F$.  
  On small $\R^3 \times S^1$, this provides a reliable mechanism of the chiral symmetry breaking and 
 a  mechanism to induce the constituent quark mass for fermions.

\vspace{3mm}
We can combine the pure flux part of the monopole operators into a matrix field, which can be interpreted as the {\it chiral field } of the chiral Lagrangian: 
 \begin{align}
 \Sigma(x) =  \begin{bmatrix}
  e^{i \alpha_1\cdot  \sigma} & 0 &&  \\
  0&e^{i \alpha_2 \cdot  \sigma} &  &   \\
 &  &  \ddots &     \\
&&  & e^{i \alpha_{N_f}\cdot  \sigma} \\
  \end{bmatrix}
  \end{align}
and the effective field theory at arbitrary long distances takes the form: 
\begin{align}
S= \int_{\R^3 \times S^1}  \frac{f_\pi^2}{4}  \tr |\partial_{\mu} \Sigma|^2 
\label{chiral-lag1}
\end{align}
which is nothing  but the chiral Lagrangian at small $\R^3  \times S^1$, and  $\Sigma \in {\bf T}^{N_f-1}$, the maximal torus. 
Note that the effect of quarks is to render $ \Sigma(x)$ chirally charged, by transmuting  the chiral charge of the fermions into dual photons. In this sense, the effect of the fermions are present in EFT, but the fermions themselves 
acquire a real mass in the $\Omega, \Omega_F$ background and   do not appear in long distance EFT.

\vspace{3mm}
Turning on a soft chiral symmetry breaking mass for quarks, $M_\psi$ in the matrix form, lifts the zero modes of the monopole-operators \eqref{mon-op} and induce a mass gap for the system. The long distance EFT is described as 
\begin{align}
S= \int_{\R^3 \times S^1}  \left[  \frac{f_\pi^2}{4} \tr |\partial_{\mu} \Sigma|^2  - c  \tr ( M_\psi ^{\dagger} \Sigma   + {\rm h.c.})
\right]
\label{chiral-lag2}
\end{align}
This is again nothing but chiral Lagrangian with a soft mass term for pion fields. 
At this stage, our original promise is fully realized and we have three comments. 
\begin{itemize}
\item { Recall that in  $ {\cal N}=1$  SYM and many other SQCD theories, 
 one can get  access to the ground state structure of the theory on $\R^4$ via  supersymmetry preserving $\R^3 \times S^1$ compactifications   through graded partition function  $\tr (-1)^F  e^{-\beta H}$ \cite{Seiberg:1996nz,Katz:1996th,Aharony:1997bx, Davies:1999uw, Davies:2000nw}.  The merit of the graded partition function (supersymmetric index)  is  that the state sum  is  not contaminated by higher states even at arbitrarily small $\beta $ thanks to Bose-Fermi symmetry. Supersymmetry  guarantees the absence of phase transitions.   In  QCD(F/adj) with $m_\lambda \gg 
\Lambda$ for which the infrared physics is  same as QCD(F),    we showed that one can determine the ground state structure of the theory on $\R^4$ via  $\R^3 \times S^1$   compactification through graded partition function $ \tr \Big[ e^{-\beta H} (-1)^F e^{i \epsilon_0 Q_0}   \prod_{a=1}^{N_f} e^{i \epsilon_a Q_a}  \Big] $. This  provides  a realization of the adiabatic continuity idea in QCD(F).  \footnote{
Adiabatic continuity between weak coupling confined phase and strong coupling confined phase cannot currently be proven analytically. However, it can be shown numerically in some non-trivial QFTs. These include  lattice  simulations  of $\N=1$ SYM \cite{Bergner:2018unx}  
and deformed Yang--Mills theory \cite{Bonati:2018rfg}. In QCD(adj) with massive fermions with periodic boundary conditions, one can show the existence of the   weak coupling confined phase at small $\R^3 \times S^1$  in QCD(adj)   \cite{Cossu:2009sq} with an intermediate deconfined regime as $\beta$ is increased. }}

\item {Despite the lack of supersymmetry in our QFT, we were able to generate powerful enough global symmetry induced cancellations in the graded state sum over the Hilbert space,  such that ${\rm Distill}[\cal H]$ does not seem to lead to any phase transition as $\beta $ is dialed.  More precisely, at small-$\beta $, the long distance  EFT is  the chiral Lagrangian \eqref{chiral-lag1}, which is nothing but  dimensional reduction of the QCD-chiral lagrangian described at large-$\beta $ in the   $\Omega_F^0$  flavor holonomy background, providing a  Hilbert space  interpretation of our earlier result \cite{Cherman:2016hcd}. }

\item{
 We will prove in the next section that the mixed anomalies that control the ground state structure of the theory on   $\R^4$  
\eqref{poly}  and on  $\R^3 \times S^1$  (with  $\Omega_F^0$ background)  given in \eqref{compano} are the same.  
 } 
 
 \end{itemize}

\vspace{3mm}
\noindent
{\bf Fermion bi-linear condensate in  statistical interpretation:} 
At leading order in semi-classics,  the vacuum is  a dilute gas of $N_c$ types of monopole-instantons  
 each with complex fugacity 
\begin{align}
\zeta_i =  e^{-S_0} e^{i \delta_i}, i= 1, \ldots, N_c
\end{align}
The magnitude of the fugacity is the density of monopole-instanton events. In statistical interpretation, 
 the vacuum expectation value $ \langle   \psi_{Ri} \psi_{L}^{ j}   \rangle$  is the average of the condensate over the space
$ V_{\R^3 \times S^1} $. 
  This receives contributions from the lumps of monopole events. Therefore, the chiral condensate will be proportional to $e^{-S_0}$.

\vspace{3mm}
In the large $m_\lambda $ limit  while staying center-symmetric    \eqref{dnd}, the theory is analytic continuation of CFC symmetric   QCD(F) to small-$\beta $ regime. 
In this case, $m_\lambda$ can be made large-enough so that it disappears from the renormalization group $\beta$-function. Adjoint fermion $\lambda$ does not enter to the RG-flow of the coupling constant at  energy  scales smaller  than 
$ \frac{(g^2N_c)^{1/2} }{ \pi \beta N_c} $.  However, $\lambda$ still    enters to the holonomy potential which is valid for 
$ \beta \Lambda \lesssim 1$, and for the purpose of holonomy potential,  $m_\lambda$ is light provided 
 $(\beta m_\lambda N_c) \lesssim  1$. This is a  strange state of affairs described around   \eqref{dnd}, which exhibits decoupling from IR-physics  and non-decoupling from holonomy potential. 

 \vspace{3mm} 
 There are two  striking similarities  between $N_f=N_c$ QCD(F), and $\N=1$ SYM, which allows us to determine the chiral condensate in $N_f=N_c$ QCD(F). First, 
there is a kinematic accident that takes place between QCD(F) with $N_f=N_c$, and $\N=1$ SYM concerning renormalization group $\beta$-function, explained below. Furthermore,  in   the 
$\Omega_F^0$-background flavor holonomy,  the fermionic zero mode structure of the monopole-instantons is identical in 
these two theories, see \eqref{index-2}. Each monopole has exactly two-fermionic zero mode,    $( \psi_{Ri} \psi_{\beta }^{i})$  in QCD(F) and $(\alpha_i \cdot \lambda)^2$ in $\N=1$ SYM.   
 
  \vspace{3mm} 
 Concerning the renormalization group  $\beta$-functions,   the universal two-loop $\beta$ function coefficients are: 
 \begin{align}
&{\rm QCD(F)}, \; \; N_f= N_c:  &&  \qquad \beta_0= 3N_c, \qquad  \beta_1= 7N_c^2 \cr
& \N=1 \;   {\rm SYM}:  &&  \qquad \beta_0= 3N_c,  \qquad \beta_1= 6N_c^2 
 \end{align}
 The two-loop strong scale of a gauge theory is given by   $\Lambda= \mu e^{-\frac{8 \pi^2 }{g^2 \beta_0} }(g^2)^{-\frac{\beta_1}{2 \beta_0^2}  }$, which implies that 
  \begin{align}
&{\rm QCD(F)}, \; \; N_f= N_c:  &&  \qquad    \Lambda_{\rm QCD}= \mu   \left( \frac{ 16 \pi^2}{3 N_c g^2} \right)^{7/18} e^{-\frac{8 \pi^2 }{3g^2N_c } } 
 \cr
& \N=1 \;   {\rm SYM}:  &&  \qquad   \Lambda_{\rm SYM} = \mu  \left( \frac{ 16 \pi^2}{3 N_c g^2} \right)^{1/3}  e^{-\frac{8 \pi^2 }{3g^2 N_c} }  
\label{strong}
 \end{align}
In other words, the one-loop strong scales of the two theories are precisely the same, but there is a small difference at  
two-loop order.

 \vspace{3mm} 
In the center-symmetric holonomy background, the   flux-part of the monopole operator condense as \eqref{flux-condensation},  spontaneously breaking chiral symmetry to its vector-like subgroup: 
In the vacuum of the theory,  we can evaluate the fermion bilinear condensate  $\langle   \psi_{Ri} \psi_{L}^{ j}  \rangle$ as follows. First, since the fermi zero mode structure of the monopole-zero modes is diagonal in flavor,  and the zero modes 
can absorb  the bilinear, the non-zero vev will be proportional to $\delta^{j}_{i}$.  The condensate will receive contributions from monopole-instanton lumps and hence, its  value  is proportional to the density of monopole-instantons $e^{-S_0}$.  
Therefore, 
the chiral condensate in  QCD(F) with $N_f=N_c$ is given by 
\begin{align} 
\langle   \psi_{L}^{i} \psi_{R j}  \rangle    \sim \delta^i_j  \beta ^{-3}  e^{-S_0}  e^{i \delta_i}= \delta^i_j   \Lambda^3 e^{i \delta_i} 
\label{chiral-cond}
\end{align}
Similar to $\N=1$ SYM, 
 we expect this quantity to agree between the small and large-$\beta $ theory. 

 \vspace{3mm} 
In other words, we claim that the mechanism of chiral symmetry breaking, and  the way  fermion bilinear condensate forms 
 is same in $\N=1$ SYM and QCD(F) with $\Omega_F^0$-twist.  In fact, the density of monopole-instantons are also equal in the vacuum of these two theories on $\R^3 \times S^1$. 
Despite this almost identical nature of the two,    
the fluctuations around the respective  condensates are   quite different.  
The fluctuations in QCD(F)  are described by gapless NG modes, and the fluctuations in  $\N=1$ SYM  are gapped. In other words, the magnetic bions   in  
$\N=1$ do induce a mass gap for dual photon, while they do  not generate any potential or   mass gap  for gauge fluctuations in QCD(F) as described around \eqref{m-bions}

 \vspace{3mm} 
The  maximal abelian subgroup  \eqref{max-ab-2} is an exact symmetry at any radius, 
and within the semi-classical description in the weak coupling regime, we proved that  it is broken down to its vector-like subgroup. 
 \begin{align}
[U(1)_V \times U(1)_A]^{N_f-1} \times U(1)_V   \longrightarrow  [U(1)_V]^{N_f}
 \label{MAG-break}
 \end{align}
 This is indeed the expected behavior at large-$\beta $ that comes out from chiral Lagrangian in the background of the flavor holonomy $\Omega_F^0$.  Therefore, it is reasonable to expect that the weak coupling and strong coupling regimes are continuously connected in the sense of realization of global symmetries, i.e, chiral and  CFC symmetries.

  \subsection{$\N=1$ SYM:   What breaks chiral symmetry on $\R^3 \times S^1$?}
\label{sec:what}
${\bf m_\lambda=0, m_\psi} \in (0, \infty] :$ Once a  mass term for fundamental  fermion is added, the theory resembles more closely to the $\N=1$ SYM. In the decoupling limit $ m_\psi \rightarrow  \infty$,  it reduces  to the supersymmetric theory. 
 In this case, one looses  all the continuous  axial chiral symmetries,  and ABJ anomaly reduces  $U(1)_{A_\lambda}$ down to $\Z_{2N_c}$. 
 \begin{align}
 {\bf G} = \Z_{2N_c},  
 \label{DCS}
 \end{align}
This is  also manifest in the  instanton amplitude
$\mathcal{I}_{4d}   \sim   e^{- \frac{8 \pi^2}{g^2}}  (\tr \lambda \lambda)^{N_c}$ which is only invariant  $\Z_{2N_c}$ discrete subgroup of $U(1)_{A_\lambda}$. 
In $\N=1$ SYM,  on $\mathbb R^4$,  it is believed that the discrete $\chi$S is broken down to $\Z_2$ 
 by fermion bilinear condensate $ \langle \tr \lambda \lambda \rangle $
 and there are $N_c$ vacua.  Indeed, the analysis on  $\R^3 \times S^1$ demonstrates analytically  the discrete chiral symmetry breaking  \cite{Davies:1999uw, Davies:2000nw}.  
 Recently, similar techniques are also used to understand  broader class of vector-like and chiral supersymmetric gauge theories  on $\R^3 \times S^1$   \cite{  Poppitz:2009kz, Poppitz:2011wy, Poppitz:2013zqa, Anber:2017ezt,   Cherman:2016jtu, Anber:2015kea,  Csaki:2014cwa,Amariti:2015kha, Csaki:2017mik,Csaki:2017cqm, Lee:2019hzq}.

   \vspace{3mm}
 Although the  chiral symmetry breaking in  $\N=1$ SYM is well-known in literature both on $\R^4$ as well as 
 $\R^3 \times S^1$, there is a  subtle  issue that we would like to highlight concerning the mechanism of chiral symmetry breaking, that is not  
 clear in earlier works.

  \vspace{3mm}
On  $\R^3 \times S^1$, 
 since each monopole operator  (\eqref{mon-op} third line)  has two-fermi zero modes and $\Z_{2N_c}$  is anomaly free, there is again  intertwining of the discrete chiral symmetry with a discrete subgroup of topological shift symmetry:
\begin{align}
 \Z_{2N_c}:  \left\{ \begin{array}{ll}
  (\alpha_i \cdot  \lambda)^2 &\rightarrow e^{i \frac{2 \pi k }{N_c}}    (\alpha_i \cdot  \lambda)^2 ,  \cr 
 e^{ \alpha_i \cdot z}  &\rightarrow e^{ - i \frac{2 \pi k }{N_c}}   e^{ \alpha_i \cdot z} 
\end{array} \right.
    \label{shift-2}
\end{align}
Therefore, on  $\R^3 \times S^1$,  as it is the case in QCD(F), QCD(F/adj), there are two-types of order parameters for the chiral symmetry:  the magnetic  flux part of monopole operator  $e^{ \alpha_i \cdot z}$   and  fermion bilinear  $\tr \lambda \lambda$.

  \vspace{3mm}
Standard interpretation in literature is that  $\chi$S is broken on small $\R^3 \times S^1$  due to chiral condensate $\tr \lambda \lambda$ acquiring a vev.  This story is   actually more subtle than often stated.  On $\R^4$, 
    $\tr \lambda \lambda$  can indeed acquire a vev and break chiral symmetry without breaking supersymmetry. This is  because it is the lowest component of chiral multiplet,  
  \begin{align}
     \epsilon^{\alpha \beta}  \tr W_{\alpha} W_\beta =   \tr \lambda \lambda + \theta \ldots
\end{align}
 hence it cannot be expressed  as 
\begin{align}
    \tr \lambda \lambda   \neq \{Q,  \cdot \}  \qquad (\rm microscopic \;  theory)
\end{align}
  Therefore,     $\tr \lambda \lambda$ can  acquire  a vev without   
clashing  with supersymmetry \cite{ Witten:1981nf}. 

  \vspace{3mm}
 On     $\R^3 \times S^1$,  Cartan components of $\lambda_{\alpha}$ are  {\it not} lowest component of the supersymmetric multiplet  in long-distance EFT,  rather    3d $\N=2$ supersymmetric multiplets that enters to EFT can be written as: 
\begin{align}
 Z= z+ \theta \lambda+ \ldots,   
 \label{multiplet}
 \end{align} 
 where the lowest component of the multiplet is  $z$, given in \eqref{z-def}. However, in 
supersymmetric EFT with  the scalar multiplet,   
\begin{align}
 z   \neq \{Q,  \cdot \}  \qquad  \lambda   = \{Q,  z \}  \qquad {\rm EFT \; on } \; \R^3 \times S^1
\end{align}
Therefore, in the EFT based on monopoles,  it is in fact  only $z$  which can obtain a 
vacuum expectation value  without  breaking supersymmetry! 
%
 The essence of this argument, without this particular application,    is  explained  in Ref.~\cite{ Witten:1981nf}. 
  Hence,     $  e^{   \alpha_i \cdot  z  } $  can acquire a vev without breaking supersymmetry, and indeed, this is what is dictated by the  affine Toda superpotential \eqref{Superpot},  as described below.  
 In the chirally broken vacuum   
 \begin{align}
 \langle  {\rm VAC}_k |e^{   \alpha_i \cdot  z  }   | {\rm VAC}_k \rangle = e^{-\frac{8 \pi^2}{g^2 N_c} + i \frac{\theta}{N_c}} e^{i \frac{2\pi k}{N_c}}, \qquad k=1, \ldots, N_c
\label{condensate}
 \end{align}  
 and these are the super-selection sectors for EFT on $\R^3 \times S^1$. 
 
   \vspace{3mm}
 At this level, discrete chiral symmetry is {\it already} broken by the condensation of magnetic flux operator \eqref{condensate}.  This is also the interpretation that is consistent with QCD(F) and QCD(F/adj) in the non-supersymmetric cases.
 We can also see the same effect by studying the EFT in the small  $\R^3 \times S^1$ theory.

\vspace{3mm}
In  $\N=1$ SYM theory, since each monopole-instanton  has two fermi zero modes, it can induce a superpotential  $\cal W$  \cite{Affleck:1982as, Seiberg:1996nz,  Katz:1996th, Davies:1999uw, Davies:2000nw},   which can be expressed in terms of 
 3d $\N=2$ supersymmetric multiplet $Z$ given in \eqref{multiplet}.  The  super-potential  is given by 
\begin{align}  
 { \cal W}_{\R^3 \times S^1}  (Z)= \sum_{i=1}^{N_c-1} e^{\alpha_i\cdot Z} + e^{ 2\pi i \tau}  e^{\alpha_N \cdot Z},  
 \label{Superpot}
 \qquad  
\end{align}
where $\tau= \frac{\theta}{2\pi} + i \frac{4\pi}{g^2}$   and $e^{ 2\pi i \tau}  = e^{-\frac{8 \pi^2}{g^2} + i \theta}$. The 
long-distance EFT on $\R^3 \times S^1$ is given by: 
\begin{align} 
\label{EFT}
{\cal L} = \half |\partial_{\mu} z|^2  +i  \bar \lambda  \sigma_\mu \partial_\mu \lambda+  \half   \sum_{i} \left|\frac{\partial W}{\partial z_i}\right|^2  +  \sum_{i,j}  \left(\frac{\partial^2 W}{\partial z_i \partial z_j} \lambda_i \lambda_j   + {\rm c.c.}  \right)
\end{align}

\vspace{3mm}
The supersymmetric vacua  of the theory are  located at the minimum of the bosonic potential  at which 
${\partial W}/{\partial z_i} =0$.  
There are  $N_c$ isolated minima, at which  
\begin{align}
\label{monopoles-condense}
  \langle e^{   \alpha_i \cdot  z  } \rangle =  e^{-\frac{8 \pi^2}{g^2 N_c} + i \frac{\theta}{N_c}} e^{i \frac{2\pi k}{N_c}}, \qquad k=1, \ldots, N_c
  \end{align}
  Again, EFT tells us that,   at this stage,  the discrete chiral symmetry  is already broken   because  the magnetic  flux part of the monopole operator acquires a vev.    
 It is an independent  question if the fermion bi-linear acquires a vev in these vacua or not.    For example, in the QCD(F/adj) with $N_f=N_c$, four-fermi operator acquires a vev but not the fermion bilinears as described around  \eqref{four-fermi}.  In $ \N=1$  SYM, the fermion bilinear acquires a vev in a
given vacuum 
 \begin{align}
   \langle  {\rm VAC}_k | \frac{1}{N_c}   \tr \lambda \lambda   | {\rm VAC}_k \rangle = 
   \Lambda^3 e^{i \frac{2\pi k + \theta}{N_c}} 
\label{chiral-cond-2} 
\end{align}

\subsection{Why  the quark condensate can  form at weak coupling?} 

Two standard beliefs    concerning  how quark condensate forms in QCD are  following:  a) The vacuum of space must be  populated by quark-anti-quark pairs.  b)  Quark condensate  in QCD cannot  occur in weakly coupled theory.   
See Tong's lectures on QFT for an up to date   review of chiral symmetry breaking.\footnote{ http://www.damtp.cam.ac.uk/user/tong/gaugetheory/5chisb.pdf }.  
The relation between these beliefs and our analytic construction is as follows.\footnote{Another analytic approach to chiral symmetry breaking  appears 
in \cite{Murayama:2021xfj} by employing  anomaly mediated supersymmetry breaking of SQCD.   }

   \vspace{3mm}
The graded partition function demonstrates that the vacuum of the theory is  indeed populated by quark-anti-quark pairs in the small $\R^3 \times S^1$ regime, see  Fig.~\ref{fig:mongas} in the calculable regime.
Yet, quark-antiquark  pairs are all chirally charged, either under continuous or discrete  (non-anomalous) chiral symmetry. An aspect that did not appear at 
all in the older literature in QCD is that the chiral charges of fermions can actually be absorbed  by gauge fluctuations, i.e, magnetic flux part of monopole operators    is capable to soak up the   chiral charge, hence, the combination is chirally neutral.   Therefore, gauge fluctuations also end up transforming under chiral symmetry. 
 It is the condensation of the monopole-flux operators that induces chiral symmetry breaking mass terms  in QCD as well as $\N=1$ SYM.    

   \vspace{3mm}
The statement that the formation of quark condensate cannot occur at weak coupling is not correct, as we have seen in detail in this work.
  The standard (and incorrect)  lore behind this is that   the condensate itself is proportional to $\Lambda_{\rm QCD}^3$, 
 where  $\Lambda_{\rm QCD}$ is the strong length scale of QCD.  Hence  in QCD,   chiral symmetry breaking must be a strong coupling phenomenon. 
and it is impossible to study the formation of chiral condensate at weak coupling.

   \vspace{3mm}
On $\R^3 \times S^1$, at small-$\beta$, the theory is indeed weakly coupled per asymptotic freedom at the scale of compactification, where 
CFC symmetry is unbroken. Yet, the density of monopoles $\zeta_i = {\cal N}_i/V_{\R^3 \times S^1}$ of type $i$, which sources chiral condensate  is: 
 \begin{align}
\zeta_i = \beta^{-3} e^{-S_0} \equiv \beta^{-3} e^{- \frac{ 8\pi^2}{ (g^2(\beta)N_c)}} 
 \end{align}
  which is non-perturbative due to appearance of $e^{-8\pi^2/(g^2 N_c)}$ factor.   It is one of those lucky moments that one realizes 
  \begin{align}
  \beta^{-3} e^{- \frac{8\pi^2}{ (g^2(\beta)N_c)}}= \Lambda_{\rm QCD}^3
  \end{align}
  using the one-loop dimensional transmutation. This is due to the fact that the one-loop beta function of  $N_f=N_c$   QCD(F) is  $\beta_0= 3N$, just like    $\N=1$ SYM theory. 
   This is nothing but the  non-perturbative density of quark-anti-quark pairs in vacuum!  In other words, we learn that the  quark condensate  of order 
   $\Lambda_{\rm QCD}^3$ can arise from weak coupling semi-classical construction naturally.

\section{Persistent mixed anomalies vs. quantum distillations}

The  idea of quantum distillation of Hilbert space, and corresponding   path integrals in the flavor holonomy background 
is  related in interesting ways  to  mixed  persistent anomalies (when the latter exists).   However, they are certainly not the same concepts.  Roughly, quantum distillation describe (mostly imperfect) cancellations within a given superselection sector and persistent mixed anomalies are tied with exact cancellations among superselection sectors.

\vspace{3mm}
  It is recently understood that mixed anomalies involving discrete 1-form symmetries   continues to persist upon compactification \cite{Gaiotto:2017yup,Komargodski:2017keh,Gaiotto:2017tne}.   
 Even if the theory does not have a 1-form symmetry, as it is the case in 
QCD(F/adj)  and QCD(F),  it is possible to prove the following remarkable result: 
\cite{Tanizaki:2017bam,Kikuchi:2017pcp,Tanizaki:2017qhf, Shimizu:2017asf,Cherman:2017dwt}. 
\begin{itemize} 
\item[]
{\bf Persistent mixed anomaly in the absence of 1-form symmetry:}  Assume the theory has only  0-form symmetry (but not a 1-form symmetry)
 $G_1 \times G_2$ where $G_1= \widetilde G_1/\Gamma$ and gauging $ \widetilde G_1 $ turns $\Gamma$ into a 1-form symmetry. Then, a triple-mixed anomaly  that  is present on $\R^4$   
 persists upon compactification on $\R^3 \times S^1$, and imposes equally powerful constraints on  long distance physics of compactified theory. 
 \end{itemize}  
Below, we will take advantage of this result.

\subsection{Mixed Anomalies in QCD(F) and   $SU(N_c) \times SU(N_f)$ quiver  theory on $\R^4$}

\begin{figure}[t]
\begin{center}
\vspace{-1.5cm}
\includegraphics[width = 1.0 \textwidth]{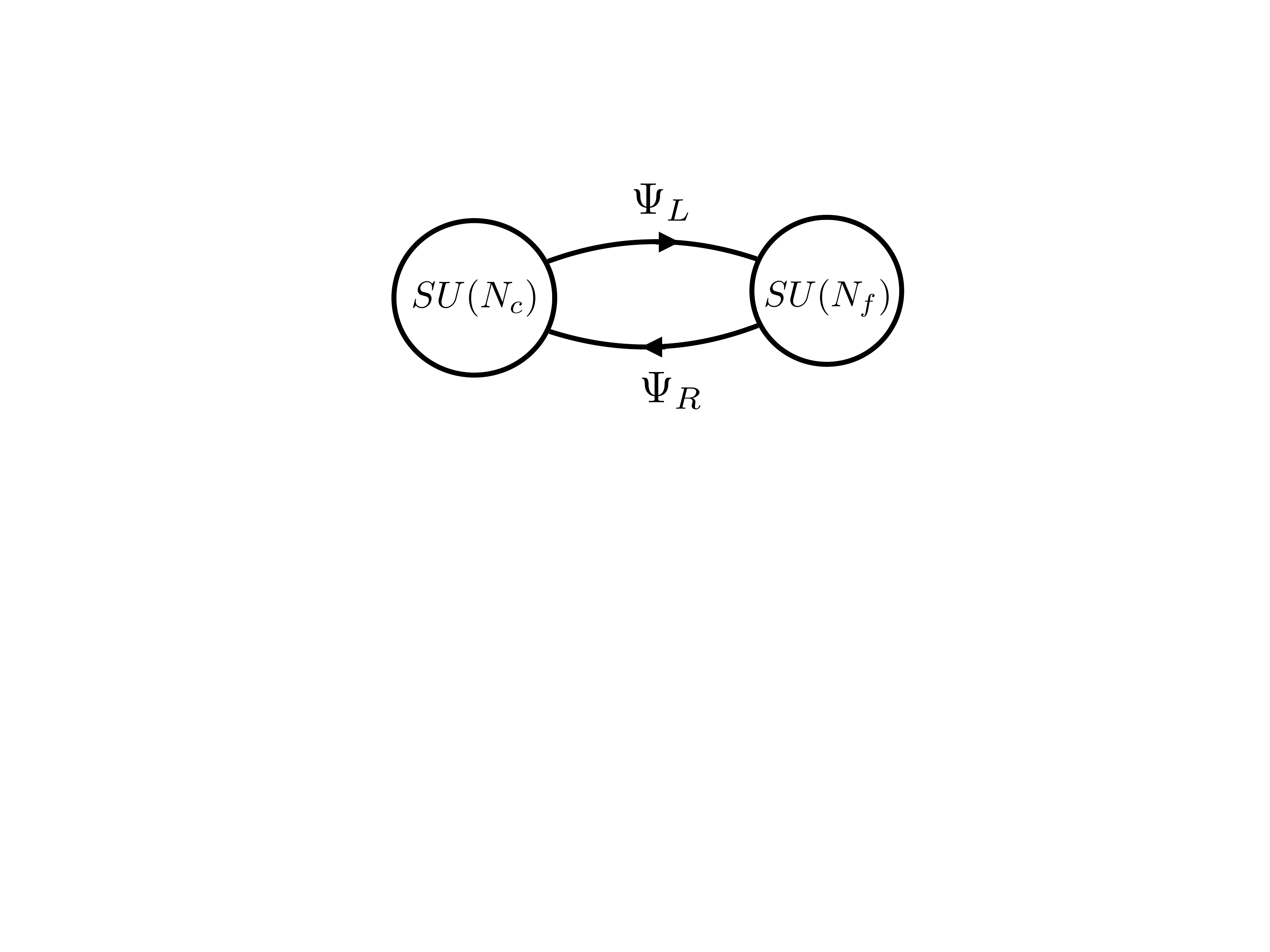}
\vspace{-7.0cm}
\caption{Upon gauging $SU(N_f)_V$, QCD(F)  becomes a two site quiver theory (QCD(BF)) with 1-Dirac fermion, $\Z_{2 {\rm gcd}(N_f, N_c)} $ chiral and $\Z_{ {\rm gcd}(N_f, N_c)}^{[1]} $ 1-form  symmetry. We must  gauge 
 $\Z_{ {\rm gcd}(N_f, N_c)}^{[1]} $ 1-form symmetry in order to  turn on a background or gauge the faithful symmetry of the microscopic theory. 
 There are intimate relations between the anomalies of the original  multi-flavor QCD(F) theory and  auxiliary  QCD(BF) theory. 
 }
\label{fig:quiver}
\end{center}
\end{figure}

Consider turning on mass for adjoint fermion $m_\lambda >0$  in QCD(F/adj). If desired, the mass can be taken large and the theory reduces to QCD(F). 
The faithful vector-like symmetry of the theory  has a subgroup:
\begin{align}
SU(N_f)_V/ \Z_{{\rm gcd}(N_f, N_c)} 
\label{faitful}
\end{align}
where  $ \Z_{{\rm gcd}(N_f, N_c)}$ part of the $SU(N_f)_V$ is a gauge redundancy and is not a genuine symmetry. Therefore, it is modded out.   Under an $ SU(N_c) \times SU(N_f)_V $ gauge $\times$ global   transformation, the quarks transform as 
\begin{align}
\psi \rightarrow  g_c(x) \psi g_f^{\dagger}
\label{global}
\end{align}
 and hence, a global  flavor transformation residing in $\Z_{{\rm gcd}(N_f, N_c)} \subset \Z_{N_f} \subset SU(N_f)$    can be undone by a gauge transformation 
 $\Z_{{\rm gcd}(N_f, N_c)} \subset   \Z_{N_c} \subset SU(N_c)$. We will use this symmetry  to describe a mixed anomaly of the theory on $\R^4$ and $\R^3 \times S^1$ following and  slightly generalizing  
 \cite{Tanizaki:2017qhf}.

\vspace{3mm}
Recall that QCD(F) has a  $\Z_{2N_f}$ chiral symmetry,  as can be seen 
from the non-invariance of the fermion  integration measure. Under a $U(1)_A$ action,  the measure transforms as: 
\begin{align}
d\mu_{\rm fermion}^{ \rm QCD(F)}  \rightarrow e^{ i \alpha 2N_f    \times  {1 \over 8\pi^2} \int 
\tr( F\wedge F) } \;  d\mu_{\rm fermion}^{ \rm QCD(F)}    
\end{align} 
Therefore, this is a symmetry only when $\alpha= \frac{2 \pi }{2 N_F}k $  and  $U(1)_A$ axial is  explicitly broken down to  
$\Z_{2N_f}$. 
  The  $\Z_{N_f}$ subgroup of      $\Z_{2N_f}$   is actually a part of the continuous chiral symmetry $\bf G$.
   Consider $ \psi_{Ra} \psi_{L}^{b}  $ which rotates under a $\Z_{2N_f}$ discrete chiral rotation into $ e^{2 \pi i  k /N_f}  \psi_{Ra} \psi_{L}^{b}  $.  The same discrete transformation  can be achieved  with a continuous rotation by an $SU(N_f)_L$ matrix or a transformation in its  maximal torus $(U(1)_L)^{N_f-1}$:
$U_L(\delta)= {\rm diag} \left( e^{ i \delta}, \ldots,   e^{ i \delta} , e^{ -i (N_f-1) \delta} \right),  \; $ by continuously varying 
$\delta \in  [0,  2 \pi i  k /N_f]$.  
Therefore, the $\Z_{2N_f}$ transformation   is part of a continuous symmetry $SU(N_f)_A$ or its maximal torus, i.e., 
\begin{align}
\Z_{N_f} \subset   {\bf G}_{\rm max-ab} \subset   {\bf G}_{\rm non-ab} =SU(N_f)_A
\end{align}
Therefore, in QCD(F),  any order parameter 
that is charged under discrete chiral symmetry  (which is not an independent symmetry)  will automatically be charged under continuous chiral symmetry, and the breaking of the former implies the breaking of the latter.\footnote{\label{Stern}Note that reverse  statement is not true. Any order parameter that  is charged under the 
 continuous chiral symmetry  is not necessarily charged under discrete chiral symmetry. An example is four fermi operator  
 that appears in the discussion of Stern phases \cite{Tanizaki:2018wtg}.   More precisely, 
if a  Euclidean QFT on 
$\R^d$  is compactified to  $\R^{d-1} \times S^1$ 
where $ d-1 >2$, then, 
discrete chiral symmetry  (which is a subgroup of continuous chiral
symmetry) breaking implies continuous chiral symmetry breaking.  However, 
in a theory compactified to 1+0 QFT (quantum mechanics)  or 1+1 QFT, since  it is not possible to break continuous global symmetries in $d \leq 2$ dimensions,  the theory  may have  multiple minima due to mixed anomalies involving discrete chiral symmetry without breaking continuous global symmetry.  }

\vspace{3mm}
Below, 
we  first describe the mixed anomaly between $SU(N_f)_V/ \Z_{{\rm gcd}(N_f, N_c)} $ and $\Z_{2N_f}$  on $\R^4$  generalizing   
 Ref.\cite{Tanizaki:2017qhf} to arbitrary $(N_f, N_c)$  for which  $ {\rm gcd}(N_f, N_c) $ is non-trivial  and then, find the conditions under which the  mixed anomaly  persists under compactification. 

\vspace{3mm}
\noindent
{\bf Quiver theory:}
First, gauge  $SU(N_f)_V$ flavor symmetry. We obtain a two-site quiver gauge theory, in which 
$N_f$ fundamental Dirac fermions  $\psi_a$  turn  into {\it one} bi-fundamental  $\Psi$  which transform under  $SU(N_c) \times SU(N_f) $ gauge transformation as  
\begin{align} 
\Psi \rightarrow  g_c(x) \Psi g_f^{\dagger}(x)
\label{local}
\end{align}
This is a local gauged version of \eqref{global}. 
The covariant derivative takes the form 
\begin{align}
D_{\mu} (a, A) \Psi= \partial_{\mu} \Psi + i a_{\mu} \Psi - i \Psi A_{\mu} \qquad a_{\mu} \in SU(N_c), \;\;   A_{\mu} \in SU(N_f)
\end{align} 
  We call this theory with product gauge group structure the  quiver  theory. Such non-supersymmetric quiver theories are examined in the context of large-$N_c$ orbifold equivalence as well as mixed  anomalies, see e.g. \cite{Tong:2002vp, Kovtun:2005kh,  Tanizaki:2017bam, Karasik:2019bxn, Kan:2019rsz}.
  
  \vspace{3mm}
The $SU(N_c) \times SU(N_f) $ quiver gauge theory has an {\it exact} 
 $  \Z_{   {\rm gcd}(N_f, N_c) }^{[1]}$ 1-form center-symmetry  and  $ \Z_{2   {\rm gcd}(N_f, N_c) }$ 0-form chiral symmetry. 
 The gauge redundancy that has been modded out in \eqref{faitful} became an exact 1-form symmetry upon the gauging of $SU(N_f)_V$.  Needless to say, this is also exactly the color-flavor-center (CFC) symmetry  \eqref{CFC}, \eqref{CFC-2}  which is  promoted to  a full 1-form center-symmetry in the quiver theory.  
 The 0-form symmetry   $ \Z_{2   {\rm gcd}(N_f, N_c) }$ lives in the $\Z_{2N_f}$  axial symmetry of QCD(F) and 
 we  must gauge 1-form center-symmetry to determine the anomaly structure. 

\vspace{3mm}
Let us first determine how the $  \Z_{2 {\rm gcd}(N_f, N_c)}$ chiral symmetry arises in the 2-site quiver theory.   Since one-Dirac fermion is  in bi-fundamental of $SU(N_c) \times SU(N_f)$, the 
non-invariance of the fermion  integration measure can now be phrased as: 
\begin{align}
d\mu_{\rm fermion}^{\rm quiver}  \rightarrow \exp \left \{ i \Upsilon  \left[ (2N_f)    \times  {1 \over 8\pi^2} \int 
\tr( F_1 \wedge F_1)   
+    (2N_c)    \times  {1 \over 8\pi^2} \int 
\tr( F_2 \wedge F_2)   \right]   \right \}  \;  d\mu_{\rm fermion}^{\rm quiver} 
\label{ups1}
\end{align} 
where $F_1, F_2$ are gauge field strengths for $SU(N_c) \times SU(N_f)$.     The measure can be rewritten as:
\begin{align}
d\mu_{\rm fermion}^{\rm quiver}  \rightarrow \exp \left \{ i 2 \Upsilon  \left[ N_f  m_1   
+    N_c  m_2      \right]   \right \}  \;  d\mu_{\rm fermion}^{\rm quiver} 
\end{align} 
where $m_1, m_2 \in \Z$ are topological charges in the corresponding gauge group factors.  To find the discrete remnant of $U(1)_A $ chiral symmetry, use  Bezout identity in elementary number theory which states that    there exists $m_1, m_2 \in \Z$ such that 
\begin{align} 
N_f  m_1   
+    N_c  m_2 = {\rm gcd}(N_f, N_c)
\end{align} 
 Therefore, 
\begin{align}
d\mu_{\rm fermion}^{\rm quiver}   \rightarrow \exp \left \{ i 2 \Upsilon     {\rm gcd}(N_f, N_c)  
 \right \}  \;  d\mu_{\rm fermion}^{\rm quiver}  
\end{align} 
and this is a symmetry only when 
\begin{align}
\Upsilon  = \frac{2 \pi }{  2 {\rm gcd}(N_f, N_c)   }k  
\label{ups2}
\end{align}
and the discrete chiral symmetry of the  auxiliary quiver theory is   $ \Z_{2   {\rm gcd}(N_f, N_c) }$. 

\vspace{3mm}
To summarize, in quiver theory,   we have  the global symmetries: 
\begin{align}
{\rm quiver \; theory\;   symmetries:}  \left\{
\begin{array}{ll}  
 \frac{U(1)_V \times   \Z_{2   {\rm gcd}(N_f, N_c)}}{\Z_2},  & \;\; 0-{\rm form }     \cr \cr
    \Z_{   {\rm gcd}(N_f, N_c) }^{[1]},  & \;\;  1-{\rm form,   \; center}  
\end{array}  \right.
\end{align} 

  \vspace{3mm}
\noindent
{\bf Gauging center:}
Now, we can  gauge  $  \Z_{   {\rm gcd}(N_f, N_c) }^{[1]}$  1-form center on $\R^4$  and describe, under what conditions   we loose   parts of $ \Z_{2   {\rm gcd}(N_f, N_c) }$ 0-form chiral symmetry. It is also possible to proceed oppositely, gauge $ \Z_{2   {\rm gcd}(N_f, N_c) }$  0-form symmetry and show that we loose the 1-form  symmetry partially. 
Gauging  $  \Z_{   {\rm gcd}(N_f, N_c) }$  amounts to 
 introducing  a  $(B, C)$  pair where 
$ B$ is a 2-form $U(1)$ gauge field, $C$ is a 1-form gauge field, and the pair must  satisfy   \cite{Gaiotto:2014kfa, Gaiotto:2017yup}
\begin{align}
 {\rm gcd}(N_f, N_c) B + dC=0.
\end{align} 
The constraint obeys the gauge invariance, 
\begin{align} 
 B \rightarrow B - d \lambda, \qquad C \rightarrow C +  {\rm gcd}(N_f, N_c) \lambda 
 \label{trans}
\end{align}
 where $\lambda$ is a 1-form gauge transformation.   We now define the $U(N_c)$ and $U(N_f)$ gauge fields by using  
  a common $U(1)$ 1-form gauge field $C$. 
 Express 
 \begin{align}
 \widetilde a = a+ \frac{1}{ {\rm gcd}(N_f, N_c) } C {\bf 1 }_{N_c},  \cr 
   \widetilde A = A+ \frac{1}{ {\rm gcd}(N_f, N_c) } C {\bf 1}_{N_f}, 
 \end{align}
  along with their respective gauge field strengths 
   \begin{align}
F_c' = d  \widetilde a  + i  \widetilde a  \wedge  \widetilde a  , \qquad 
F_f'= d  \widetilde A  + i  \widetilde A  \wedge  \widetilde A  
 \end{align}
 Clearly, $F_c'$  and $F_f'$ are not gauge invariant under the 1-form gauge transformation because of their common $C$ components,  $F_c' \rightarrow  F_c'  + d\lambda$ and $F_f' \rightarrow  F_f'  + d\lambda$. But the combinations 
\begin{align} 
 F_c'  + B {\bf 1}_{N_c}, \qquad F_f'  + B {\bf 1}_{N_f} 
 \end{align}
 are gauge-invariant under one-form gauge transformations \eqref{trans}.

  \vspace{3mm}
Under an    $ h \in  \Z_{2   {\rm gcd}(N_f, N_c) }$ discrete  chiral transformation, 
(which used to be an invariance of the theory   before  the  $ \Z_{  {\rm gcd}(N_f, N_c) }^{[1]} $  1-form center-symmetry was gauged),   using \eqref{ups1} and \eqref{ups2}, one observes that 
the fermion measure  under the background 2-form field $B$ fields transforms into 
 \begin{align}
d\mu_{\rm fermion}^{\rm quiver}  \rightarrow   
 \exp  i  &   \left[ \frac{ (2N_f) }{2 {\rm gcd}(N_f, N_c)}     {1 \over 4\pi } \int 
\tr( (F_c'  +  B  {\bf 1}_{N_c} )  \wedge (F_c' +B  {\bf 1}_{N_c} ) )    \right. \cr
& \left.   +    \frac{ (2N_c) }{  2 {\rm gcd}(N_f, N_c)}    {1 \over 4 \pi} \int 
\tr( (F_f' +B  {\bf 1}_{N_f} )  \wedge (F_f' +B  {\bf 1}_{N_f} ))   \right]    \;  d\mu_{\rm fermion}^{\rm quiver}  
\end{align} 
Therefore, the non-invariance  of partition function in the $(A, B)$ background under a chiral transformation   can be expressed as 
 \begin{align}
\Z_{2   {\rm gcd}(N_f, N_c) }:  \; \; {\cal Z}(A, B) ) &  \mapsto   \exp      \left[  - i  \frac{ (4 N_f N_c)  }{2 {\rm gcd}(N_f, N_c)}     { 1 \over 4\pi } \int 
  B  \wedge B     \right ]  {\cal Z}((A, B) )  \cr
&  = 
 \exp       \left[  - i  { 2 {\rm lcm}(N_f, N_c)   \over 4\pi } \int 
  B  \wedge B    \right]  {\cal Z}((A, B) )    \cr
  & =  \exp       \left[  - i  2 \pi  { 2 {\rm lcm}(N_f, N_c)   \over   \left( {\rm gcd} (N_f, N_c) \right)^2  } \right] {\cal Z}((A, B) )   
  \label{poly}
 \end{align}  
We used 
$\int  B  \wedge B   = 
   \left( \frac{1}{ {\rm gcd} (N_f, N_c)} \right)^2 
 \int 
  dC  \wedge dC   =     8 \pi^2  \left( \frac{1}{ {\rm gcd} (N_f, N_c)} \right)^2 $ in the last step. 
The phase   is non-trivial (mod $2\pi$) provided 
 \begin{align}
 \frac{  2 {\rm lcm}(N_f, N_c)}  { \left( {\rm gcd} (N_f, N_c) \right)^2 }\in   {\mathbb Q \backslash \mathbb Z}, 
 \label{ano}
 \end{align}
 For example, for $N_f=N_c$,    there is a mixed anomaly provided $ \frac{2 }{N_c} \in   {\mathbb Q \backslash \mathbb Z}$.  This  agrees with the existence of mixed anomaly for all $N_c =N_f \geq  3$ theories   \cite{Tanizaki:2017qhf}.    \footnote{By turning on an extra $U(1)_V$ background, it is also possible to make  the non-invariance of the action $  {\cal Z}(A_1, A_2, B)  \mapsto      \exp       \left[  - i    \frac{ 2\pi}{N_c} \right]   {\cal Z}((A_1,A_2.  B) ) $ in the $N_f=N_c$ case.  Then, an anomaly will also persist in the $N_c=2$ theory, and also  impose stronger constraint on the $N_c =N_f \geq  3$ theories. The general idea of turning on all possible backgrounds to obtain a stronger constraint on IR-physics is discussed in detail \cite{Anber:2019nze}.}
 
 
Assuming  \eqref{ano} is satisfied and anomaly exists (clearly, there are many such cases), it  has implications for   two related class of theories: 
\begin{itemize}
\item $\bm {SU(N_c) \times SU(N_f)} $ {\bf quiver theory}: There is a mixed anomaly between   $  \Z_{   {\rm gcd}(N_f, N_c) }^{[1]}$ 1-form center-symmetry and   $  \Z_{  2  {\rm gcd}(N_f, N_c) }$ 0-form chiral-symmetry. 
\item  $\bm {SU(N_c)} $ {\bf QCD(F)   with $\bm {N_f}$  flavors of massless  fermions:} There is a mixed anomaly between 
  $SU(N_f)/ \Z_{  {\rm gcd}(N_f, N_c) }$  symmetry and $  \Z_{  2 N_f }$  chiral symmetry. 
  \end{itemize}
The anomaly polynomial corresponding to both cases  is given by \eqref{poly}. This implies that a unique gapped  (trivial)  ground state in  $ SU(N_c)$  QCD(F)  and $ {SU(N_c) \times SU(N_f)} $ { quiver theory} are  impossible.   We list the  possibilities for ground states  after discussing the conditions under which this anomaly persists upon compactification.

 \subsection{Conditions for persistence of 0-form mixed anomalies  on $\R^3 \times S^1$}
 \label{sec:persistent}
A mixed anomaly between a  1-form symmetry  and 0-form symmetry persists upon compactification, e.g.  $\R^3 \times S^1$, see  \cite{Gaiotto:2017yup} for center and time reversal mixed anomaly at $\theta=\pi$ and center and discrete chiral in QCD(adj) in \cite{Komargodski:2017smk, Shimizu:2017asf}. This concept is sometimes called persistent order. 

  \vspace{3mm}
  As we stated at the beginning of this section,  if the theory does not  possess a  1-form symmetry, then, a  mixed anomaly involving two 0-form symmetries  $G_1 \times G_2$ does  not  impose a constraint on IR-physics in an obvious way. 
However, if the 0-form symmetries are of the form $G_1 \times G_2$ , where $G_1= \widetilde G_1/\Gamma$ and  gauging $ \widetilde G_1 $ turns $\Gamma$ into a 1-form symmetry, then   a triple-mixed anomaly  may persist upon compactification \cite{Tanizaki:2017qhf}.

  \vspace{3mm}
Let us now investigate the condition under which the anomaly persists. 
 This will bring 
$SU(N_f)_V$ flavor twist that we called $\Omega_F^0$ as the hero of the story.  

  \vspace{3mm}
Consider the partition function in the $\Omega_F^0$ background ${\cal Z}_{\Omega_F^0 }$. This is equivalent to imposing  
 $\Omega_F^0$  twisted boundary conditions on fermions \eqref{tbc-f}  
 as described in Section \ref{CFC-sec}. 
  Under an aperiodic  center-transformation,    $\Omega_F^0 \mapsto  \omega \Omega_F^0$ and partition function maps to  ${\cal Z}_{ \omega \Omega_F^0 }$, hence ${\cal Z}_{\Omega_F^0 }$    is not  invariant. However, the partition function can be made invariant by invoking 
  a transformation, $S \in  \Gamma_S \subset  SU(N_f)_V$, provided it obeys 
  \begin{align}
  S\Omega_F^0S^{-1} = \omega \Omega_F^0
  \end{align}
 This is the symmetry of  ${\cal Z}_{\Omega_F^0 }$. The solution to this algebra is unique up to conjugations. 
 As described around \eqref{MAG}, the choice of ${\Omega_F^0 }$ introduces a  0-form color-flavor center (CFC)-symmetry under which Polyakov loop is charged.  But  at the same time, it explicitly breaks the flavor symmetry down to 
 maximal Abelian subgroup ${\bm G}_{\rm max-ab}$. Crucially, the faithful vector-like flavor symmetry 
 of compactified theory becomes  
  \begin{align}
  K= \widetilde K/\Gamma \equiv U(1)^{N_f-1}/\Z_{{\rm gcd}(N_f, N_c)}.
  \label{K}
 \end{align}
 
   \vspace{3mm}
 Introducing the background gauge field  for $K=U(1)^{N_f-1}$,  $\Z_{{\rm gcd}(N_f, N_c)} $ emerges as a 1-form symmetry. We also introduce a 2-form field $B^{(2)}$ and 1-form field  associated with 1-form and 0-form part of center-symmetry 
$ \Z_{{\rm gcd}(N_f, N_c)}$,  and decompose the the 2-form field $B$ on 
\begin{align}
B=   B^{(2)} +   B^{(1)}  \wedge \beta^{-1} dx^4
\label{decom}
\end{align}
The partition function in the $(A_K,   B^{(2)}, B^{(1)})$ background is not invariant under a discrete chiral transformation 
$h \in \Z_{2 {\rm gcd}(N_f, N_c)}  $   and the   anomaly polynomial on 
  $\R^3 \times S^1$  can be obtained as:
\begin{align} 
{\cal Z}_{\Omega_F^0 } ( h(A_K,   B^{(2)}, B^{(1)})) &=   \exp \left[  - i  { 2 {\rm lcm}(N_f, N_c)   \over 2 \pi } \int 
  B^{(2)}  \wedge B^{(1)}     \right]  {\cal Z}_{\Omega_F^0 } ( A_K,   B^{(2)}, B^{(1)})   \cr
   & =  \exp       \left[  - i  2 \pi  { 2 {\rm lcm}(N_f, N_c)   \over   \left( {\rm gcd} (N_f, N_c) \right)^2  } \right] 
    {\cal Z}_{\Omega_F^0 } ( A_K,   B^{(2)}, B^{(1)})  
    \label{compano}
  \end{align}
The anomaly polynomial can also be deduced  from \eqref{poly}  with the substitution  \eqref{decom} 
Therefore, there is a triple  mixed anomaly between shift symmetry $\Gamma_S  \subset  SU(N_f)_V$, abelianized flavor symmetry $U(1)^{N_f-1}/\Z_{{\rm gcd}(N_f, N_c)}$,  and the discrete chiral symmetry   $\Z_{2N_f}$ provided \eqref{ano} holds. This is indeed  the same condition as in $\R^4$.

\subsection{Implication of mixed anomalies  on $\R^4$ and  $\R^3 \times S^1$} 
\label{sec: possibilities}
The  mixed anomaly on $\R^4$ is between  $SU(N_f)_V/ \Z_{  {\rm gcd}(N_f, N_c) }$  symmetry and $  \Z_{  2 N_f }$. 
We make two remarks.  {\bf 1)} There exists no order parameter which is charged under the discrete $\chi$S $  \Z_{  2 N_f }$  but not under continuous $\chi$S. \footnote{Opposite statement is not true. See Footnote \ref{Stern}.}  Therefore, the spontaneous breaking of the discrete symmetry implies spontaneous breaking of $SU(N_f)_A$ \cite{Cherman:2017dwt}.
  {\bf 2)} By a  theorem in  Ref.~\cite{Vafa:1983tf}, in vector-like theories, vector-like global symmetries cannot be spontaneously broken  as long as one assures  positivity of the path integral measure (which is the case in our theory with $\Omega_F^0$ twist.)
 
      \vspace{3mm}
  The existence of the mixed anomaly implies that the ground state of QCD(F) on $\R^4$ cannot be 
unique, gapped (trivial) state. In the light of above statements, there are two options on    $\R^4$:

\begin{itemize}
\item $SU(N_f)_A$ chiral symmetry is spontaneously broken and there are massless NG-bosons. 
\item Low energy theory is a CFT. 
\end{itemize} 
  
       \vspace{3mm}
On $\R^3 \times S^1$, the triple  mixed anomaly is between shift symmetry $\Gamma_S \subset SU(N_f)_V$, abelianized flavor symmetry $U(1)_V^{N_f-1}/\Z_{{\rm gcd}(N_f, N_c)}$,  and the discrete $\chi$S   
$  \Z_{  2 N_f }  \subset  U(1)_A^{N_f-1} $.  The options that can saturate the  anomaly are: 
\begin{itemize}
\item  $U(1)_A^{N_f-1} $  chiral symmetry is spontaneously broken and there are  NG-bosons.  
\item The color-flavor center (CFC) symmetry  $\Z_{{\rm gcd}(N_f, N_c)}$ is spontaneously broken. 
\item Both are spontaneously broken. 
\item Low energy theory is a CFT. 
\end{itemize} 

     \vspace{3mm}
By using rigorous semi-classics on small $\R^3 \times S^1$ and provided $m_\lambda < m_\lambda^*$, we showed that the first option, spontaneous breaking of $U(1)_A^{N_f-1} $  chiral symmetry  is realized on small circle. At large-$S^1$, we also expect to see spontaneous breaking of chiral symmetry. However, anomaly  allows other mixed anomaly respecting intermediate phases or anomaly respecting phase transitions which is not associated with the change in symmetry realization.\footnote{As discussed around \eqref{options},   for  $m_\lambda=0$ theory,  $U(1)_A^{N-1}$ and CFC realization in small and large circle are certainly the same, but $U(1)_{A_D}$ realization may change.    Let us assume that there exists a phase transition associated with $U(1)_{A_D}$, and it is broken at large-$\beta$. Turning on a soft  $m_\lambda>0$ breaks $U(1)_{A_D}$ symmetry explicitly. 
If the transition at $m_\lambda=0$ is first order,   it will continue to persists even for small  $m_\lambda>0$, but  the transition  will not be associated with a change in symmetry realization. Furthermore,  the first order line may end up with a second order critical point. }
 This is a realization of the persistent order idea in QCD.  Indeed, in the next section, we will show that for $m_\lambda > m_\lambda^*$, there exists intermediate phases  in which only   CFC symmetry is spontaneously broken. 

     \vspace{3mm}
For $0< m_\lambda < m_\lambda^*$, there is strong reasons to believe that the theory exhibits adiabatic continuity in the $(\beta, m_\lambda)$ plane,  i.e, small-circle  ${\bf G}_{\rm max-ab} $ broken phase is continuously connected to the strong coupling  ${\bf G}_{\rm max-ab} $ broken phase.  In this sense,  adiabatic continuity is a refined version of  the persistent order or  mixed 't Hooft anomaly in which  only one  mode of the anomaly constraint is  realized at arbitrary  $\beta$. 
 We believe providing a proof of this statement would be  a tremendous progress in the understanding of strongly coupled QCD.

\section{Mixed anomaly permitted phase transitions on $\R^3 \times S^1$ }

As described in Sec.~\ref{sec: possibilities},  the theory can only be in a phase that is compliant with  mixed anomaly.  Phase transitions are possible, but different phases must be a realization of a  mode of the anomaly constraint. This is called persistent order. The ground state is never a trivial gapped phase. 

\begin{figure}[t]
\begin{center}
\includegraphics[width = 1.03\textwidth]{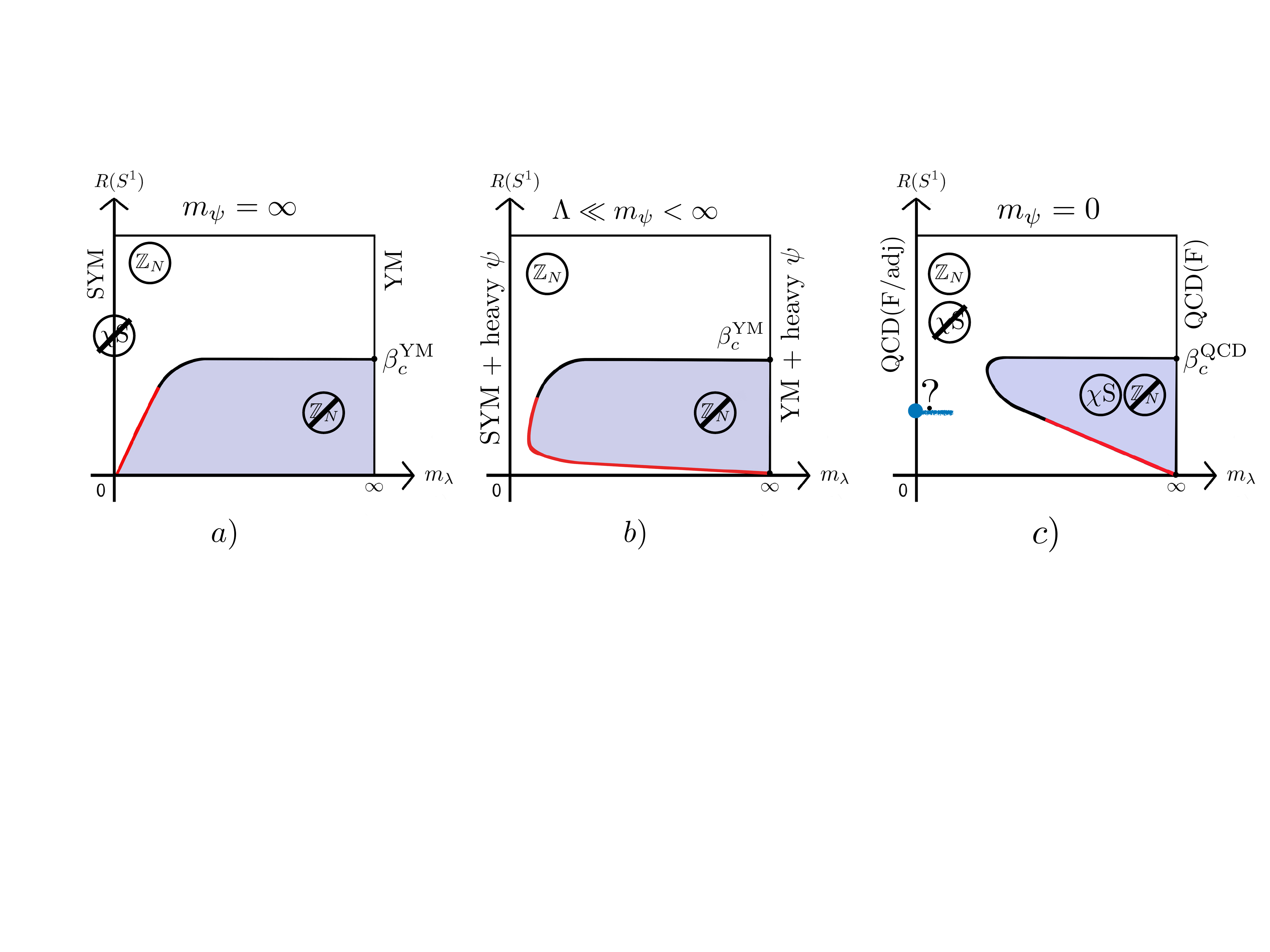}
\vspace{-5.5cm}
\caption{Analytically calculable (red) and incalculable (black) $\Z_N$ center or CFC symmetry changing   phase transitions in gauge theories on $\R^3 \times S^1$  
 for various values of $m_\psi$ on $(\beta, m_\lambda)$ plane.   a) $\Z_{N_c}$ center-symmetry changing phase transition in 
 mass deformed SYM. This is studied in detail in  \cite{Poppitz:2012nz} and presented here for completeness.  b) For heavy fundamental quark with $\Omega_F^0$-twist,  the  $\Z_{N_c}$  CFC symmetry is stabilized for any $m_\lambda < m_{\lambda}^{*}$. This theory exhibits   adiabatic continuity for Yang--Mills theory (with very heavy adjoint and fundamental fermions) between small and large $S^1$ regimes. c) This theory also  exhibits   adiabatic continuity for QCD(F)  (with very heavy adjoint  fermions) between small and large $S^1$ regimes. See text concerning the blue line. 
  }
\label{fig:phase}
\end{center}
\end{figure}

\vspace{3mm}
Throughout this section, we consider the phases of $N_f=N_c$ QCD(F/adj) 
as a function of the parameters $(m_\lambda, m_\psi, \beta)$. For  $m_\psi= \infty$, $\Lambda \ll m_\psi < \infty$, $m_\psi=0$, we determine phase diagram in the $(m_\lambda,  \beta)$ plane. There are  
both calculable and incalculable phase transitions.  The main outcome is shown in Fig.~\ref{fig:phase} which we discuss below.

\subsection{Phases of  $m_{\psi} =0 $ theory in the  $  (m_\lambda, \beta)$ plane }  
 Turning on a soft mass term for adjoint fermion, the balance between the center-destabilizing gauge fluctuations and center-stabilizing 
 adjoint fermion breaks down. Hence, at one-loop order, a center-destabilizing one-loop potential do get induced 
  \cite{Unsal:2010qh, Poppitz:2012sw}
  \begin{align}
 V^{\rm  gauge}_{\rm 1-loop} + V^{\rm \lambda}_{\rm 1-loop}   &= 
 \frac{2}{\pi^2 \beta^4}  \sum_{n=1}^{\infty}  \frac{ \left[ -1+ \frac{1}{2} (n\beta m_\lambda)^2 K_2 (n \beta m_\lambda) \right] }{n^4}  |\tr (\Omega^n)|^2     \cr
&= -\frac{m_\lambda^2}{2 \pi^2 \beta^2}  \sum_{n=1}^{\infty} \frac{1}{n^2}   |\tr (\Omega^n)|^2  + O(m_\lambda^4) 
\label{one-loop-desta}
\end{align}
where  $K_2$ is modified Bessel function.  However, this potential is parametrically small compared to 
generic one-loop potentials by a  factor  $(m_\lambda  \beta)^2 $.  At two-loop order, there exists a center-stabilizing term due to fundamental fermions  (with $\Omega_F^0$-boundary conditions) given in \eqref{V2psi-min}. Despite the fact that the   two-loop potential    is suppressed compared to {\it generic} one-loop terms with a parameter $(g^2N_c)$, it can compete with \eqref{one-loop-desta} which is already  parametrically suppressed.

\vspace{3mm}
\noindent
{\bf Non-commutativity of limits-1:} For sufficiently small $(m_\lambda  \beta)^2  \ll g^2N_c $, the  center-stabilizing two loop potential  dominates.  
For sufficiently small $ g^2N_c  \ll (m_\lambda  \beta)^2$, the center-destabilizing one-loop potential dominates. 
Therefore, the two limits are non-commuting and we obtain:
\begin{equation}
 \left\{ 
\begin{array}{ll l }
m_\lambda^{-1} \rightarrow 0 &\qquad  \beta= {\rm fixed}  & \qquad  \qquad  \Z_{N_c} \;  \text{ broken} \cr
 \beta \rightarrow 0, & \qquad  m_\lambda^{-1} =  {\rm fixed} & \qquad  \qquad  \Z_{N_c} \;  \text{symmetric}
\end{array}  \right\}
\end{equation}
 In the weak coupling regime, this  implies that the 
center-symmetry changing phase transition can be studied analytically along a line emanating from  $(m_\lambda^{-1}, \beta)=(0,0)$ point, shown in Fig.\ref{fig:phase}c.

\vspace{3mm}
In this regime, center-symmetry changing phase transition is  due to a competition between a  parametrically suppressed one-loop potential (with parameter $m_\lambda \beta $) and two-loop potential. The two contribution become parametrically comparable for 
\begin{align}
\beta_c \sim \frac{  (g^2N_c)^{1/2} }{m_\lambda}
\label{crit-2}
\end{align}
 where a phase transition is expected. This will be  further quantified below.

\vspace{3mm}
As described earlier, the combination of the one- and two-loop potentials for fundamental fermions $\psi^a$ with twisted boundary condition has both  single-trace terms  of the form $ \tr \Omega^{N_c k} $ which do  not play important role in center-symmetry realization and double trace terms which prefer  center-symmetric vacuum, see \eqref{V2psi-min}.  The combined potential which determines the center-symmetry realization is: 
 \begin{align}
 V^{\rm  gauge}_{\rm 1-loop} + V^{\rm \lambda}_{\rm 1-loop} + V^{\rm \psi}_{\rm 2-loop, \Omega_F^0}
&= 
 \frac{2}{\pi^2 \beta^4}  \sum_{n=1}^{\infty}  \frac{ \overbrace{ \left[ -1+ \frac{1}{2} (n\beta m_\lambda)^2 K_2 (n \beta m_\lambda)  + \frac{g^2N_c}{16 \pi^2} \right]  }^{a_n} }{n^4}  |\tr (\Omega^n)|^2     \cr 
 & \approx   \frac{2}{\pi^2 \beta^4}  \sum_{n=1}^{\infty}  \frac{ \left[ -  \frac{1}{4} (n\beta m_\lambda)^2   + \frac{g^2N_c}{16 \pi^2} \right] }{n^4}  |\tr (\Omega^n)|^2     
%
\end{align}
where in the second formula, we used small-$z$ asymptotic  of the Bessel function $K_2(z)$.   
$a_n$ is the effective mass square  $m_n^2$ for the winding number $n$ Polyakov loop $\tr \Omega^n$. If the effective mass is positive for all $n \leq \lfloor \frac{N_c}{2} \rfloor$,  the minimum of the effective potential lies at a 
  center-symmetric point, invariant under $\Z_{N_c}$ transformation:
  \begin{align}
 \Z_{N_c} \;\;  {\rm stability}:   a_1>0, \; a_2 >0, \; \ldots, a_{N_c/2}>0:  \Longrightarrow  \qquad  (\beta m_\lambda)   <  \frac{(g^2N_c)^{1/2} }{  N_c \pi} 
 \label{f-stability}
  \end{align}
Since $m_n^2$ is monotonically    decreasing function of its argument, if $a_1$ is negative, then so are $a_n, n\geq 2$.  In this case,  $ \Z_{N_c}$ is completely broken: 
  \begin{align}
 \Z_{N_c} \;\;  {\rm fully \; broken}:   a_1<0:  \Longrightarrow    \qquad  ( \beta m_\lambda)   >  \frac{(g^2N_c)^{1/2} }{ 2 \pi}
  \end{align}

\vspace{3mm}
If $N_c$ is large, there exists intermediate ranges where center symmetry is partially broken. The instability point for the 
Wilson line with winding number $k$ is, $a_k=0$, corresponding to 
\begin{align}
 \beta_k  =  \frac{(g^2N_c)^{1/2} }{ 2 \pi  m_\lambda  }  \frac{1}{k} \equiv \beta_1\frac{1}{k} 
 \label{sequence}
\end{align}
When the compactification radius lies between $\beta \in [\beta_k, \beta_{k-1}]$, the partial center symmetry that is preserved is $\Z_k$,  corresponding to a configuration of eigenvalues in which eigenvalues form $k$-clump each of which possess $N_c/k$ coincident eigenvalues. 

\vspace{3mm}
So, in the range 
 $ \beta   <  \frac{(g^2N_c)^{1/2} }{  N_c \pi m_\lambda}, $   $ \Z_{N_c}$  is  fully restored. This range shrinks to zero as $N_c$ increases. We also expect that for  $ \beta   \gtrsim c \Lambda^{-1} $ where $c$ is a pure number, the center to be fully stabilized. This is shown in Fig.~\ref{fig:phase}c 
 but intermediate phases with partial center breaking are not detailed in there. 
 \begin{align} 
  \left\{ 
\begin{array}{ll  }
\beta< \beta_{N_c/2}   & \qquad  \qquad  \Z_{N_c} \;  \text{ symmetric} \cr 
\beta_k < \beta < \beta_{k-1} & \qquad  \qquad { (\approx  \text {or exactly}) }   \;\;\;  \Z_{k} \;  \text{ symmetric} \cr
\beta_1< \beta<  \beta^*=c\Lambda^{-1}    & \qquad  \qquad  \Z_{N_c} \text{  fully broken } \cr
\beta>\beta^*  & \qquad  \qquad  \Z_{N_c} \;  \text{ symmetric} 
\end{array}  \right\}
\end{align}

 \vspace{3mm}
Note that this analysis is  almost identical   to the one loop analysis of the calculable phase transition in  QCD(adj) with $1 \ll N_f \leq N_f^*$ flavors  in the $(m_\lambda,\beta)$ plane,  see \cite{Unsal:2010qh}. The reason 
  is that in QCD(F/adj), fundamental fermions,   in their contribution to holonomy potential, 
behave as  $x O (g^2N_c)$ many adjoint fermions.  Hence, the system behaves as if it has 
$1+ x O (g^2N_c)$ adjoint fermions, where $x = N_f / N_c $ is finite. 

 \vspace{3mm}
Finally, we remark on the blue line in Fig.~\ref{fig:phase}c. As described around \eqref{options}, the  theory  for $m_\lambda = 0$ on small circle  
exhibits   ${  G \rightarrow   G_{\rm V} \times U(1)_{\rm A_D}  } $,  but on large $S_1$ , both  ${  G \rightarrow   G_{\rm V} \times U(1)_{\rm A_D}  } $ and  ${  G \rightarrow   G_{\rm V}} $ are reasonable possibilities according to anomaly consideration. As explained around  \eqref{four-fermi}, the second possibility seems more likely if we trust semi-classical EFT at the boundary of its region of validity. If true, there has to be a phase transition associated with the change of  $U(1)_{\rm A_D}$. The rest of axial symmetry  is already broken both at large and small circle.   Turning on a small $m_\lambda$, it is plausible that the phase transition, which is now not associated with any exact symmetry,  may persist for a while. If the transition at $m_\lambda =0$ is first order, this is certainly expected to be the case. But ultimately, the phase transition line is expected to  end by a second    
 order critical point. The interesting thing about this phase transition is that the symmetry realization is same both above and below the phase transition line: ${  G \rightarrow   G_{\rm V}} $  and $\Z_N$  CFC is unbroken. 
  It is again conceivable that these two regimes are adiabatically connected.\footnote{Thanks to Ofer Aharony for explanations at this point.}

 \subsection{Phases of  $m_\psi =\infty $ theory in the  $  (m_\lambda, \beta)$ plane }  
 The $m_\psi =\infty $  limit of QCD(adj/F) is $\N=1$ SYM  theory.  The phase diagram of this theory in the  $  (m_\lambda, \beta)$ plane  is  investigated in depth in the literature \cite{Unsal:2010qh, Poppitz:2012sw, Poppitz:2012nz}. In fact, this example provides the first  semi-classically calculable realization  of center-symmetry changing phase transitions  in gauge theory on $\R^3 \times S^1_{\beta}$. 
   For completeness of the  phase diagram  in the $(m_\lambda, m_\psi, \beta)$-domain,   we briefly remind the phase transition in this limit.  
 
  \vspace{3mm}
 Turning on a soft mass term for adjoint fermion,  as discussed earlier \eqref{one-loop-desta},  leads to   a center-destabilizing one-loop potential. In the $m_\psi =\infty$ limit, since the fundamental fermions decouple, their help to center-stabilization is lost. In fact, in the $m_\lambda=0$ theory,   the holonomy direction is a moduli-space  classically and to all orders in perturbation theory due to  $\N=1$ supersymmetry. 
At this level,  the eigenvalues of Wilson line do not interact.

   \vspace{3mm}
   Due to non-perturbative effects, the moduli space is lifted in favor of a center-symmetric minimum. 
 This happens   at second order is semi-classics.  $\N=1$ SYM has center stabilizing neutral bions with amplitudes 
 $[{\cal M}_i  \bar {\cal M}_{i}] = e^{- 2  \times \frac{4 \pi}{g^2}   \alpha_i \cdot  ( v+ \phi)}  \sim  e^{- 2  \times \frac{4 \pi}{g^2}    (v _{i+1} - v_i) }$ where in the latter form, we set the fluctuations to zero.   The effect of the neutral bions is to induce a repulsive potential between the eigenvalues $v_i$  of Wilson line. 
  It should be noted that in the presence of $m_\lambda >0$ fermions, the fermionic zero modes of the monopoles do get lifted, and hence, it becomes possible for the monopoles to contribute to holonomy potential as well. But they do not contribute to holonomy potential at $m_\lambda =0$. 
 So, there are three competing effects: 
 \begin{itemize}
\item  $O (m_\lambda^2 \beta^2)$:  perturbative one-loop center-destabilizing, 
 \item $O (m_\lambda \beta e^{-S_0})$: NP,  semi-classic first order,   center-destabilizing, 
 \item $O ( e^{-2S_0})$: NP, semi-classic second order,   center-stabilizing, 
 \end{itemize}
 where  $e^{-S_0} = (\Lambda \beta)^3$ in $\N=1$ SYM. 
 Below, we briefly review this  phase transition, because momentarily, we  will show that this phase structure changes in crucial ways once massive fundamental fermions with   $ m_\psi <\infty$ and  $\Omega_F^0$ twisted boundary condition is introduced. 
 
 \vspace{3mm}
 \noindent
{\bf Non-commutativity of limits-2 and phase transition:}  
First, let us parametrize the holonomy potential as: 
\begin{align}
V[\Omega] =  \frac{1}{\beta^4} \sum_{n=1}^{\lfloor \frac{N_c}{2} \rfloor} M_n^2  \frac{1}{n^4} \tr |\Omega^n|^2
\end{align} 
The (dimensionless) mass square  $ M_n^2$  for the Wilson line with winding number one (or few)  receives  contributions from perturbative fluctuations \eqref{one-loop-desta},   monopole-instantons and neutral bions, of the form $-O (m_\lambda^2 \beta^2)
$,  $-O (m_\lambda \beta (\Lambda \beta)^3)$ and $+O( (\Lambda \beta)^6)$. In the vicinity of  the phase transition scale, 
the perturbative term is down by three powers of $\log(\frac{1}{\beta N_c \Lambda})$ and can be neglected. Therefore, the center-symmetry changing phase transition boils down to a competition between neutral bion effect and monopole-instanton effect. The mass square for the Wilson line around the phase transition scale 
\begin{align} 
M_1^2 \big|_{m_\psi=\infty} \sim ( -(m_\lambda \beta )(\Lambda \beta)^3 +   (\Lambda \beta)^6 )  \qquad {\rm for } \;\;  \beta \sim \beta_{c1}  \sim \Lambda^{-1} \sqrt {\frac{m_\lambda}{\Lambda}}
 \end{align}
 where 
 \begin{align}
\beta_{c1} \sim \Lambda^{-1} \sqrt{\frac{m_\lambda}{\Lambda} }
\label{Lc1}
\end{align}
 is the parametric value of the center-symmetry changing phase transition. 
  
 For sufficiently small mass  $m_\lambda  \beta  \ll e^{-S_0} = (\Lambda \beta)^3  $, the  center-stabilizing neutral bion  potential  dominates.  
For  $  e^{-S_0} \ll m_\lambda  \beta $, the center-destabilizing monopole-instantons  dominate.  As $\beta \rightarrow 0 $, the mass square is dominated by the perturbative  $-O (m_\lambda^2 \beta^2)$ term. 
 \begin{align}
M_1^2 \big|_{m_\psi=\infty} \sim - (m_\lambda^2 \beta^2) <0
 \qquad    \beta \rightarrow 0 
\end{align}
and center symmetry is broken for all   $ 0 < \beta< \beta_{c1}$.
 Therefore,  the theory may land on two different phases in the  vicinity of $(m_\lambda, \beta)=(0,0)$:
\begin{equation}
 \left\{ 
\begin{array}{ll}
m_\lambda \rightarrow 0 &\qquad  \beta= {\rm fixed}   \qquad \;\;   \Z_{N_c} \;  \text{ symmetric}  \cr
 \beta \rightarrow 0, & \qquad  m_\lambda =  {\rm fixed}  \qquad  \Z_{N_c} \;  \text{ broken} 
\end{array}  \right\}
\end{equation}  
The mass square is always negative definite for  $\beta< \beta_{c1}$.  A sketch of the mass-square for Wilson line  is shown in Fig.\ref{fig:massholonomy}, left panel.

 \subsection{Absence of phase transition for  $   m_\lambda=0,  0 \leq m_\psi  < \infty$}

  In this domain, center-destabilizing effect of the gauge fluctuations is cancelled by the massless adjoint fermion  
  to all orders in  perturbation theory.  There are two-center stabilizing effects. One is perturbative two-loop potential  due to $\Omega_F^0$-twisted boundary conditions  for  fundamental fermions  
  \eqref{V2psi-min} and    the other is non-perturbative neutral bion effects  $[{\cal M}_i  \bar {\cal M}_{i}] = 
   e^{- 2  \times \frac{4 \pi}{g^2}    (v _{i+1} - v_i) }$. 
 
  \vspace{3mm}
 In the whole $m_\lambda=0$ subspace of the $(m_\lambda,  \beta, m_\psi)$ space, CFC-symmetry is always stable. At small $m_\psi$, stability is due to two-loop fundamental fermion contribution and in the   $m_\psi = \infty$,  it is due to neutral bion effects in $\N=1$ SYM.  The QCD(F/adj) with $N_f= N_c$   in the  $m_\lambda=0$ plane is free of any center-symmetry changing phase transition, and for   $0 < m_\psi < \infty, 0 <\beta < \infty$,  it is free of any phase transitions assuming expected behavior on $\R^4$.

  \subsection{Phases of  large $ m_\psi $  theory:  Two calculable phase transitions }  
 There is one extremely  interesting corner of the phase diagram, in which  one can  analytically show  that: 
  \begin{align} 
  & {\rm for } \; \; m_\lambda <  m_\lambda^* : \qquad 
 \left\{ 
 \begin{array}{ll}
0 < \beta<  \infty  & \qquad  \qquad  \Z_{N_c}  \;\;   {\rm symmetric }  
 \end{array} \right.   \cr   \cr
& {\rm for } \; \; m_\lambda >  m_\lambda^* : \qquad 
 \left\{ 
 \begin{array}{ll}
  \beta>  \beta_{c1}  & \qquad  \Z_{N_c}    \;\;   {\rm  symmetric } \cr
   \beta_{c1} > \beta >  \beta_{c2} &  \qquad    \Z_{N_c}    \;\;   {\rm  broken}  \cr
 \beta_{c2}>\beta  & \qquad  \Z_{N_c} \;\;   {\rm symmetric }  
 \end{array} \right.  
  \end{align} 
Namely, there exists a critical  $m_\lambda^*$  such that for $m_\lambda  \in [(0,   m_\lambda^* )$, the small-circle regime is 
$\Z_{N_c}$   center-symmetric.  and chiral symmetry \eqref{MAG} is broken to vector-like subgroup.  
 If we make standard assumptions about the dynamics concerning $\R^4$, this regime realizes the same symmetry realization as large $S^1$ and is likely continuously connected to it.  
 
 \vspace{3mm}
 For   $ m_\lambda > m_\lambda^*$  there  are  two center-symmetry changing phase transitions. 
 If  $ m_\lambda  $ is slightly above $m_\lambda^*$,  then both of these phase transitions are analytically calculable, 
 see Fig.~\ref{fig:phase}b. If $m_\lambda  > \Lambda_{QCD}$, the  phase transition at  $ \beta_{c1} $    turn into a semi-classically incalculable  phase transition.

 \vspace{3mm}
 There can be partially broken phases for $ \beta_{c2} < \beta <  \beta_{c1}$, but our primary concern is the points where center is fully restored. 
 The realization of center-symmetry is  determined by the first $\lfloor \frac{N_c}{2} \rfloor$ terms in the holonomy  potential. Below, we would like to treat $N_c$ as an order one  number, and give order of magnitude estimates for the phase transition scales. 
 
  \vspace{3mm}
 \noindent
 {\bf Non-commutativity of limits-3:} Does heavy fermion always decouple from dynamics?  Assume $m_\psi < \infty $ and  large. Normally, one  would think that the heavy fermion should decouple from the dynamics. Indeed,  in the  one- and two-loop potentials $V^{\rm \psi}_{\rm 1-loop, \Omega_F^0}+  V^{\rm \psi}_{\rm 2-loop, \Omega_F^0}$, $m_\psi  $ 
  appear through the combinations  such as $ \half (m_\psi \beta n)^2 K_2(m_\psi \beta n)$. At finite-$\beta$, as $m_\psi  \rightarrow  \infty$, 
 the effect of the fundamental fermion will disappear exponentially $e^{-m_\psi \beta n}  $ by the large-argument asymptotic of Bessel function.  
 However, at finite- $m_\psi$, and as $\beta  \rightarrow 0 $, despite the fact that fermion is heavy, it will behave as if it is massless, and will follow small-argument asymptotic of Bessel function.  
 This  has interesting impacts on phase transitions. 
\begin{figure}[t]
\begin{center}
\includegraphics[width = .80\textwidth]{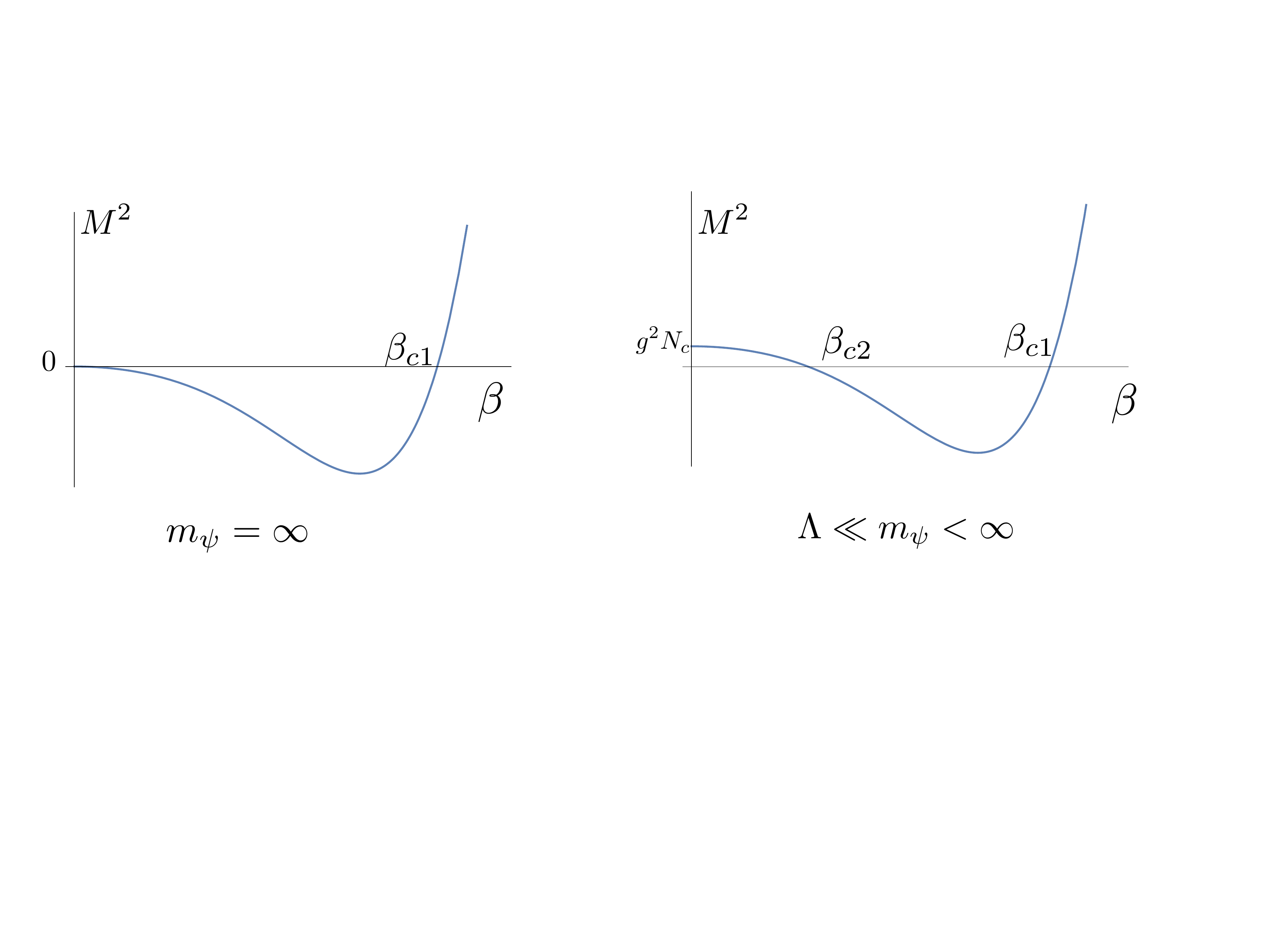}
\vspace{-4cm}
\caption{The mass square for gauge holonomy. 
$m_\psi=\infty,  \; 0<m_\lambda \ll \Lambda $ of QCD(F/adj) is $\N=1$ SYM with soft supersymmetry breaking mass term. The center-symmetry changing phase transition is analytically calculable, and takes place at $\beta_{c1}$.  When  $ m_\psi <\infty$,  the fundamental fermions with $\Omega_F^0$ twist always gives a center-stabilizing contribution, and  the center-symmetry always restores at arbitrarily small-$\beta$. This gives an example of two calculable phase transitions.  
  }
\label{fig:massholonomy}
\end{center}
\end{figure}

   \vspace{3mm}
Consider  the $(m_\lambda,  \beta, m_\psi)=(0,0, \infty)$ corner of phase diagram. Move off the corner slightly in a generic direction. 
There are four competing effects  which can determine  the realization of center symmetry. 
  \begin{itemize}
\item  $O (m_\lambda^2 \beta^2)$:  perturbative one-loop center-destabilizing from  $ V^{\rm  gauge}_{\rm 1-loop} + V^{\rm \lambda}_{\rm 1-loop} $
 \item $O (m_\lambda \beta e^{-S_0})$: NP,  semi-classic first order (monopole-instantons)   center-destabilizing, 
 \item $O ( e^{-2S_0})$: NP, semi-classic second order (neutral bions),   center-stabilizing, 
\item $ O((g^2 N_c)K_2(m_\psi \beta)) $: perturbative two-loop center-stabilizing from $V^{\rm \psi}_{\rm 2-loop, \Omega_F} $
 \end{itemize}

 \vspace{3mm}
\noindent
${\bf {m_\psi< \infty}:}$
 As  long as  $m_\psi \beta_{c1} \gg1 $, the  $V^{\rm \psi}_{\rm 2-loop, \Omega_F} $ term has no effect on the phase transition taking place at $\beta= \beta_{c1}$. However, if  we take $\beta \rightarrow 0$ limit such that $m_\psi \beta \lesssim 1 $, the fundamental fermion induces  $V^{\rm \psi}_{\rm 2-loop, \Omega_F^0} $ term,  in which  the contribution to  mass-square  for the Wilson lines   is always positive definite! 
 
  \vspace{3mm}
 Therefore, with the inclusion of finite mass fundamental fermion, the mass-square in the $ \beta \rightarrow 0 $ limit, as per \eqref{V2psi-min},  becomes 
positive: 
 \begin{align}
M_1^2 \big|_{m_\psi<\infty} \rightarrow +  \frac{(g^2N_c)}{8 \pi^4}  >0   \qquad    \beta \rightarrow 0 
\label{positive-3} 
\end{align}
which implies that the center-symmetry must be stabilized as $\beta\rightarrow 0$ limit. 
This shows  that  there must exist an   $\beta_{c2} \in (0, \beta_{c1}) $ such that  Wilson line is non-tachyonic for   $\beta<\beta_{c2}$ and  center symmetry is restored.  A sketch of the mass-square for Wilson line for this case   is shown in Fig.\ref{fig:massholonomy}, right panel.

\vspace{3mm}
The phase transition at $\beta_{c2}$ is a result of the competition between    $O (m_\lambda^2 \beta^2)$   one-loop center-destabilizing effects from  $ V^{\rm  gauge}_{\rm 1-loop} + V^{\rm \lambda}$ and  perturbative  
two-loop center-stabilizing effects from $V^{\rm \psi}_{\rm 2-loop, \Omega_F^0} $.  We can use this to determine the position of $\beta_{c2}$.  
 However, currently, we do not have a closed form expression for $V^{\rm \psi}_{\rm 2-loop, \Omega_F^0} $ for massive fundamental fermions.    Note that this does not interfere with the argument above about the existence of center-symmetric 
 phase for $\beta < \beta_{c2}$ and the positivity of the mass term for Wilson line in the $\beta \rightarrow 0 $ limit.  To prove this positivity, we needed the expression in the $m_\psi \beta \rightarrow 0 $ limit, which we could deduce analytically.

\vspace{3mm}
If we would like to estimate the parametric form of  $ \beta_{c2}$, we  need the  holonomy potential \eqref{V2psi-min}  for $\Omega_F^0$ twisted massive fundamental fermions, in particular, the center-stabilizing double trace operator. 
In perturbation theory, a holonomy dependence involving double-trace term  arises  when  
one gluon or adjoint quark propagator goes around the $S^1$ circle. However, we are considering  \eqref{V2psi-min},   which is sourced by fundamental quarks. At two loop order, a quark and anti-quark  that goes around the circle $S^1$ can emulate 
an adjoint matter and induce a double-trace term. This term is  roughly: 
$  \frac{ (g^2 N_c)} {\beta^4 8  \pi^4} \sum_{n=1}^{\infty}  \half (n\beta (2 m_\psi))^2 K_2 (n \beta(2 m_\psi))   \frac{1}{n^4} |\tr \Omega^n|^2 $.  Therefore,  to estimate the    the phase transition point closer to the   $\beta \rightarrow 0$  limit,  we inspect the  the mass  square for the Polyakov loop order parameter: 
\begin{align} 
M_1^2 \big|_{m_\psi< \infty} \sim  - \frac{1}{2 \pi^2}  m_\lambda^2 \beta^2  + 
\frac{ (g^2N_c)} {8 \pi^4}  ( 1-m_\psi^2 \beta^2)  \qquad {\rm for } \;\;  \beta \lesssim \beta_{c2}
\end{align}
Hence, the critical radius  parametrically takes the form:
\begin{align}
\beta_{c2} \sim \frac{ \frac{ (g^2N_c)^{1/2}}{2 \pi}  }{\left( m_\lambda^2 +  \frac{ (g^2N_c) }{4 \pi^2} m_\psi^2 \right)^{1/2} }
\label{crit-3}
\end{align}
For $m_\psi = \infty$, the $\beta_{c2} \rightarrow 0$  and the center symmetry is broken for all $\beta < \beta_{c1}$, which is the result for $\N=1$ SYM.  
For $m_\psi = 0$, we obtain   $\beta_{c2} \sim \frac{  (g^2N_c)^{1/2} }{  2 \pi m_\lambda } $, in agreement  with \eqref{crit-2}.

\section{Large-$N_c$ volume independence in the Veneziano type  limit}
In this section, we describe briefly the relation between the quantum distillation, $\Omega_F^0$ twisted boundary conditions and large-$N_c$ volume independence in QCD(F/adj). 

A sub-class of  large-$N_c$ gauge theories,    when studied in  toroidal  compactification    of $\R^d$   (or its latticy  version)   have properties independent of compactification radius. This property is called large-$N_c$ volume independence \cite{Kovtun:2007py}, which is a special case of large-$N$ orbifold equivalence.  The extreme version of volume independence, where a space-time lattice $L^d$ is reduced to one-site   $1^d$ lattice  is called Eguchi-Kawai reduction or large-$N_c$ reduction \cite{EguchiKawaiOriginal, GonzalezArroyo:1982hz, Bhanot:1982sh}. 
  The necessary and sufficient conditions  for the validity of the volume independence are 
\begin{itemize}
\item Translation symmetry of lattice  $L^d$  is not spontaneously broken. 
\item  $(\Z_{N_c})^d$  center symmetry is not spontaneously broken.
\end{itemize}
Volume independence applies  to the  expectation values and connected correlators of topologically trivial Wilson loops 
at leading order in $N_c$.  The  sector of the theory neutral under center transformations is called 
{\it neutral sector}.  For example, string tensions, mass gap, spectrum of the theory,  free energies, pressures  are in the  neutral sector  and volume independence applies to them.  Polyakov loop expectation values and correlators are part of the non-neutral 
sector observables. 

\vspace{3mm}
The  original proposal   \cite{EguchiKawaiOriginal} actually fails to satisfy volume independence below certain critical size due to center symmetry breaking.  But  there are other versions or theories in which it works. The   cleanest examples to volume independence are new versions of twisted Eguchi-Kawai (TEK) models \cite{GonzalezArroyo:2010ss, Perez:2014sqa} and 
QCD with   $ N_f $ adjoint Weyl fermions endowed  with periodic boundary conditions   \cite{Kovtun:2007py}. 
 QCD(adj), if studied with 
periodic boundary conditions for fermions in path integral formulation or equivalently via  $\tr [(-1)^F e^{-\beta H}]$ in Hamiltonian formulation,  obeys  volume independence  \cite{Kovtun:2007py}. The fact that volume independence works means the theory avoids all possible phase transition and Hagedorn singularities. An important point  that started to emerge 
fairly recently is that at the root of working versions of   volume independence, there must be 
profound spectral  cancellations \cite{Basar:2013sza, Cherman:2018mya,Sulejmanpasic:2016llc} or an extreme version of quantum distillation. Otherwise, generically,  there will be phase transitions. 
In QCD(adj), this manifests itself  as Bose-Fermi cancellations in the absence of supersymmetry discussed in 
 \cite{Basar:2013sza, Cherman:2018mya}.  The spectral cancellation and quantum distillation of Hilbert space  must also occur in the  twisted Eguchi-Kawai model \cite{ GonzalezArroyo:1982hz, GonzalezArroyo:2010ss, Perez:2014sqa}.

\vspace{3mm}
 Volume independence on $\R^3 \times S^1$ translates to  temperature independence of the  neutral sector observables in the leading order in large-$N_c$ limit. 
In $SU(N_c)$ Yang--Mills theory, large $N_c$ volume independence holds as long as all compactification radii are larger than a critical radius 
$\beta_c \sim \Lambda^{-1}$ and  fails below $\beta_c$ due to a center-symmetry breaking phase transition \cite{Kiskis:2003rd}.   
In QCD  with $N_f \sim N_c$ fundamental fermions (namely, the Veneziano limit),  one does not even discuss volume independence, because there is not even a center-symmetry to begin with. 

\vspace{3mm}
In fact, it is  historically believed that QCD(F) with  $N_f \sim N_c$ fundamental fermions   manifestly disobeys  volume  (or temperature) independence due to lack of center-symmetry.  For example,  assuming chiral symmetry is broken, this theory has $(N_f^2 -1)$ NG bosons and the pressure is  
$p  (T)   =  \frac{\pi^2}{45}   (N_c^2-1) T^4  $ at leading order in large-$N_c$ for $N_f=N_c$.  Clearly, this is temperature dependent at leading order and large-$N_c$ temperature independence fails to hold. 

\vspace{3mm}
However, as described in  Section \ref{CFC-sec}, the theories with $N_f= kN_c$ possess an exact color-flavor center symmetry, which is in the diagonal  of  center of gauge group  and  cyclic permutations $\Gamma_S \in SU(N_f)$ living  in flavor rotations.  The Polyakov loops winding around the $S^1$ circle are charged under the $  \Z_{N_c} $  CFC symmetry. 
Therefore, one can meaningfully talk about the realization of this CFC-symmetry. 

\vspace{3mm}
The existence of $  \Z_{N_c} $   CFC symmetry manifests  itself in the  loop potential for holonomy. The potential is only a function of $\tr (\Omega^{N_c k})$ and $ \big| \text{Tr} (\Omega^k) \big|^2, k \in \Z$ which are  singlet under the CFC symmetry.  In particular, terms such as $\tr (\Omega^{q}), q \neq 0  $ (mod $N_c$),     which explicitly break the CFC symmetry does not  appear in the holonomy potential. 
Whether the $  \Z_{N_c} $  symmetry is  spontaneously broken or not depends on the  dynamics.  In QCD(F), CFC is unbroken for $\beta> \beta_c \sim \Lambda^{-1}$  and is broken for  $\beta< \beta_c$. Therefore, volume and temperature independence 
hold in QCD(F)  for $\beta> \beta_c \sim \Lambda^{-1}$   provided it is studied via the graded partition function \eqref{graded} or equivalently, $\Omega_F^0$-twisted boundary conditions.

\vspace{3mm}
In QCD(F/adj) with  $N_f= kN_c$, which is a slight  generalization of  Veneziano limit to mixed representation matter,  the situation is very  intriguing. 
At large-$N_c$ limit, the combination of one and two loop potential for QCD(F/adj) acquires a simple  form. 
Remarkably, the one-loop potential actually vanishes. So, in perturbation theory at one-loop order, this class of theories 
has a moduli-space, which allows all possible realizations of center-symmetry.    At two-loop order, the theory prefers a center-symmetric minimum. At $N_c=\infty$, 
\begin{align} 
 V_\text{1-loop} & =  V^{\rm  gauge}_{\rm 1-loop} + V^{\rm \lambda}_{\rm 1-loop}  +  V^{\rm \psi}_{\rm 1-loop, \Omega_F} =0 \cr   \cr
V_{\text{2-loop}, \Omega_F^0} &=   +  x  \frac{(g^2N_c) }{ 8 \pi^4 \beta^4}
	\sum_{n=1}^{\infty}\frac{\big| \text{Tr} (\Omega^n) \big|^2}{n^4}  \,, 
	\label{V12largeN-1} \qquad x= \small{\frac{N_f}{N_c} }
\end{align} 
 First of all, the $  \Z_{N_c} $ CFC  symmetry is stable at large-$N_c$.  (It is in fact stable at any $N_c$). 
 The $N_f$  fundamental Dirac fermions with flavor twisted boundary conditions  \eqref{flavor-hol-0} behaves as if  $(g^2N_c) x $ adjoint Weyl fermions with periodic boundary conditions for the purpose of holonomy potential!

\vspace{3mm}
The two-loop stability of center-symmetry cannot be altered by three or higher loop orders or non-perturbative contributions. 
The three or higher loop appear at order $(g^2N_c)^p, p=2,3, \ldots$ respectively  and at weak coupling, it  cannot alter the implications of two-loop order.\footnote{Once the realization of center-symmetry is determined at $p$ loop order, the 
 $(p+1)$ and higher loop orders as well as non-perturbative corrections cannot alter this result. $(p+1)^{\rm th} $ order is needed if and only if  
$(p)^{\rm th} $ order exhibits degeneracy. In our analysis, there is degeneracy at one-loop order  $p=1$,  hence we need $p=2$. In supersymmetric QFTs, all orders in perturbation theory do not lift  degeneracy, hence one needs non-perturbative terms to determine center-symmetry realization.} 

\vspace{3mm}
One intriguing implication of volume-independence in the Veneziano-type limit is emergent CFC symmetry. 
For finite $N_f, N_c$,  the exact 
CFC symmetry is $\Z_{{\rm gcd}{(N_f, N_c)}}$.  If $N_f$ and $N_c$ are co-prime, formally 
there is no center-symmetry. However, despite being formally correct, this does not reflect the truth sufficiently well. For example, with the center-symmetric $\Omega_F^0$ and in the Veneziano limit, all  single trace terms in the potential of the 
form $\tr (\Omega^{N_f k})$ disappear. Normally, these terms transform non-trivially under $\Z_{N_c}$ and breaks it explicitly to  $\Z_1$. However, the prefactor of these terms is $\frac{1}{N_f^3}$ and they vanish  in the Veneziano large-$N_c$ limit.   In the two-loop expression, the only term that is left is the $\Z_{N_c}$  center-symmetric double-trace term. 
And hence, for co-prime $N_f, N_c$, in the Veneziaono limit, despite the fact that center-symmetry is formally $\Z_1$,   the minimum of the holonomy potential is still at a 
$\Z_{N_c}  $ center-symmetric point.  This is similar to  emergent center-symmetry previously appeared in the context of two-index representation fermions in the large-$N_c$ limit \cite{Armoni:2007kd}.   For fundamental fermions in 't Hooft limit,  it is a trivial result as the fermionic degrees of freedom are suppressed as $\frac{N_fN_c}{N_c^2} \sim \frac{N_f}{N_c}$. For fundamental fermions in the Veneziano  limit with $\Omega_F^0$ twist,  it is a non-trivial  result. Fermion effects on the holonomy potential  is suppressed by quantum distillation with a factor  $\frac{N_c/N_f^3} {N_c^2} \sim \frac{1}{N_c^4}$ as can be deduced from \eqref{V2psi-min}, more so than the loop suppression of the quarks in the 't Hooft limit. 

\vspace{3mm}
The fact that the theory  satisfies volume independence implies that there are very powerful Bose-Fermi, Fermi-Fermi and Bose-Bose cancellation over the state sum. These cancellations follow a very similar pattern to pheonomenological description given in  Sec.~\ref{sec:pheno}, and powerful enough to demolish both  the Hagedorn growth as well as standard Boltzmann growth of the density of states.  As conjectured in Sec.~\ref{sec:pheno},  the 
relative density of states on a curved 3-manifold  grow   similar to the one of 2d QFT. 
 Non-supersymmetric  QCD(F/adj)  theory acts in a similar way to supersymmetric theories on curved spaces \cite{DiPietro:2014bca},  and 
QCD(adj)\cite{Cherman:2018mya}.

 \subsection{Color-flavor-momentum transmutation}
If a gauge theory is compactified on a circle with size $\beta$ in the presence of trivial  gauge holonomy, 
the Kaluza-Klein modes of the periodic  fermion  fields $\psi$ gets  quantized in units of 
$\frac{2\pi}{\beta}$.  The KK-decomposition of the fermions take the form
\begin{align}
\bar \psi_{aj} (n)  \left( \gamma_i \partial_i + \gamma_4 \frac{2\pi n }{\beta}   \right) \psi^{aj} (-n), \qquad j= 1, \ldots, N_c, \qquad 
a= 1, \ldots, N_f,
\label{standard}
\end{align}
where $j$ is color,  $a$ is  flavor  and $n \in \Z$ is Kaluza-Klein momentum index.  Clearly, the  KK-decomposition does not quite care about the color and flavor structure. 
 In contradistinction, if the dynamical  gauge holonomy background $\Omega$ and non-dynamical flavor holonomy  $\Omega_F^0$ are center-symmetric, 
 the fermion kinetic term and KK-decomposition is refined as: 
 \begin{align}
&\bar \psi_{aj}  (n)  \left( \gamma_i \partial_i + \gamma_4  \left( \frac{2\pi j }{\beta N_c}  +  \frac{2\pi a }{\beta N_f} +  \frac{2\pi n }{\beta }  \right)  \right) \psi^{aj} (-n) \cr 
= &\bar \psi_{aj}  (n)  \left( \gamma_i \partial_i + \gamma_4   \frac{2\pi q }{ {\rm lcm}(N_c,N_f) l }    \right) \psi^{aj} (-n) 
\label{CFT}
\end{align}
where 
\begin{align}
q(j, a, n)= \frac{  {\rm lcm}(N_c,N_f)}{N_c}  j  +  \frac{  {\rm lcm}(N_c,N_f)}{N_f}  a +   {\rm lcm}(N_c,N_f) n \in \Z
\end{align}
 and lcm stands for least common multiple. 
 We refer to this effect as  {\it  color-flavor-momentum transmutation}. It  is  a  generalization  of Gross-Kitazawa color-momentum
 transmutation which  incorporates flavor \cite{Gross:1982at}, and has some similarity to Ref.~\cite{Kazakov:1982zr}. However, unlike Ref.~\cite{Kazakov:1982zr, Gross:1982at},  center-symmetry is dynamically stable in our construction in the  $\Omega_F^0$ flavor holonomy background, and volume independence is valid even at arbitrarily small-$\beta $.  
 The KK modes for fermions are now quantized in units of 
 \begin{align}
 \frac{2\pi q }{{\rm lcm}(N_c,N_f) \beta }  
 \end{align} 
 For $N_f=N_c$ theory, this is  quantization of KK-momentum modes  in units of  $\frac{2\pi}{\beta N_c} $, 
 which is an imprint of large-$N_c$ volume independence.  In other words, 
 for quarks, it is as if the effective  space size is $ {\rm lcm}(N_c,N_f) \beta $. For gauge bosons in center-symmetric background, of course, the space size is effectively, $\beta N_c$, and same for the full theory.

 \vspace{3mm}
 In the standard Kaluza-Klein decomposition, infinite volume can only be captured  by taking $\beta  \Lambda \gg  1$, hence there are many KK-modes below the strong scale $\Lambda$, $\frac{1}{\beta } \ll \Lambda$. This is how lattice gauge theory formulated in a finite volume can capture the properties of the QFT on $\R^4$. 
 
  \vspace{3mm}
 With the $\Omega_F^0$ twisted  boundary conditions and  graded partition function, in the $N_c \rightarrow \infty$ limit, regardless of value of $\beta $, 
 the refined KK modes form a continuum as if the theory is on $\R^4$.  In perturbation theory, this is how infinite volume limit is captured at arbitrarily small-$\beta $ in the large-$N_c$ limit, similar to  the discussions in   \cite{Gross:1982at,GonzalezArroyo:1982hz, Unsal:2010qh}.

   \vspace{3mm}
 For example, one can derive the  renormalization group $\beta$ function of the theory on 
$ \R^4$  even by studying with the theory at  $ \R^3 \times S^1$, as Gross and Kitazawa did with the matrix model reduction  \cite{Gross:1982at}. 
 It is easy to show that the reduced 3d theory  produces, to all orders in perturbation theory, the  standard Feynman diagrams for invariant Green functions of the theory on $ \R^4$. Of course, this equivalence is not only restricted to perturbation theory, and all the neutral sector observables must  agree between the reduced theory and the theory on $\R^4$ non-perturbatively. 
  
     \vspace{3mm}
  Lattice simulation of the QCD(F/adj)  is possible \cite{Bergner:2020mwl}.  In particular, the flavor twist $\Omega_F$ does not induce a sign problem in Euclidean path integral formulation. Such a construction may be useful to learn further about non-perturbative properties of the theory and  test adiabatic continuity  (at finite $N_c$) and volume independence  (at large $N_c$)    in the strong coupling domain.


\section{Summary} 

{\bf Color-flavor center symmetry:}
The presence of fundamental matter fields explicitly breaks  one-form center-symmetry $\Z_N^{[1]}$. Thus, one may be tempted to think, as it is commonly accepted, that Polyakov loops  on $\R^3 \times S^1$  can never be 
 genuine order parameters in theories with fundamental matter. 
   In  \cite{Cherman:2017tey},  it was  realized  that 
 a diagonal subgroup of the center of $SU(N_c)$ and a cyclic permutation subgroup of $SU(N_f)_V$,   $(\Z_{{\rm gcd}(N_f, N_c)})_D$, the color-flavor center (CFC) symmetry,  can remain as a true symmetry of the theory, and Polyakov loops can in fact be good order parameters  provided  ${\rm gcd}(N_f, N_c) >1$.   
 
 The existence of CFC symmetry explains the  sharp phase transition observed in  lattice simulations with flavor-twisted background \cite{Iritani:2015ara}.  In this work, we provided an interpretation for the results of \cite{Iritani:2015ara}  from the viewpoint of quantum distillation of Hilbert space. We argued that the distilled Hilbert space of QCD(F) with $\Omega_F^0$ distillation carries characteristic features of the 
 the Hilbert space of pure Yang--Mills theory as described in \eqref{largeE-spec} and \eqref{largeE-spec-2}.
Ref.~\cite{Iritani:2015ara} indeed showed that CFC is spontaneously broken in QCD(F) at small-$\beta $ and restored at large-$\beta $. The reason for the breaking of the CFC in QCD(F) is in essence the same as in Yang--Mills theory. The growth of the density of  flavor singlet hadronic states is powerful enough to induce a phase transition.  

    \vspace{3mm}
     \noindent
{\bf Preservation of the CFC-symmetry at small-$\beta $:}
 The story takes an even more interesting form if we consider QCD(F/adj).  Fundamental fermions with the $\Omega_F^0$ twisted boundary conditions in fact  favor a center-symmetric minimum, just like adjoint fermions with periodic boundary conditions! 
In QCD(F/adj) with $N_f=N_c$ where the center-breaking effect of gauge fluctuation is undone by one adjoint fermion, the ultimate decision is given by fundamental fermion, which 
lead to the stability of  $\Z_{N_c}$ color-flavor center symmetry. 
This opens the prospect of adiabatic continuity in QCD(F/adj)  and QCD(F) between 
$\R^4$ where these theories are strongly coupled and  $\R^3 \times S^1$ where they become weakly coupled and calculable.

       \vspace{3mm}
        \noindent
{\bf Power of  quantum distillation: }
   The Hilbert space distillation in QCD(F/adj)  is induced by the insertion of $(-1)^F \prod_{a=0}^{N_f} e^{\frac{2 \pi i} {N_f}  Q_k} $ in the operator formalism. 
What is left from the Hilbert space of QCD(F/adj)  ${\cal H}_{\rm QCD(F/adj)} $ after all the cancellation  is quite small, the effective density of states after all cancellations grows as the one of 2d QFT in the large-$N_c$ limit, in a similar way to supersymmetric theories on curved spaces \cite{DiPietro:2014bca}. Although the cancellation is milder than what 
  supersymmetric  index achieves on flat space, e.g. 
  $   \rm Distill [{\cal H}_{\N=1 SYM}]= \textrm \{Ground \;  states\} $, it is powerful enough to  avoid  Lee--Yang singularities  and phase transitions in certain cases,  providing generalized partition functions which are smooth functions of $\beta$. 

     \vspace{3mm}
        \noindent
{\bf Chiral symmetry breaking by monopole-operators  at weak coupling on  $\R^3 \times S^1$:} Perfect quantum distillation tells us that the graded partition function 
is saturated by only a few states, so that as one interpolates between large $S^1$ and small $S^1$,
the ground state  may remain  adiabatically connected. 
In  QCD(F/adj) with one heavy adjoint fermion and  $N_f = N_c$ massless fundamental fermions in the CFC symmetric regime, the continuous chiral symmetry must be  spontaneously broken even at weak coupling \cite{Cherman:2016hcd}.   The most interesting and unconventional outcome of this analysis is that the  chiral field $\Sigma(x)  $ that appears in the chiral Lagrangian is the collection of the monopole-flux operators  
$ \Sigma(x) = \rm Diag \left(   e^{i \alpha_1\cdot  \sigma}, e^{i \alpha_2 \cdot  \sigma}, \ldots,  e^{i \alpha_{N_f}\cdot  \sigma} \right)$, where $\sigma$ is the dual photon associated with gauge fluctuations.   Due to  Nye--Singer index theorem for Dirac operators in monopole-backgrounds 
\cite{Nye:2000eg, Poppitz:2008hr},  and the fact that the Cartan subgroup of axial chiral symmetry is non-anomalous,   the dual photons must and does acquire  a chiral charge. 
The condensation of $\Sigma$, rather than condensation of the fermion bilinear, leads to chiral symmetry breaking in a weak coupling regime of QCD(F), 
and dual photons acquire an interpretation as Nambu--Goldstone bosons. In a given superselection sector, determined by the VEV of monopole operators, we can calculate the expectation value of the fermion bilinear condensate. Remarkably, it is  $\langle  \psi_R \psi_L  \rangle = \Lambda_{\rm QCD}^3 $, and the strong scale emerges naturally as described around \eqref{chiral-cond}, almost exactly  as in $\N=1$ SYM. 
The condensation of the  monopole-flux operators induce a chiral symmetry breaking mass term for massless fermions. 

       \vspace{3mm}
        \noindent
{\bf Persistent anomaly upon compactification on  $\R^3 \times S^1$:} The mixed anomaly which dictates possible ground state structures on $\R^4$  persists upon compactification on  $\R^3 \times S^1$ provided ${\rm gcd}(N_f, N_c)  >1 $  and  an appropriate $\Omega_F$  twisted background for 
$SU(N_f)_V$ is used.   Phase transitions exist between various anomaly respecting phases. Remarkably,  a number of these phase transitions are calculable on $\R^3 \times S^1$. 

       \vspace{3mm}
        \noindent
{\bf Adiabatic continuity:} We showed that  $N_f=N_c$ QCD(F) with one massive adjoint fermion on small   $\R^3 \times S^1$ does not break CFC and the Polyakov loop expectation value is strictly zero.  But it does break continuous chiral symmetry \eqref{MAG} at weak coupling. This is the  expected behavior of this theory on large $\R^3 \times S^1$ at strong coupling. 
It seems highly plausible that these two regimes are continuously connected, but we cannot prove this statement. 
The best we can do is to prove that the possible ground state structures are controlled by the same mixed  't Hooft anomalies, given in \eqref{poly} and \eqref{compano}, which is still a remarkable persistent order, but not a proof of adiabatic continuity.

       \vspace{3mm}
  \noindent  
 {\bf Summary of general construction:}
  \begin{align}
  { \scriptsize
\begin{tabular}{ |l|l|} 
 \hline
Hamiltonian {\bf H } &  Hamiltonian {\bf H } + \; grading \; operator   \cr 
\hline 
Hilbert space ${\cal H}$ &  Distill[${\cal H}$] \cr   
\hline
Thermal state sum over ${\cal H}$  & Graded state sum  over ${\cal H}$ \cr 
${\cal Z}(\beta )=\tr \Big[ e^{-\beta  H}  \Big]   $ & ${\cal Z}_{\Omega_0}(\beta)= \tr \Big[ e^{-\beta H} (-1)^F   e^{i \pi Q_0}     e^{-i \frac{\pi}{N_f}  Q_0}    \prod_{a=1}^{N_f} e^{i \frac{2\pi a}{N_f}  Q_a}  \Big],  $ \cr 
\hline \hline 
Path integral with thermal b.c. & Path integral with $(-1)^F$ and $\Omega_F^0$  flavor twisted b.c.  \cr 
Gauge-holonomy potential $V[\Omega]$ &  Gauge-holonomy potential   in the presence of flavor holonomy $V_{\Omega_F}[\Omega]$  \cr
\hline  \hline  
Mixed anomaly on $\R^4$ between $SU(N_f)/\Z_{N_c}$ and $(\Z_{2 N_f})_A$ & Mixed anomaly on $\R^4$ is  persistent  on $\R^3 \times S^1$.  \cr 
Does not persists   on $\R^3 \times S^1$ with thermal compactification.  & Does persists   on $\R^3 \times S^1$ with  
$\Omega_F^0$ compactification.  \cr
\hline  \hline 
 Thermodynamics: Thermodynamic worth of ${\cal H}$  & Graded Thermodynamics: 
 Thermodynamic worth of Distill[${\cal H}$] 
  \cr
Free energy  & Graded free energy (or twist free-energy or flavor-holonomy potential)  \cr 
Pressure & Graded Pressure \cr
\hline
All incalculable phase transitions &  
Adibatic continuity   and/or persistent order \cr 
No anomaly constraint on phase transitions & Many calculable  phase transitions, mixed anomaly respected
\cr \hline
$\rho(E)$,  density of hadronic states &  $\rho_{\rm distill}(E)$, density of hadronic states  corresponding to    Distill[${\cal H}$]  \cr  
\hline
Volume dependence at large-$N_c$ &  Volume independence at   large-$N_c$  \cr   
\hline
\hline
\end{tabular} 
\label{summary} }
\end{align}

  \noindent  
 {\bf Remarks}
 
         \vspace{3mm}
          \noindent
{\bf Not all   sign problems are bad.} The insertion of the operator $ \prod_{a=0}^{N_f} e^{\frac{2 \pi i} {N_f}  Q_k}   $   
into the trace  induces a ``sign problem'' in the state sum based on the Hamiltonian formulation.  In other words, quantum distillation is actually a sign problem, albeit a good one! 
The tremendous cancellations in the graded state sum are a consequence of destructive interference due to concerted phases attached to physical states. Interestingly, these phases do not induce a sign problem in the 
Euclidean path integral formulation.\footnote{The situation is reversed for the inclusion of a chemical potential, which does not induce a sign problem in the Hamiltonian formulation but does induce one in the path integral formulation.} The sign problem in the state sum over Hilbert space may lead 
to sufficient cancellations which in turn lead to the absence of phase transitions as $\beta$ is dialed.

          \vspace{3mm}
           \noindent
{\bf SQCD vs. QCD(F/adj):} Consider $N_f=N_c$  SQCD with a soft mass for the fundamental scalar.    Ref.~\cite{Aharony:1995zh} showed that with a soft mass $m_{q_a}$, the chiral symmetry breaking pattern \eqref{pattern2} holds where the $U(1)_{A_D}$ part of the chiral symmetry is unbroken.  It is not known if this phase persists in the decoupling limit 
$m_{q_a} =\infty$, where SQCD reduce to QCD(F/adj). 
Remarkably, in QCD(F/adj) with $m_\lambda=0$ we demonstrated at small $\R^3 \times S^1$ that the chiral symmetry breaking pattern that takes place is same as in supersymmetric theory with soft SUSY breaking term  \eqref{pattern2}. Therefore, it seems very likely that provided 
\begin{align}
{\rm Max} (\beta \Lambda, m_{q_a} \Lambda^{-1}) \ll 1 
\label{domain}
\end{align}
there will be adiabatic continuity between SQCD and QCD(F/adj).  In particular, the $m_{q_a} \Lambda^{-1}\ll 1, \beta  \Lambda=\infty$ limit seems to be adiabatically connected to the $m_{q_a} \Lambda^{-1}=\infty, \beta  \Lambda \ll 1 $ regime, as depicted in Fig.~\ref{fig:continuity}. 
Turning on a small mass for adjoint fermion,  the continuity may persist beyond the domain \eqref{domain} and to the whole $(m_{q_a}, \beta)$ plane.

   \vspace{3mm} 
    \noindent
{\bf Possibility of  dualities and interesting IR behaviors in QCD(F/adj):}
We have shown on $\R^4$  that all the interesting mixed anomalies of SQCD for $N_f=N_c$ theory are also satisfied 
by QCD(F/adj). Of course, this is not an accident. What enters into the traditional 't Hooft anomalies \cite{tHooft:1979rat} of SQCD are just the currents of global symmetries that act on fermions. And the microscopic fermionic content and associated currents of these theories are the same. Hence, the  UV anomalies of SQCD and QCD(F/adj) are the same for any suitable  $N_f$.  
The interesting thing is  in  the IR,  Fradkin-Shenker complementarity \cite{Fradkin:1978dv} and the fact that the existence of just one adjoint fermion changes the story of QCD(F/adj) dramatically. 
Thanks to complementarity  \eqref{FS-comp} which substitutes elementary scalars with adjoint/fundamental  fermion bilinears with identical   local and global symmetry quantum numbers, 
we can express all the fermionic components of composite mesons and baryons in the same way as in SQCD,  mimicking   Seiberg's analysis \cite{Seiberg:1994bz, Seiberg:1994pq}. Therefore, we believe that non-trivial anomaly matching that occurs in the context of SQCD (either in dual formulations or as interesting IR-behaviors)  must  have an image in QCD(F/adj).

    \vspace{3mm} 
 It is becoming clear that QCD(F/adj) is a class of non-supersymmetric theories that is intermediate between QCD and supersymmetric QCD. 
 It is in this   unfamiliar realm that  we find new possibilities.

 \acknowledgments  I am  thankful to  O. Aharony, P.~Argyres, P.~Draper,  M. Nguyen,   D. Gross,  A. Behtash, T.~Sulejmanpasic, A. Cherman, Y. Tanizaki, 
 Y. Guo, and S. Gukov  for discussions.  I am especially grateful to    T. Kanazawa    for discussions,  collaboration in an earlier related work,    his  help with some of the  figures, and numerical extremization of the the  holonomy potential.   
 This work  was supported by the
U.S. Department of Energy, Office of Science, Office of Nuclear Physics under Award Number DE-FG02-03ER41260.

\appendix 

\section{Elementary  examples of symmetry graded state sums}
First, we provide two  simple examples of symmetry graded state  sums and cancellations  in  quantum mechanics. These example are  presented for pedagogical reasons. Then, we will use exactly the same symmetry grading in  non-trivial  asymptotically free quantum field theories in 2d, a vector model $\mathbb {CP}^{N-1}$ and a matrix model, the principle chiral model.  In the bulk of this paper, we implemented these types of symmetry grading to QCD(F/adj) in 4d. 

\noindent
\subsection{$N$-dimensional harmonic oscillator}
Consider $N$-dimensional simple harmonic oscillator with Hamiltonian ${H}=\sum_{j=1}^{N} \omega \widehat{a}^{\dagger}_j \widehat{a}_j  $. The global symmetry is $U(N)$.  States are totally symmetric representations of $SU(N)$. 
 The partition function of this system is 
\begin{align}
{\cal Z}(\beta) = \tr (e^{-\beta H})
= \sum_{k=0}^{\infty}  {\rm deg}(k)
\mathrm{e}^{-\beta \omega  k},  \qquad {\rm deg}(k) =  { N+k-1 \choose k}
\label{partitionf} 
\end{align}
where the degeneracy increases extremely rapidly.

 \vspace{3mm} 
Now, construct the symmetry graded state sum   
\begin{align}
 {\cal Z}_{\Omega_F^0} (\beta) = \tr (e^{-\beta H} 
\prod_{j=1}^{N} e^{ { i {2\pi\over N}j    \widehat Q_j} } )
\end{align}
    where $Q_j = \widehat{a}^{\dagger}_j \widehat{a}_j $ is the number operator for the $j^{\rm th}$ oscillator.   In the graded  state sum,  
  many degenerate states cancel among themselves due to phases attached to them.    After cancellations, graded degeneracy factors reduce into 
\begin{align}
{\rm deg}(k)  \mapsto  \left\{ \begin{array}{ll}
  1 & \qquad k=0 \;\;  {\rm mod}  \; N  \cr 
0 & \qquad   k=0 \;\;  {\rm mod} \; N 
\end{array} \right.
\end{align}
instead of the growth \eqref{partitionf}. As a result, the graded partition function of the $N$-dimensional oscillator with frequency $\omega$ is equivalent to thermal  partition function  of a 1-dimensional oscillator   with frequency $N\omega$ (or 1-dimensional oscillator with frequency $\omega$ but inverse temperature $\beta N$)
\begin{align}
{\cal Z}_{\Omega_F^0} (\beta) =  \prod_{j=1}^{N}  \left( \frac{1}{1-e^{ -\beta  \omega +   i {2\pi\over N}j  } } \right) = \frac{1}{1-e^{ -\beta N \omega} }
\end{align} 
This indicates dramatic reduction in the density of states. In particular, in the $N\rightarrow \infty$ limit, 
\begin{align}
\lim_{N  \rightarrow \infty} {\cal Z}_{\Omega_F^0} (\beta) =1
\label{gss}
\end{align}
indicating that only the ground state contributes to the state sum. 

 \vspace{3mm} 
At this stage, the graded state sum in purely bosonic theory achieves something quite remarkable, similar to what supersymmetric index achieves in supersymmetric quantum mechanics \cite{Witten:1982df}. 
 The distilled Hilbert space is just the ground state 
of the bosonic $N\rightarrow \infty$ dimensional oscillator and only one-state contributes to the state sum.

 \vspace{3mm} 
The reader may think that this is a trivial non-interacting system in quantum mechanics. Remarkably, the same phenomena does  occur in the non-trivial asymptotically free QFT in 2d and higher dimensions. 

 \vspace{3mm} 
\noindent
\subsection{Hydrogen atom}
Consider the hydrogen atom and ignore the spin of the electron.  The discrete part of the spectrum is 
$E_{n} = - E_1/n^2$ and the degeneracy factor is ${\rm deg}(n)=n^2$.  The simplest way to understand the degeneracy is to 
note that the global symmetry of the hydrogen atom is  
\begin{align}
G= \frac{SU(2) \times SU(2)}{\Z_2}
\end{align}
which are ultimately related to  angular momentum and Laplace-Runge-Lenz vector. 
 The states  in the Hilbert space fill   the bi-product of spin-$j$ irreducible representation with its conjugate,  where  $j =  (n-1)/2, n=1,2, \ldots $  and the degeneracy factor is given by 
 \begin{align}
 j \otimes  j  \in G \qquad \qquad \deg[(j \otimes j ) ]= (2j+1)^2=n^2
\end{align}
%
Now, we can construct a symmetry graded state sum by the insertion of the operator $e^{i \epsilon (J_{1z} + J_{2z} )}$ into the trace. This modifies  the partition function as 
\begin{align}
 \sum_{n=1}^{\infty}  {\rm deg}(n)
\mathrm{e}^{-\beta E_n}   \Rightarrow  \sum_{n=1}^{\infty}  (\chi_{\frac{n-1}{2}}(\epsilon))^2
\mathrm{e}^{-\beta E_n} 
\label{partitionfH} 
\end{align}
where  $\chi_{j}(\epsilon) = \frac{ \sin \frac{(2j+1) \epsilon}{2}}{\sin \frac{\epsilon}{2}}$.  The best quantum distillation is achieved at $\epsilon=\pi$ and hence, a useful object to consider is  $\tr (e^{-\beta H}  e^{i \pi (J_{1z} + J_{2z} )})$, 
leading to modification in the state sum 
\begin{align}
{\rm deg}(n)=n^2   \mapsto \left\{ \begin{array}{ll}
  1 & \qquad n=1, 3, 5,  \ldots    \cr 
0 & \qquad   n=2,4,6, \ldots  
\end{array} \right.
\end{align}
The  state sum is modified as 
\begin{align}
 \sum_{n=1}^{\infty} n^2  \; 
\mathrm{e}^{-\beta E_n} \mapsto   \sum_{n=1,3, \ldots }^{\infty} 1  \; 
\mathrm{e}^{-\beta E_n} 
\end{align}

 \vspace{3mm} 
\subsection{${\mathbb {CP}}^{N-1}$ model in 2d} 
\label{app3}

The global symmetry of the  ${\mathbb {CP}}^{N-1}$  model is 
\begin{align}
G=PSU(N)= \frac{SU(N)}{\Z_N}.
\end{align}
This is because $\Z_N$ is part of the $U(1)$ gauge redundancy of the theory and $G$
is the symmetry that acts faithfully in the Hilbert space $\cal H$.  
This means   only  the representations of  $SU(N)$ that also happen to be representations of 
$PSU(N)$ appear in the Hilbert space. This amounts to adjoints and products thereof and singlets.  If we denote the elementary  ${\mathbb {CP}}^{N-1}$  field as $z_a(x)$, the states in the Hilbert space are generated by operators such as 
\begin{align}
\bar z^a(x) z_b(x), \qquad    \bar z^b(x) e^{i \int_{x}^{y} a } z_b(y), \ldots 
\end{align}
which are faithful realizations of the symmetry. 

 \vspace{3mm} 
In the large-$N$ limit,  with the use of regular thermal partition function,  and periodic boundary conditions, 
\begin{align}
z_a(x_1, x_2+ \beta)= z_a(x_1, x_2) 
\end{align}
 ${\mathbb {CP}}^{N-1}$  is known to have a zero temperature $(\beta=\infty)$ phase transition shown by Affleck \cite{Affleck:1979gy}.  With the $\Omega_F^0$ twisted boundary condition  using flavor rotation  \eqref{flavor-hol-1} 
\begin{align}
z_a(x_1, x_2+ \beta)= \Omega^0_{ab} z_b(x_1, x_2) 
\end{align}
 in path formulation, the story is opposite.  There are no phase transition at any finite $\beta$.   The theory satisfies volume independence at $N=\infty$ and adiabatic continuity at finite-$N$ \cite{Dunne:2012zk}.  Ref.~\cite{Sulejmanpasic:2016llc} interpreted this result of the path integral formalism in terms of state sums.  The Hilbert space interpretation of the twisted boundary condition is  analogous with  what we have written down for $N$-dimensional oscillator, ${\cal Z}_{\Omega_F^0} (\beta) = \tr (e^{-\beta H} 
\prod_{j=1}^{N} e^{ { i {2\pi\over N}j    \widehat Q_j} } )$  where now ${\widehat Q_j}$ are charges associated with Cartan subgroup of $G$.  The implication of this insertion is that it assigns phases $e^{i \frac{2\pi (a-b)}{N}}$ to adjoint representation states, and modifies the state sum as  $(N^2-1)e^{-\beta E_{\rm adj}} \rightarrow (-1)  e^{-\beta E_{\rm adj}} $. 
At large-$N$ limit,   since the singlet  $\bar z^a(x) e^{i \int_{x}^{y} a } z_a(y)$ is  degenerate with the adjoint  \cite{Witten:1978bc},   the symmetry graded state sum leads to cancellation among the lightest $N^2$ physical particles in the spectrum.   
\begin{align}
(N^2-1)e^{-\beta E_{\rm adj}}  + 1 e^{-\beta E_{\rm sing.}} \rightarrow (-1)  e^{-\beta E_{\rm adj}}   +    e^{-\beta (E_{\rm adj} +O(1/N^2))}  \underbrace{\rightarrow}_{N \rightarrow \infty}  0  
\end{align}
This process continues for the other states as well, and in the large-$N$ limit, the only states that contribute to the graded state sum is the ground state.

 \vspace{3mm} 
In Affleck's analysis of partition function, it is the large multiplicity of the first excited state  $N^2 e^{-\beta E_{\rm adj}} $ that forces the system to a phase transition immediately at $\beta  =\infty$  \cite{Affleck:1979gy}.  In our graded partition function, the fact that the first excited state contribute $0 \times  e^{-\beta E_{\rm adj}} $ (similarly for higher states)  is the reason that there is no phase transition all the way down to  $\beta  =0$\cite{Dunne:2012zk}.
 In contrast to  simple $N$-dimensional oscillator example,  this is a  non-trivial strongly coupled quantum field theory and yet, 
\begin{align}
{\rm Distilled}[\cal H]= \{\rm ground \; state(s)\} 
\label{DHS}
\end{align}
The quantum distillation  in this purely bosonic QFT achieves  what   the supersymmetric index achieves in supersymmetric QFTs. This is ultimately the reason for the adiabatic continuity and volume independence observed in Ref.\cite{Dunne:2012zk}.  In particular, in the $N \rightarrow \infty$ limit, only the ground state survives in  the graded  partition function for $\theta\neq \pi$ and two-ground states survive  for $\theta= \pi$:
 \begin{align} 
&\lim_{N \rightarrow \infty} {\cal Z}_{\Omega^0} (\beta, \theta)=1 e^{-\beta E_0(\theta)},  \qquad \theta\neq \pi  \cr 
&\lim_{N \rightarrow \infty} {\cal Z}_{\Omega^0} (\beta, \theta=\pi)=2 e^{-\beta E_0(\theta=\pi)}, 
\end{align}

 \vspace{3mm} 
\noindent
\subsection{Principle chiral model  in 2d} 
\label{app4}
The bosonic PCM is an asymptotically free matrix field theory in $d=2$ dimensions. 
Let $U(x)$ denote the principal chiral field. The global symmetry of the theory is 
\begin{align}
G=\frac{SU(N)_L \times SU(N)_R}{\Z_N}
\end{align}
 The states  in the Hilbert space fill   the bi-product of $k$-index antisymmetric irreducible representations with its conjugate,  and the degeneracy factor is. given by 
 \begin{align}
 V_k \otimes \bar V_k  \in G \qquad \qquad \deg[(k, \bar k) ]= \left[{N \choose k}\right]^2
\end{align}
Imposing 
\begin{align}
U(x_1, x_2+ \beta)=\Omega^0 U(x_1, x_2) \bar \Omega^0
\end{align}
 boundary conditions in path integral correspond to 
${\cal Z}_{\Omega^0} (\beta) = \tr (e^{-\beta H}  \prod_{j=1}^{N} e^{ { i {2\pi\over N}j    \widehat Q_j} } )  $  in path integral where   ${\widehat Q_j}$ are charges associated with Cartan subgroup of the vector-like symmetry subgroup $SU(N)_V$.  Clearly, this assignment  forces  the contributions of all excited states with  $k \neq 0 \mod N $ to vanish, i.e., in the state sum, we have the replacement 
\begin{align}
\deg[(k, \bar k) ] \Rightarrow 0, \qquad  \forall k \neq 0 \mod N 
\end{align}
In the large-$N$ limit, it is again only the ground state that contributes to the state sums \eqref{DHS}.  This is again the reason for volume independence for $N=\infty$ limit  and adiabatic continuity for finite $N$ \cite{Cherman:2014ofa}.

\bibliographystyle{JHEP}
\bibliography{small_circle}
\end{document}